# Antiferromagnetic spintronics

V. Baltz[1,a], A. Manchon[2,b], M. Tsoi[3], T. Moriyama[4], T. Ono[4], and Y. Tserkovnyak[5]

[1]SPINTEC, Univ. Grenoble Alpes / CNRS / INAC-CEA, F-38000 Grenoble, France
[2]King Abdullah University of Science and Technology (KAUST), Physical Science and Engineering Division (PSE), Thuwal 23955-6900, Saudi Arabia
[3]Department of Physics, The University of Texas at Austin, Austin, Texas 78712-0264, USA
[4]Institute for Chemical Research, Kyoto University, 611-0011 Uji, Kyoto, Japan
[5]Department of Physics and Astronomy, University of California, Los Angeles, California 90095, USA

[a] vincent.baltz@cea.fr
[b] aurelien.manchon@kaust.edu.sa

Antiferromagnetic materials could represent the future of spintronic applications thanks to the numerous interesting features they combine: they are robust against perturbation due to magnetic fields, produce no stray fields, display ultrafast dynamics and are capable of generating large magneto-transport effects. Intense research efforts over the past decade have been invested in unraveling spin transport properties in antiferromagnetic materials. Whether spin transport can be used to drive the antiferromagnetic order and how subsequent variations can be detected are some of the thrilling challenges currently being addressed. Antiferromagnetic spintronics started out with studies on spin transfer, and has undergone a definite revival in the last few years with the publication of pioneering articles on the use of spin-orbit interactions in antiferromagnets. This paradigm shift offers possibilities for radically new concepts for spin manipulation in electronics. Central to these endeavors are the need for predictive models, relevant disruptive materials and new experimental designs. This paper reviews the most prominent spintronic effects described based on theoretical and experimental analysis of antiferromagnetic materials. It also details some of the remaining bottlenecks and suggests possible avenues for future research. Our review covers both spin-transfer-related effects, such as spin transfer torque, spin penetration length, domain wall motion and 'magnetization' dynamics; and spin-orbit related phenomena, such as (tunnel) anisotropic magnetoresistance, spin Hall and inverse spin galvanic effects. Effects related to spin-caloritronics, such as the spin Seebeck effect, are linked to the transport of magnons in antiferromagnets. The propagation of spin waves and spin superfluids in antiferromagnets are also covered.

# CONTENTS

# I.    INTRODUCTION

## A.    Current challenges

In the field of spintronics much effort is being deployed to reduce device power consumption and scale (Duine 2011; Sinova and Žutić 2012). Antiferromagnetic materials have great potential in this regard, which makes them outstanding candidates for the next generation of spintronic applications. Ultimately, antiferromagnets could replace ferromagnets as the active spin-dependent element on which spintronic devices are based. Antiferromagnetic materials, through their robustness against perturbation due to magnetic fields, absence of production of parasitic stray fields, ultrafast dynamics and generation of large magneto-transport effects, have a number of interesting properties. For instance, synthetic antiferromagnets [i.e., two ferromagnets coupled antiparallel usually by Ruderman-Kittel-Kasuya-Yoshida interactions (Parkin 1991)] are currently used to overcome device malfunction associated with ferromagnetic stray fields when lateral dimensions are reduced (e.g. crosstalk in magnetic random access memories: mutual influence of neighboring cells which are supposed to be isolated from one another). However synthetic antiferromagnets never entirely compensate, and small, but non-zero stray fields persist. With antiferromagnetic materials, the net compensation is intrinsic except for a very small proportion at the interface. To build a functional '(ferro)magnet-free' device, it is first necessary to determine whether and how spin transport can be used to *write* the antiferromagnetic order and *read* subsequent variations through the development of predictive models, relevant disruptive materials and new experimental designs. Several teams at different points in the globe are already studying the theoretical and experimental aspects of the subject. To begin with, we review the antiferromagnetic materials suitable for antiferromagnetic spintronics and their fundamental properties. **S**elected current topics are then dealt with in different sections, depending on whether they relate to spin transfer electronics, spin-orbitronics or spin-caloritronics. The contributions of these different sub-fields to solutions to write (through spin torque) and read (via magnetoresistance) the antiferromagnetic order is also discussed. Interested readers are encouraged to complement their knowledge on specific points by consulting focused reviews, e. g. (Haney et al. 2008; MacDonald and Tsoi 2011; H. V. Gomonay and Loktev 2014; Jungwirth et al. 2016; Sklenar et al. 2016; Fina and Marti 2017).

## B.    Materials survey

In the context of ferromagnetic/antiferromagnetic exchange bias magnetic interactions (see section I.C.3), antiferromagnetic materials have been the subject of intense research for no less than 60 years. In this framework, several authors (Berkowitz and Takano 1999; Nogués and Schuller 1999; Umetsu et al. 2003; Coey 2009; Máca et al. 2012) extensively reviewed the properties of antiferromagnetic materials focusing on composition, atomic structure, spin structure, stoichiometry range and critical temperatures. While the antiferromagnetic properties required for ferromagnetic/antiferromagnetic exchange bias are reasonably well established, those for antiferromagnetic spintronics are still under investigation. In this paragraph, we briefly discuss some antiferromagnetic materials and their properties from the perspective of spin transport (TABLE 1-3). The spintronic effects identified will be further detailed throughout the review. The materials listed in the tables are split into three categories depending on their metallic, insulating or semiconducting/semimetallic nature. Given their wide variety, we restricted our tables to selected materials either because they have a bulk Néel temperature above room temperature or because they played a key role in the development of antiferromagnetic spintronics. TABLE 1-3 illustrate the vast numbers of materials available, while also listing the corresponding crystallographic and spin structures. This multitude of options opens numerous pathways for the investigation of spintronics with antiferromagnetic materials.

## 1. Metals

Metallic antiferromagnets (TABLE 1) comprise Mn-based alloys such as IrMn, FeMn, and PtMn. These types of materials are most often produced by sputter deposition, and are by far the most widely used in industrial applications (mostly for exchange bias). Examples of applications include read heads in hard disk drives and magnetic memories. The possibility to switch from an alloy containing a light element (like Fe) to heavier elements (like Ir and Pt) is essential for the exploration of spin transfer effects [e. g. spin penetration length (Acharyya et al. 2010; Acharyya et al. 2011; Merodio, Ghosh, et al. 2014) (see II.B), magnetic order manipulation by spin transfer torque (Wei et al. 2007) (see II.A.2), giant magnetoresistance (Y. Y. Wang, Song, Wang, Miao, et al. 2014) (see II.C) and enhanced spin pumping near the Néel temperature (Frangou et al. 2016) (see I.C.2)], spin-orbit effects [e. g. the inverse spin Hall effect (Mendes et al. 2014; W. Zhang et al. 2014) (see III.C.2.)], and their subsequent use for the deterministic reversal of ferromagnets by spin Hall torque [(Fukami et al. 2016; Lau et al. 2016; Brink et al. 2016; Oh et al. 2016) (see III.D.4.)]. In addition, the $3d$ shell of the Mn transition metal offers large spontaneous moments, while the $5d$ shell of the noble metals (Pt in PtMn, Ir in IrMn and Au in $Mn_2Au$) provides large spin-orbit coupling. This combination is ideal for handling strong magnetic anisotropy phenomena [e. g. tunnel anisotropic magnetoresistance (Shick et al. 2010; B. G. Park et al. 2011; Y. Y. Wang et al. 2012), (see III.B.) and anisotropic magnetoresistance (Galceran et al. 2016; H.-C. Wu et al. 2016), (see III.A.)]. Moreover, the non-collinearity of the spin texture, found for example in γ-FeMn (Shindou and Nagaosa 2001) and $IrMn_3$ (H. Chen, Niu, and MacDonald 2014), breaks the invariance under the combination of time-reversal symmetry with a crystal symmetry operation resulting in a finite anomalous Hall effect (see III.C.1.). In the specific case of $Mn_2Au$, inverse spin galvanic effects have been predicted, with current-induced staggered spin accumulation matching the staggered spin texture (Železný et al. 2014) (see III.D.2.). Metallic antiferromagnets also comprise archetypal materials such as Cr with an intriguing spin density wave configuration (Fawcett 1988). This material has been thoroughly studied in Cr/MgO-based multilayers (Leroy et al. 2013) (see II.C). Metallic metamagnets with an antiferromagnetic to ferromagnetic transition (like FeRh) offer the possibility to indirectly operate the antiferromagnetic order via iterative steps consisting in manipulating the ferromagnetic order and undergoing the magnetic phase transition. In the case of FeRh, a small (non-crystalline) anisotropic magnetoresistance was also detected in the antiferromagnetic phase and quasistatic write-read operations were demonstrated (Marti et al. 2014; Moriyama, Matsuzaki, et al. 2015; Clarkson et al. 2016). Other important materials include, for example, Gd alloys, such as GdSi, GdGe, and $GdAu_2$ (Tung et al. 1996). These alloys offer the possibility to exploit rare-earth-based antiferromagnetic metals; while combination with a heavy noble metal in the case of $GdAu_2$ will potentially produce interesting spin-orbit properties. Other materials may also be interesting to consider for future fundamental studies on antiferromagnetic spintronics, such as TiAu (Svanidze et al. 2015), an itinerant antiferromagnet without magnetic constituents, or $CrB_2$ (Brasse et al. 2013) which potentially combines antiferromagnetic spintronics and superconductivity.

| Electrical category | Crystallographic and spin structures (bulk) | Material | $T_{N,bulk}$ (K) | $T_{N,Finite Size}$ (K) |
|---|---|---|---|---|
| | 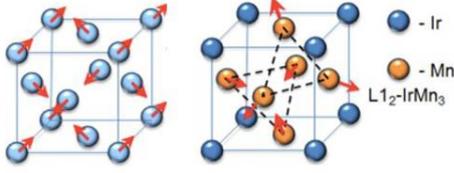 γ-phase = disordered L1$_2$ From (A. Kohn et al. 2013). | Ir$_{20}$Mn$_{80}$ | 690 | (Frangou et al. 2016; Petti et al. 2013) |
| | | Fe$_{50}$Mn$_{50}$ | 490 | n/a |
| | 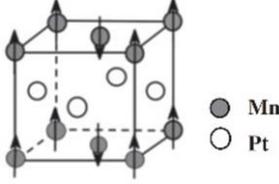 S$_{Mn}$ ∥ c (e.g. PtMn) or a (e.g. NiMn) From (Umetsu, Fukamichi, and Sakuma 2006). | Ir$_{50}$Mn$_{50}$ | 1150 | n/a |
| | | Ni$_{50}$Mn$_{50}$ | 1070 | n/a |
| | | Pt$_{50}$Mn$_{50}$ | 970 | n/a |
| | | Pd$_{50}$Mn$_{50}$ | 810 | n/a |
| Metal | 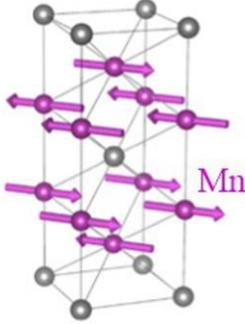 From (Barthem et al. 2013). | Mn$_2$Au | ~ 1500 | n/a |
| | 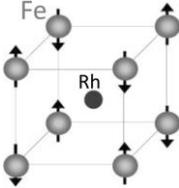 From (Ju et al. 2004). | Fe$_{50}$Rh$_{50}$ | 380[*] | (G. C. Han et al. 2013; Saidl et al. 2016) |

TABLE 1. Crystallographic structure, spin structure and Néel temperature ($T_N$, antiferromagnetic to paramagnetic transition) of some metallic antiferromagnets. The last column indicates whether studies investigated finite size effects on $T_N$ ($T_{N, Finite Size}$). [*]Data for Fe$_{50}$Rh$_{50}$ are for the specific antiferromagnetic to ferromagnetic transition.

## 2. Insulators

Insulating antiferromagnets (TABLE 2) are mostly oxides. They are ideal for studying magnonic effects [e.g. the propagation of spin waves in NiO (Hahn et al. 2014; H. Wang et al. 2014; Moriyama, Takei, et al. 2015), (see II.2 and IV.B.)] and subsequent caloritronic effects [e.g. spin Seebeck in $Cr_2O_3$ (Seki et al. 2015) (see IV.B.)]. Optical excitation of spin waves in antiferromagnets is briefly discussed in I.C.3. Tunnel anisotropic magnetoresistance (III.B.) based on antiferromagnetic CoO barriers was also demonstrated (K. Wang et al. 2015). In addition, some specific antiferromagnetic insulators show interesting multiferroicity and magnetoelectric effects (Binek and Doudin 2005; Martin et al. 2008). Typical mechanisms involve direct coupling between ferroic orders, like in perovskite $BiFeO_3$, where the ferroelectric and antiferromagnetic orders co-exist (Sando, Barthélémy, and Bibes 2014). Alternatively, the magnetoelectric effect in the antiferromagnet gives rise to electrically-induced interface magnetization which can couple to an adjacent ferromagnetic film, such as with $Cr_2O_3$/[Co/Pd] multilayers (He et al. 2010). Other perovskite antiferromagnets have been explored and may be promising materials for antiferromagnetic spintronics. For example, antiferromagnetic order can be excited in $LaMnO_3$ and $La_2CuO_4$ (Coldea et al. 2001) (I.C.3), which could be of interest when seeking to generate spin currents from antiferromagnets (II.A.4). $La_2CuO_4$ cuprate also exhibits antiferromagnetism and traces of superconductivity, which opens appealing perspectives for superconducting antiferromagnetic spintronics. Rare earth orthoferrites like $TmFeO_3$ also exhibit a (distorted) perovskite structure and canted antiferromagnetism (i.e. weak ferromagnetism). They are known to display a strong temperature-dependent anisotropy. Optical manipulation of the antiferromagnetic order in the TeraHertz range was demonstrated (Kimel et al. 2004). Finally, the properties of antiferromagnetic spinel like $ZnCr_2O_4$ and garnets have been thoroughly investigated (Belov and Sokolov 1977). In the future, combinations of optics and antiferromagnetic spintronics may produce interesting results.

Among insulators, halides like $CuCl_2$, $FeCl_2$, $MnF_2$ and $FeF_2$ deserve some mention. Halide antiferromagnets have already proven very useful for the investigation of antiferromagnetic spintronics. In particular, $MnF_2$ (Jacobs 1961) and $FeCl_2$ (Jacobs and Lawrence 1967) are typical examples of antiferromagnets displaying spin-flop and spin-flip transitions, respectively. Mechanisms related to antiferromagnetic order, like resonant mode degeneracy, can be switched on and off thanks to these transitions (Hagiwara et al. 1999). They were also recently used to demonstrate the antiferromagnetic spin Seebeck effect (S. M. Wu et al. 2016; Seki et al. 2015; Rezende, Rodríguez-Suárez, and Azevedo 2016a). Mixed halides such as $KNiF_3$ perovskite are now attracting attention for optical manipulation of the antiferromagnetic order in the TeraHertz range (Bossini et al. 2015).

These examples are only an illustration of the vast complexity of antiferromagnets and more complex frustrated systems (Balents 2010). Other exotic quantum phases include fractional spinon excitations in $CuSO_4$ (Mourigal et al. 2013), quantum criticality (Lake et al. 2005; Merchant et al. 2014) and Bose-Einstein condensation of magnons in antiferromagnetic dimers such as $TlCuCl_3$ halide (Nikuni et al. 2000; Giamarchi, Rüegg, and Tchernyshyov 2008), or spin liquid phases in antiferromagnetic kagome lattices (Han et al. 2012). The nature of these exotic phases may hold the key to high-$T_c$ superconductivity (Dai 2015). These fascinating features pertain to the physics of frustrated magnetic systems, which is beyond the scope of the present review.

| Electrical category | Crystallographic and spin structures (bulk) | Material | $T_{N,bulk}$ (K) | $T_{N,Finite\ Size}$ (K) |
|---|---|---|---|---|
| Insulator | 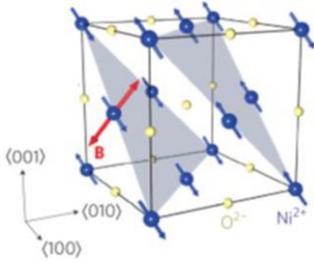<br>From (Kampfrath et al. 2011). | NiO | 520 | (Abarra et al. 1996; Qiu et al. 2016; Lin et al. 2016) |
| | | CoO | 290 | (Molina-Ruiz et al. 2011; Ambrose and Chien 1996; Abarra et al. 1996; Qiu et al. 2016; Y. Tang et al. 2003; Lin et al. 2016) |
| | 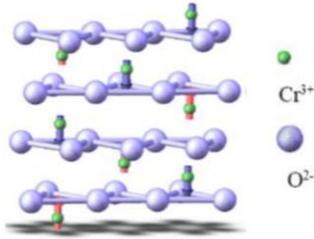<br>From (He et al. 2010). | $Cr_2O_3$ | 310 | (Pati et al. 2016) |
| | 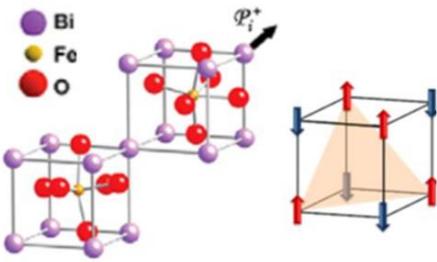<br>From (Martin et al. 2008). | $BiFeO_3$ | 653 | n/a |
| | 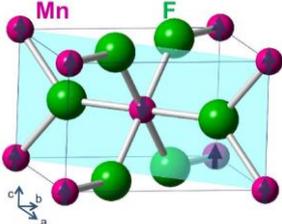<br>From (S. M. Wu et al. 2016). | $MnF_2$ | 68 | n/a |

TABLE 2. Crystallographic structure, spin structure and Néel temperature of some insulating antiferromagnets.

## 3. Semiconductors and semimetals

Although they have received less attention until now than their metallic and insulating counterparts, semiconducting and semimetallic antiferromagnets (TABLE 3) are another class of important materials for the study of spin transport. For example anisotropic magnetoresistance (see III.A.) was detected in semiconducting $Sr_2IrO_4$ (crystalline component) (Fina et al. 2014; C. Wang, Seinige, Cao, Zhou, et al. 2014), semimetallic CuMnAs (non-crystalline component) (Wadley et al. 2016) and II-VI semiconducting MnTe (non-crystalline component) (Kriegner et al. 2016). Recently, the first '(ferro)magnet-free' memory prototype with electrical writing/readout was produced with CuMnAs, exploiting the inverse spin galvanic effect for writing and the planar Hall component of the anisotropic magnetoresistance effect for reading (Wadley et al. 2016), (see III.D.2.). The vast number of antiferromagnetic semiconductors, like Mn(II)-VI, Fe(III)-V and Gd(III)-V alloys hold great promise for future research. Among classes of antiferromagnetic semiconductors which have yet to be exploited for antiferromagnetic spintronics, we can list CuFeS2 (I-IV-III-IV), which has an exceptionally high Néel temperature of 825 K, $MnSiN_2$ (II-V-IV-V, 490 K) and LiMnAs (I-II-V, 374 K). Other semiconducting antiferromagnets are listed in (Máca et al. 2012; Jungwirth et al. 2016). Recently, it was realized that antiferromagnetism may co-exist with topologically non-trivial phases of matter, such as Weyl semimetals (e.g. GdPtBi (Hirschberger et al. 2016) and $Mn_3Ge$ and $Mn_3Sn$ (H. Yang et al. 2017)), and CuMnAs (P. Tang et al. 2016). This field is expected to be the focus of significant attention in the near future (Šmejkal et al. 2017).

| Electrical category | Crystallographic and spin structures (bulk) | Material | $T_{N,bulk}$ (K) | $T_{N,Finite\ Size}$ (K) |
|---|---|---|---|---|
| | 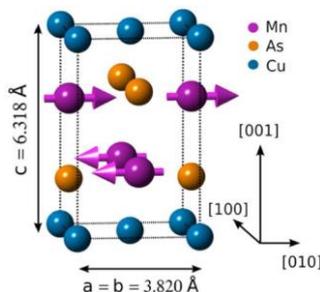 $S_{Mn} \parallel c$ (LiMnAs) or a (e.g. CuMnAs) From (Wadley et al. 2015). | CuMnAs LiMnAs | 480 374 | n/a n/a |
| | 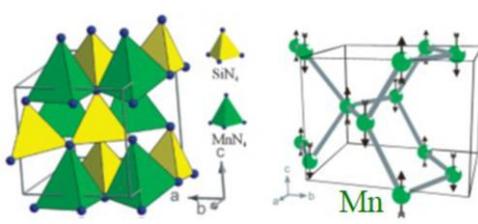 From (Esmaeilzadeh, Hålenius, and Valldor 2006) | MnSiN₂ | 490 | n/a |
| Semi conductor / Semimetal | 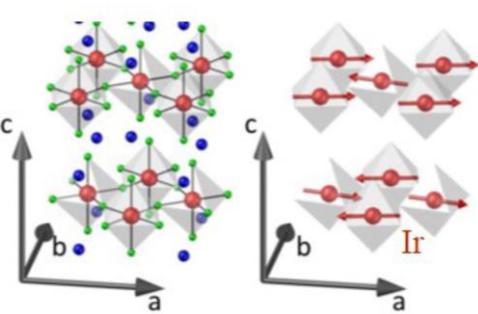 From (C. Wang et al. 2015). | Sr₂IrO₄ | 240 | (Fina et al. 2014) |
| | 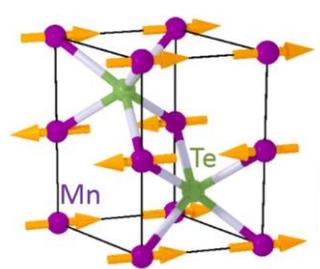 From (Kriegner et al. 2016). | MnTe | 323 | n/a |

TABLE 3. Crystallographic structure, spin structure and Néel temperature of some semimetallic and semiconducting antiferromagnets.

## C.    Basics of antiferromagnets

### 1.  Quantum aspects

***A prologue on quantum antiferromagnets***

A particularly illustrative example of the thought-provoking character of antiferromagnets concerns the ground state of antiferromagnetic chains, represented by a nearest-neighbor Heisenberg Hamiltonian

$$\hat{H} = -\sum_i J_{i,i+1} \mathbf{S}_i \cdot \mathbf{S}_{i+1} \ .$$

(1)

Here, the spin operator $\mathbf{S}_i$ is in unitless of $\hbar$ and $J_{i,i+1}$ is the exchange energy between neighboring sites. In the case of a positive exchange integral, $J_{i,i+1} > 0$, the system's ground state forms a ferromagnetic chain (such as $\left| \uparrow\uparrow\uparrow ... \right\rangle$ or $\left| \downarrow\downarrow\downarrow ... \right\rangle$), i.e., any high spin configuration. In contrast, when the exchange integral is negative, $J_{i,i+1} < 0$, nearest-neighbor magnetic moments tend to align antiferromagnetically. The antiferromagnetic Heisenberg Hamiltonian, Eq. (1), can be derived from Hubbard's model in the limit of U/t >> 1 at half filling (where U is Hubbard's parameter for electron correlation, and t is the term describing hopping between nearest neighbors (Anderson P. W. 1959; Takahashi 1977)). It would appear logical to assume that the ground state is an antiferromagnetic chain of the form $\left| \uparrow\downarrow\uparrow ... \right\rangle$. However, this is not at all the case. For instance, the ground state of a 4-site spin-1/2 antiferromagnetic chain is given by $\left| \uparrow\uparrow\downarrow\downarrow \right\rangle + \left| \downarrow\downarrow\uparrow\uparrow \right\rangle + \left| \downarrow\uparrow\uparrow\downarrow \right\rangle + \left| \uparrow\downarrow\downarrow\uparrow \right\rangle - 2\left| \uparrow\downarrow\uparrow\downarrow \right\rangle - 2\left| \downarrow\uparrow\downarrow\uparrow \right\rangle$. Therefore, the ground state of the antiferromagnetic Heisenberg Hamiltonian is *not* a collinear antiferromagnetic chain at $T$=0 K, as confirmed by experiments (see e. g. (Hirjibehedin, Lutz, and Heinrich 2006)). Hence, quantum antiferromagnets host exotic, strongly correlated excitations, and because of these, they have been the objects of intense research in condensed matter physics for more than 60 years (Giamarchi 2003; Balents 2010). Although this topic is beyond the scope of the present review, we believe that the recent observation of a spinon-mediated spin Seebeck effect in the $Sr_2CuO_3$ one-dimensional antiferromagnet might bridge the gap between antiferromagnetic spintronics and frustrated systems (Hirobe et al. 2016).

### *Electronic band structure*

Although antiferromagnets appear in a wide variety of flavors in terms of crystal symmetries and magnetic textures (TABLE 1-3), we will illustrate a few of their important aspects by addressing the properties of the simplest paradigm: a collinear bipartite antiferromagnet composed of two interpenetrating square lattices, A and B, possessing antiferromagnetically aligned moments (see FIG. 1(a) – extension to three dimensions is straightforward). This configuration is usually referred to as a G-type or checkerboard antiferromagnet. Assuming only nearest-neighbor hopping for simplicity, the tight-binding Hamiltonian reads

$$\hat{H} = \sum_{i,j} \left[ c_{i,j}^+ \left( \varepsilon_0 + \left(-1\right)^{i+j} \Delta \hat{\sigma}_z \right) c_{i,j} - t \left( c_{i,j}^+ c_{i+1,j} + c_{i,j}^+ c_{i,j+1} + c.c. \right) \right],$$

(2)

where $t$ is the hopping parameter, and $\varepsilon_0$ is the onsite energy. We assume an exchange energy between the itinerant spins and the local moments $\left(-1\right)^{i+j} \Delta$ that is positive and negative on sublattices A and B, respectively. It is convenient to rewrite this Hamiltonian in the basis $\left( \left| A \right\rangle, \left| B \right\rangle \right) \ddot{A} \left( \left| - \right\rangle, \left| ^- \right\rangle \right)$, which reduces to

$$\hat{H} = \varepsilon_0 \hat{1} \otimes \hat{1} + \gamma_k \hat{\tau}_x \otimes \hat{1} + \Delta \hat{\tau}_z \otimes \hat{\sigma}_z,$$

(3)

with $\gamma_k = -2t\left(\cos k_x a + \cos k_y a\right)$. The 2x2 Pauli matrices $\hat{\sigma}_i$, and $\hat{\tau}_i$ refer to sublattice and spin subspaces, respectively. It is straightforward to calculate the eigenstates and band structure of this system and one obtains

$$\varepsilon_k^s = \varepsilon_0 + s\sqrt{\gamma_k^2 + \Delta^2}$$

$$\psi_s^\sigma = \frac{1}{\sqrt{2}}\left(\sqrt{1 + s\sigma\frac{\Delta}{\sqrt{\gamma_k^2 + \Delta^2}}}\,|A\rangle + s\sqrt{1 - s\sigma\frac{\Delta}{\sqrt{\gamma_k^2 + \Delta^2}}}\,|B\rangle\right) \otimes |\sigma\rangle$$

(4)

where σ refers to the spin index, while *s* denotes the conduction *(s=+1)* or valence band *(s=-1)*. In collinear antiferromagnets, the spin degree of freedom remains a good quantum number, but not the sublattice degree of freedom. Therefore, the eigenstates are a spin-dependent mixture of A and B states. Their spatial profile displays a lattice dependent modulation of the density that is spin-dependent and band-dependent [FIG. 1(b) and (c)]. In other words, although the exchange $\left(-1\right)^{i+j}\Delta$ breaks time-reversal symmetry, its combination with spatial translation provides an analogous to Kramers' theorem, which produces two degenerate bands with opposite spins that *do not carry spin current*. Notice that this property may not hold true for certain non-collinear antiferromagnets, as discussed in III.C.1. An interesting consequence on this property has been pointed out by Haney and MacDonald (Haney and MacDonald 2008). At normal metal/antiferromagnet interfaces incident spins may undergo spin-flip accompanied by sublattice interchange. This mechanism governs spin angular momentum transfer at these interfaces as it results in a non-vanishing spin mixing conductance and thereby spin transfer torque and magnetoresistance in antiferromagnetic devices (section II).

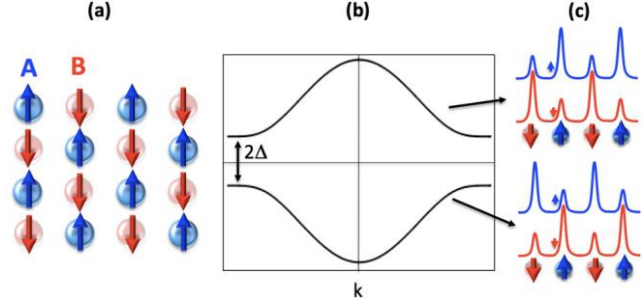

FIG. 1 (Color online). (a) Schematics of a two-dimensional G-type antiferromagnet with two sublattices referred to as A (blue) and B (red); (b) Electronic band structure and (c) spatial distribution of the corresponding density of states.

## 2. Some key parameters

### *Magnetic susceptibility and Néel temperature*

In the course of his exploration of Weiss molecular field theory, Néel (Néel 1932; Néel 1938; Néel 1971) and subsequently Bitter (Bitter 1938) and van Vleck (Van Vleck 1941) addressed the magnetic susceptibility of substances with negative local exchange integrals between collinear moments. These substances were baptized antiferromagnets by Bitter (Bitter 1938). In the mean field approximation, each individual magnetic moment is antiferromagnetically coupled to its nearest neighbors through their mean value field (extrapolation to next-nearest neighbors is straightforward). The energy of the magnetic moment at position $i$ on sublattice A reads, $H_i \approx -2J\mathbf{S}_i \cdot \sum_{j \in B} \langle \mathbf{S}_j \rangle = -2Jz\mathbf{S}_i \cdot \langle \mathbf{S}_B \rangle$ (Van Vleck 1941), where $z$ is the number of nearest neighbors and $\langle \mathbf{S}_B \rangle$ is the mean value of the magnetic moments on sublattice B. In an oversimplified treatment, where magnetic anisotropy is neglected (Kittel 1976) (a small easy-axis anisotropy is implicit in order to prevent spin-flop), the magnetization of sublattice $\partial$(=A,B) reads

$$M_\alpha = \frac{N}{2} g\mu_B S B_S\left(\frac{4zJ}{k_B T}\frac{SM_\alpha}{g\mu_B N}\right),$$

$$\text{with } B_S(x) = \left(1+\frac{1}{2S}\right)\coth\left[\left(1+\frac{1}{2S}\right)x\right] - \frac{1}{2S}\coth\left[\frac{x}{2S}\right],$$

(5)

where $B_S$ is the Brillouin function, $S$ is the on-site spin, $N$ is the number of magnetic atoms per unit volume (assuming that half of these $N$ belong to one of the two collinear sublattices), $T$ is the temperature, $g$ is the $g$-factor, and $\mu_B$ is the Bohr magneton. By taking the limit of vanishing magnetization, the Néel temperature can be defined

$$k_B T_N = \frac{2}{3} zJS(1+S).$$

(6)

The magnetic susceptibility above the Néel temperature then becomes

$$\chi_{T>T_N} = \mu_0\left(g\mu_B\right)^2 N\frac{S(1+S)}{3k_B}\frac{1}{T+T_N},$$

(7)

where $\mu_0$ is the vacuum permeability. In contrast to the Curie-Weiss law for ferromagnets, the susceptibility of antiferromagnets does not diverge at the critical ordering (Néel) temperature. Below the Néel temperature, the susceptibility depends on the direction of the field applied with respect to the magnetic order

$$\chi_{\parallel} = \mu_0 \frac{N\left(g\mu_B S\right)^2 B_{S'}\left(\dfrac{2zJ}{k_B T}\dfrac{SM}{g\mu_B N}\right)}{k_B T + 2zJS^2 B_{S'}\left(\dfrac{2zJ}{k_B T}\dfrac{SM}{g\mu_B N}\right)}$$

$$\chi_{\perp} = \mu_0 \frac{N(g\mu_B)^2}{4zJ} = \mu_0 \frac{N(g\mu_B)^2}{6k_B T_N} S\left(1+S\right).$$

$$(8)$$

$\chi_{\parallel,\perp}$ is the susceptibility when the external magnetic field is applied parallel or transverse to the order parameter. The parallel susceptibility of MnO measured by Bizette et al. (Bizette, Squire, and Tsaï 1938) was probably the first observation of an antiferromagnetic response to an external field and the demonstration of how different it was from the response of ferro- and para-magnets. The qualitative agreement between experimental data for $\chi_{\parallel}$ and $\chi_{\perp}$ and the molecular field theory is illustrated in FIG. 2 (for MnF$_2$, a collinear coplanar antiferromagnet). Qualitatively similar results were recently obtained when the molecular field theory was extended to non-collinear coplanar antiferromagnets (Johnston 2012).

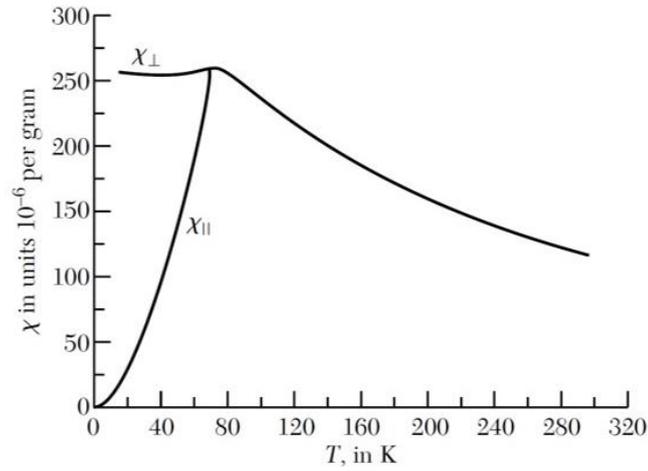

FIG. 2. Representative experimental data for magnetic susceptibility. For MnF$_2$ from (Kittel 1976).

Experimentally, considerable data is available for bulk antiferromagnets, sufficiently thick single layers or for multiply-repeated thinner layers since most techniques are volume-sensitive. In contrast, it is much more challenging to determine the magnetic susceptibility and Néel temperature for a thin film of isolated antiferromagnetic material. Despite the importance of such a basic parameter for antiferromagnetic spintronics and how finite size effects influence the magnetic susceptibility and the Néel temperature of antiferromagnetic films, very few studies to date have presented quantitative data (TABLE 1-3). This dearth of data stems from a lack of routinely available rapid measurement techniques compatible with most antiferromagnetic thin films. The next paragraph focuses on this issue.

***Magnetic phase transition and finite size scaling***

Theoretical calculations accounting for magnetic phase transitions and finite-size scaling (R. Zhang and Willis 2001) relate size effects to the loss of spin coordination when the size is reduced resulting in smaller critical temperatures. The model takes into account the finite divergence of the spin-spin correlation length near the critical temperature. As we will further detail, intense research efforts are focused on elucidating the impact of ultrathin films on spin-dependent properties, although transition temperatures have yet to be established for these systems. Extrapolating for the case of all-antiferromagnetic [i.e. '(ferro)magnet-free'] devices (Petti et al. 2013), the order-disorder Néel temperature would set the temperature threshold for data retention. This temperature is defined by the stiffness of the exchange between moments in the antiferromagnet. Sometimes, it is mistakenly confused with the blocking temperature, which is specific to ferromagnetic/antiferromagnetic exchange bias interaction (see I.C.3), whereas the Néel temperature is intrinsic to the antiferromagnet (Nogués and Schuller 1999; Berkowitz and Takano 1999). The blocking temperature can easily be determined experimentally, for example by measuring the disappearance of the hysteresis loop shift as the external temperature rises, or by using specific field-cooling protocols (Soeya et al. 1994; Baltz, Rodmacq, et al. 2010). In contrast, it is much more challenging to determine the Néel temperature for a thin film of isolated antiferromagnetic material. To the best of our knowledge, neutron diffraction (Yamaoka, Mekata, and Takaki 1971), magnetic susceptibility (Ambrose and Chien 1996), nanocalorimetry (Abarra et al. 1996), resistivity (Boakye and Adanu 1996), and optical measurements (Saidl et al. 2017) can be used to determine the Néel temperature for sufficiently thick single layers or for superlattices of thinner layers. Alternatively, the Néel temperature can be indirectly determined through experiments involving non-conventional ultrafast measurements of ferromagnetic/antiferromagnetic exchange-biased bilayers. In these bilayers, the ferromagnetic/antiferromagnetic blocking temperature increases with the ferromagnetic material's magnetization sweep-rate and can reach the antiferromagnetic intrinsic Néel temperature when measuring in the nanosecond regime (Lombard et al. 2010). However, like nanocalorimetry, this technique requires full device nanostructurations.

Some recent experimental works applied alternative methods to determine the Néel temperature (Frangou et al. 2016; Qiu et al. 2016; Lin et al. 2016). These studies found that enhanced spin pumping can be achieved by using a fluctuating spin sink close to its magnetic order transition temperature. The principle of the experiment (FIG. 3) is that the non-equilibrium magnetization dynamics of a spin injector (NiFe) pumps a spin current ($I_S$) into an adjacent layer, called the spin sink (IrMn). This spin sink absorbs the current to an extent which depends on its spin-dependent properties. To eliminate direct exchange interactions and focus only on the effects due to the interaction between the spin current and the spin sink, the injector and the sink are separated by an efficient spin conductor (Cu). The findings were corroborated by recently-developed theories which link the enhanced spin pumping into a fluctuating spin sink to interfacial spin mixing conductance (Ohnuma et al. 2014; K. Chen et al. 2016). This conductance depends on the transverse spin susceptibility of the spin sink, which is known to vary around critical temperatures (FIG. 2).

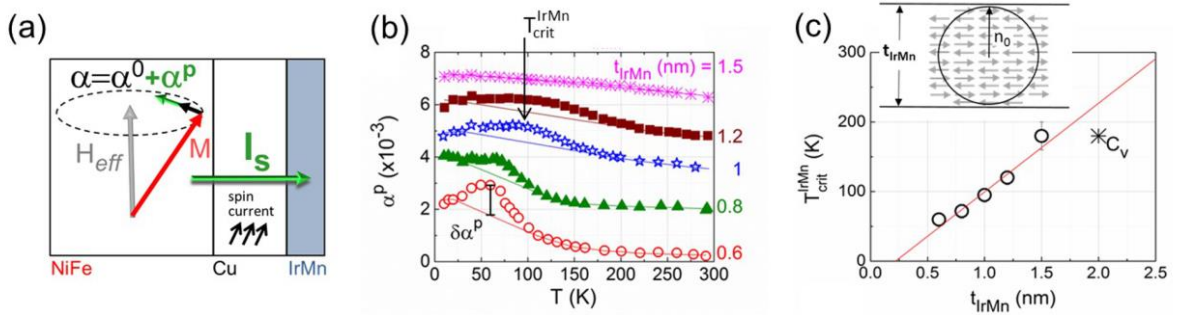

FIG. 3 (Color online). (a) Diagram representing the spin pumping experiment. (b) Temperature (T) dependence of the IrMn spin pumping contribution to NiFe damping ($\alpha^p$). To facilitate reading, data have been shifted vertically. The enhanced spin pumping occurring during the IrMn magnetic phase transition is $\delta\alpha^p$. (c) Dependence of $T_{crit}^{IrMn}$ on $t_{IrMn}$, where $T_{crit}^{IrMn}$ is the critical temperature for the IrMn magnetic phase transition. Data fitting returns the spin-spin correlation length ($n_0$). Adapted from (Frangou et al. 2016).

Unlike previous techniques (e.g. susceptibility, neutron diffraction, calorimetry) which tend to be volume-sensitive, this method is surface-sensitive. By showing that magnetic phase transitions of isolated thin films can be detected by spin pumping, these works open the possibility of further investigation of non-trivial magnetic orders, not limited to antiferromagnetism. For example, the dependence of the IrMn critical temperature on the thickness of this layer was experimentally determined by spotting the spin pumping peak (FIG. 3). This information provided access to a fundamental parameter, the characteristic length for spin-spin interactions. Until now, for isolated IrMn thin films, this parameter had been experimentally inaccessible, and it remains to be measured for a number of common antiferromagnets, including FeMn, PtMn, CuMnAs and $Mn_2Au$ (TABLE 1-3). Size effects also result in a reduction of Néel temperature from larger to smaller grains in polycrystals (even in the 8 nm range, on which the grain size distribution is typically centered). This effect results in a grain-to-grain Néel temperature distribution, which remains to be determined experimentally.

### *Spatial variability of magnetic properties*

Another important finite size effect relates to variability. Spatial variability of magnetic properties refers to how the magnetic properties are distributed when measured at different spatial locations. In spintronics this problem was raised when the very first generation of magnetic random access memory (MRAM) chips was developed and it has received considerable attention. Most studies focused on the variability of the shapes of memory bits produced during the nanofabrication process (Slaughter, Rizzo, and Mancoff 2010). As a rule, the use of nanostructures calls for statistical representations, a need which becomes even more pressing when antiferromagnetic materials are involved. This is because both polycrystalline and epitaxial antiferromagnetic films are very sensitive to spin texture faults (due to roughness, atomic stacking faults, etc.) that create randomly-spread disordered magnetic phases (Takano et al. 1997; Baltz, Rodmacq, et al. 2010; Lhoutellier et al. 2015). In addition, lateral finite size effects on antiferromagnetic materials are not straightforward: lateral size reduction affects the antiferromagnetic domain size, and it also introduces boundaries. These boundaries have two main consequences: they cut the grains located at the edges (in polycrystalline films), thus reducing their volumes (Baltz et al. 2004; Vallejo-Fernandez and Chapman 2010) and they add unstable disordered magnetic phases along the edges due to the suppression of atomic bonds (this applies to both polycrystalline and epitaxial films) (Baltz, Gaudin, et al. 2010).

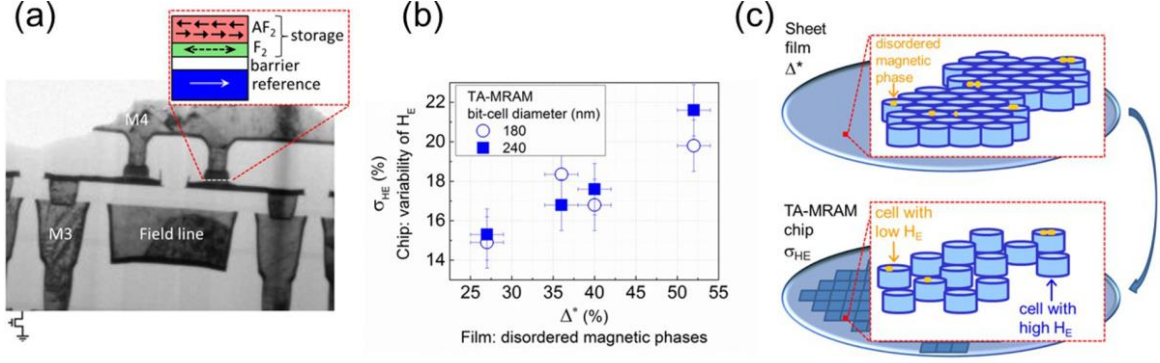

FIG. 4 (Color online). (a) Transmission electron microscopy (TEM) cross-section zoomed on two memory bits of a typical thermally assisted (TA)-MRAM chip on CMOS with a simplified stack [reference/barrier/{$F_2$/$AF_2$}]. (b) Variability of the ferromagnetic/antiferromagnetic ($F_2$/$AF_2$) bilayer exchange bias ($\sigma_{HE}$) in memory bits vs. proportion of disordered magnetic phases initially present in the film ($\Delta^*$). (c) Sketch showing how disordered magnetic phases (orange dots) are spread over a polycrystalline film and the resulting variability of exchange bias in memory bits when the film is patterned to form memory bits. Adapted from (Akmaldinov et al. 2015).

In MRAM chips, Akmaldinov et al. (Akmaldinov et al. 2015) recently experimentally demonstrated that antiferromagnetic disordered magnetic phases initially present in the film cause nanostructure-to-nanostructure dispersion of the antiferromagnet-related properties when the film is patterned, thus creating only atypical nanostructures (see FIG. 4).

It is essential to take this variability factor into account through statistical analysis when dealing with the interaction between a current and an antiferromagnetic material in a nanostructure. This analysis can help to avoid drawing conclusions based on potentially misleading individual cases. From the moment that arrays of nanostructures are involved, statistical analysis should be performed when determining spin-transfer, spin-orbit and spin-caloritronics effects.

## 3. Manipulation by magnetic field, light, and exchange bias

### *Manipulation by magnetic field*

While antiferromagnetic materials are often reported to be robust against perturbation due to magnetic fields, this does not mean that they are insensitive to magnetic fields. The fact that antiferromagnets can be driven by magnetic fields was briefly introduced in I.C.2 when we presented magnetic susceptibility. In this paragraph, we clarify how the magnetic moments in an antiferromagnetic material can be appreciably rotated in a quasistatic manner. We then discuss the dynamics of antiferromagnets and the impact of magnetic field excitation.

A simple but powerful model to describe quasistatic magnetic field manipulation of a bipartite antiferromagnet is the Stoner-Wohlfarth (Stoner and Wohlfarth 1948) or coherent rotation model. In the macrospin approximation, the energy density of one sublattice with uniaxial anisotropy (K) and subjected to a magnetic field (H) is given by

$$E = \mu_0 H_E M_s \cos(2\phi) + K \cos^2(\phi) - \mu_0 H M_s \cos(\phi),$$

(9)

where $\phi$ is the angle between **H** and the magnetic moments of the sublattice, $\mu_0 H_E = J_{AF} S^2 / \mu_B$ is the antiferromagnetic exchange field, $J_{AF}$ is the antiferromagnetic exchange energy and $M_s$ is the magnetization of one sublattice. When an external magnetic field is applied perpendicular to the easy axis, the magnetic moments on the two sublattices cant and a net magnetization **M** gradually builds up proportional to the field [FIG. 5(a)]. The net magnetization saturates when the Zeeman energy approximately equals the exchange energy ($\mu_0 H_{sat} M_s = \mu_0 H_E M_s + K$). The scenario is significantly altered when the external field applied is parallel to the easy-axis. When the anisotropy energy is small compared to the exchange energy, sublattice magnetization remains in its easy-axis state, with zero net magnetization until the magnetic field compensates the anisotropy. The sublattice magnetization is then free to rotate in directions approximately perpendicular to the easy-axis while consuming almost no energy [FIG. 5(a)]. This is known as the spin-flop transition, and it occurs at the spin-flop field $\mu_0 H_{sf} \approx \mu_0 \sqrt{H_K H_E}$, where $H_K = 2K / \mu_0 M_s$. After the transition, the sublattice magnetizations cant increasingly in the direction of the magnetic field and the net magnetization is proportional to the magnetic field [FIG. 5(a)]. When the anisotropy energy is large compared to the exchange energy, net magnetization jumps directly from zero (sublattice magnetic moments antiparallel along the easy-axis) to saturation (sublattice magnetic moments parallel along the easy-axis). This is known as the spin-flip transition [FIG. 5(a)]. Examples of antiferromagnets with spin-flop [FIG. 5(b)] and spin-flip transitions are $MnF_2$ (Jacobs 1961) and $FeCl_2$ (Jacobs and Lawrence 1967), respectively. Note that a (spin-flop) field of 9-10 T in $MnF_2$ is enough to rotate the magnetic moments by 90 degrees [FIG. 5(b)]. This reorientation of an antiferromagnetic sublattice by applying a magnetic field can be used to determine the magnetic anisotropy of antiferromagnets, but does require very large magnetic fields. The process can also be used as a means to activate and deactivate mechanisms related to antiferromagnetic order, such as resonant mode degeneracy (Hagiwara et al. 1999) [FIG. 6(a)] and the antiferromagnetic spin Seebeck effect (S. M. Wu et al. 2016; Seki et al. 2015; Rezende, Rodríguez-Suárez, and Azevedo 2016a) (FIG. 46).

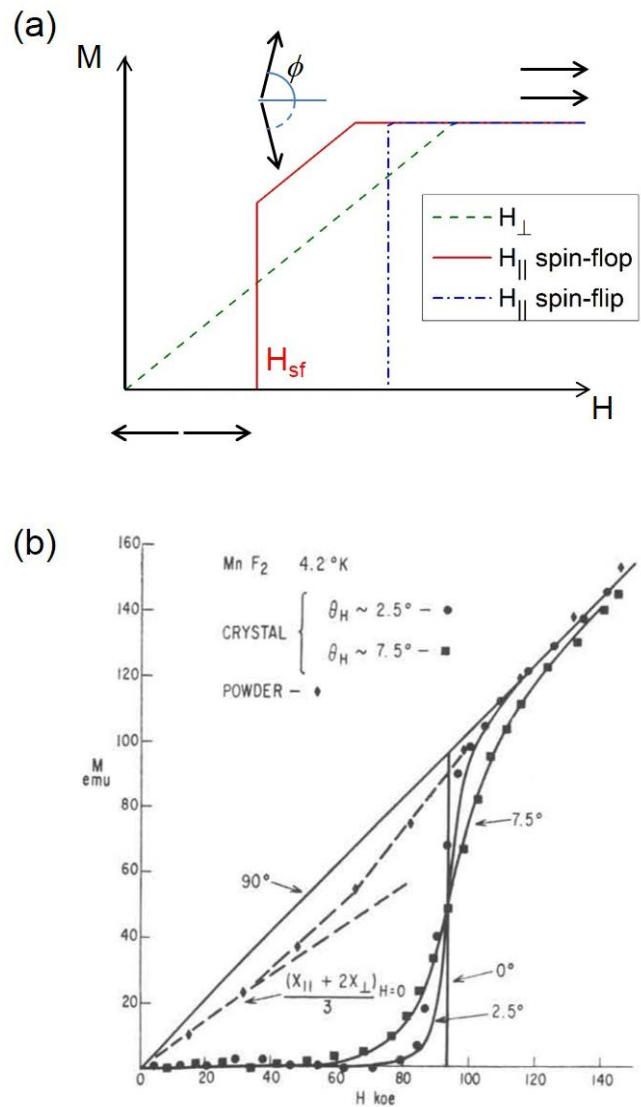

FIG. 5 (Color online). (a) Diagrammatic representation of antiferromagnetic manipulation

by magnetic field. (b) Spin-flop transition measured in MnF₂. From (Jacobs 1961).

Let us now turn our attention toward the dynamics of antiferromagnets and the impact

of a magnetic field excitation. For the simplest bipartite collinear antiferromagnet, the

classical coupled dynamics can be modeled by the phenomenological Landau-Lifshitz-Gilbert

equation

$$\partial_t \mathbf{m}_A = -\gamma \mathbf{m}_A \times \mathbf{H}_A + \alpha \mathbf{m}_A \times \partial_t \mathbf{m}_A,$$

$$\partial_t \mathbf{m}_B = -\gamma \mathbf{m}_B \times \mathbf{H}_B + \alpha \mathbf{m}_B \times \partial_t \mathbf{m}_B,$$

$$(10)$$

where $\mathbf{H}_\alpha = -\delta_{\mathbf{m}_\alpha} W / M_s$, $\mathbf{m}_\alpha$ is the spin density directional unit vector of sublattice $\alpha$ ($=A,B$), $\gamma$ ($>0$) is minus the gyromagnetic ratio, and $W$ is the magnetic energy density of the system including external applied fields, magnetic anisotropy and antiferromagnetic exchange ($\mu_0 H_E M_s \mathbf{m}_A \cdot \mathbf{m}_B$). This simple model leads to a number of important results that we outline below [e.g. see (H. V. Gomonay and Loktev 2014; Ivanov 2014)].

Let us first consider antiferromagnetic resonance and spin waves. In principle there exist as many excitation modes as basic constituents of the magnetic unit cell. Therefore, diatomic collinear antiferromagnets possess two modes that become non-degenerate in the presence of an external field or magnetic anisotropy (Kittel 1951; Keffer and Kittel 1952; Nagamiya 1951). In the simplest case where a small field, $H_z$, is applied along the uniaxial anisotropy axis, two circularly polarized modes emerge with frequency

$$\omega = \gamma \mu_0 \sqrt{H_K \left(2H_E + H_K\right)} \pm \gamma \mu_0 H_z .$$

$$(11)$$

Here, $H_K$ is the anisotropy field and $H_E$ is the exchange field. The resulting frequency is in the range of hundreds of GHz, as shown in FIG. 6(a) for MnF$_2$ ($H_K \sim 0.82$ T, $H_E \sim 53$ T).

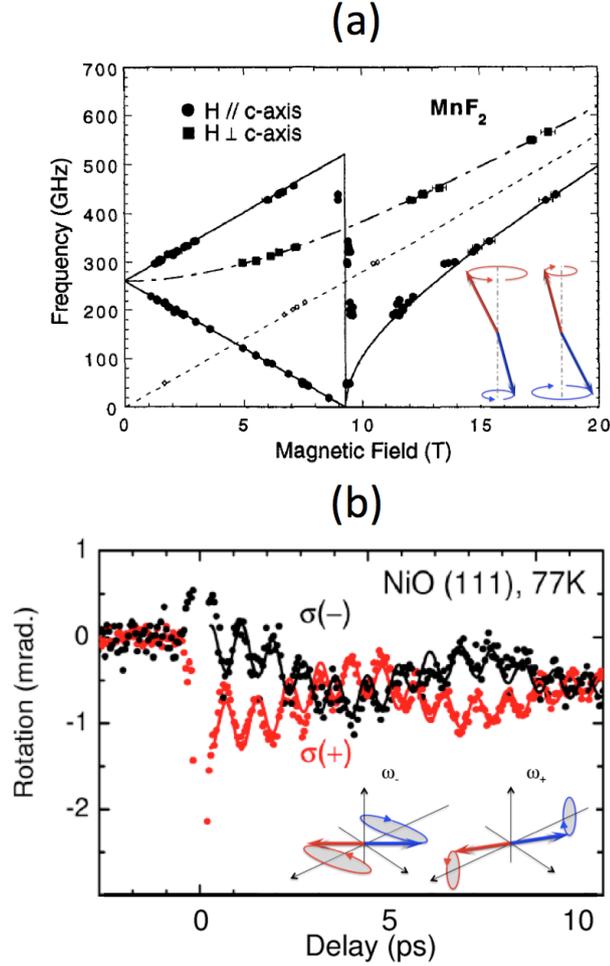

FIG. 6 (Color online). (a) Field-dependence of antiferromagnetic resonance of MnF$_2$. The mode degeneracy is lifted by applying an external field. From (Hagiwara et al. 1999). (b) Temporal evolution of two mixed spin wave modes in NiO. The rapid oscillations correspond to the out-of-plane mode, while the slow oscillations correspond to the in-plane mode. Adapted from (Satoh et al. 2010). The insets show the excitation modes.

In the case of biaxial anisotropy, such as in NiO (T$_N$ ~ 523 K), the two modes are associated with the different anisotropy constants and referred to as in-plane and out-of-plane modes with respect to the (111) planes (TABLE 2), also referred to as acoustic and optical modes. In NiO(111), the hard-axis ($\perp$) anisotropy along (111) coexists with an in-plane easy-axis ($\parallel$) anisotropy, yielding two different excitations branches, $\omega_+ = \gamma\mu_0\sqrt{H_E\left(H_\perp + H_\parallel - Aq^2\right)}$ and $\omega_- = \gamma\mu_0\sqrt{H_E\left(H_\parallel + Aq^2\right)}$ as displayed in FIG. 6(b) ( $H_\parallel \sim$

0.8 T, $H_\perp \sim 0.035$ T, $H_E \sim 937$ T), where $q$ is the wavevector.

In the absence of anisotropy, the spin wave dispersion becomes linear $\omega \sim q$ (see FIG. 7), which has important consequences in terms of spin and heat transport, as discussed further below (in section IV.B.).

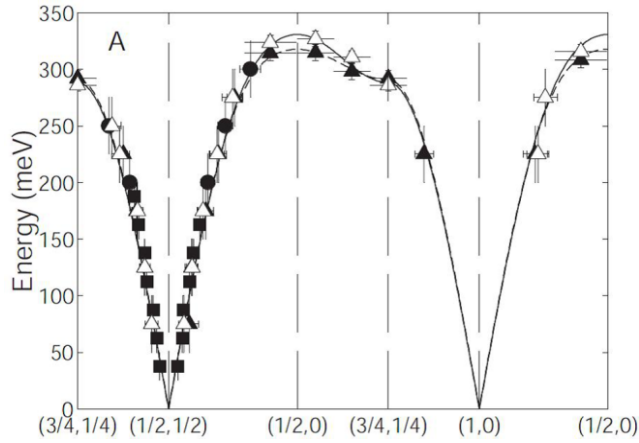

FIG. 7. Energy dispersion of antiferromagnetic spin waves in La$_2$CuO$_4$. From (Coldea et al. 2001).

Let us now consider the impact of a magnetic field on the dynamics of antiferromagnets. From Eq. (10), one obtains the coupled dynamics of the antiferromagnetic order parameter $\mathbf{l} = \left(\mathbf{m}_A - \mathbf{m}_B\right)/2$ and of the normalized spin density $\mathbf{m} = \left(\mathbf{m}_A + \mathbf{m}_B\right)/2$. In the limit of large antiferromagnetic exchange, $\mathbf{m}$ becomes a slave variable such that $|\mathbf{m}| \ll |\mathbf{l}| \sim 1$ and

$$\mathbf{m} = \frac{1}{\gamma\mu_0 H_E}\mathbf{l}\times\partial_t\mathbf{l} - \frac{1}{H_E}\mathbf{l}\times(\mathbf{H}\times\mathbf{l}).$$

$$(12)$$

In this case, one can derive the equation of motion for the antiferromagnetic order parameter only, also called *sigma model* (Ivanov 2014).

$$\partial_t^2\mathbf{l}\times\mathbf{l} = \gamma\mu_0\left[2(\mathbf{l}\cdot\mathbf{H})\partial_t\mathbf{l} - (\mathbf{l}\times\partial_t\mathbf{H})\times\mathbf{l}\right] + (\gamma\mu_0)^2(\mathbf{l}\cdot\mathbf{H})\mathbf{l}\times\mathbf{H} + \gamma\mu_0\alpha(H_E/2)\mathbf{l}\times\partial_t\mathbf{l}.$$

$$(13)$$

Anisotropies and spatial inhomogeneities are disregarded here. These are restored in Eq. (16) below. Several aspects are worth noticing. First and foremost, Eq. (13) is second order in time derivative, in sharp contrast with the ferromagnetic Landau-Lifshitz-Gilbert equation that is first order only. Hence, the dynamics of the antiferromagnetic order parameter presents similarities with the inertial dynamics of classical mechanical systems described by Newton's kinetic equation. In other words, the kinetic term $\partial_t^2\mathbf{l}\times\mathbf{l}$ acts like an acceleration, which has been demonstrated to have a dramatic impact on the dynamics of antiferromagnets, and will be discussed hereafter (e.g. (Wienholdt, Hinzke, and Nowak 2012; Cheng et al. 2015; O. Gomonay, Klaui, and Sinova 2016)).

Another noticeable aspect is that damping and anisotropy are both enhanced by the exchange field $H_E$. As a consequence, the dynamics of antiferromagnets is much faster than that of their ferromagnetic counterpart (as already noticed above) whereas the magnetic damping is also enhanced. Finally, while an external magnetic field acts on the magnetic order parameter $\mathbf{m}$ in the first order through the term $\mathbf{l}\times(\mathbf{H}\times\mathbf{l})$ in Eq. (12), its time derivative directly acts on the antiferromagnetic order parameter $\mathbf{l}$, through the term $(\mathbf{l}\times\partial_t\mathbf{H})\times\mathbf{l}$ in Eq. (13). Notice that a time-independent external magnetic field acts on the antiferromagnetic order parameter $\mathbf{l}$ in the second order through the term $(\mathbf{l}\cdot\mathbf{H})\mathbf{l}\times\mathbf{H}$, but this effect is of the order $(H/H_E)^2 \ll 1$ and is therefore generally negligible.

*Optical manipulation*

Because antiferromagnets break time-reversal symmetry, they display linear magnetic dichroism: the absorption of linearly polarized light depends on the orientation of the Néel order parameter. This property can be exploited to observe antiferromagnetic domains (section I.C.4), and to selectively heat different domains and thereby control their extension, as recently demonstrated in $MnF_2$ (Higuchi and Kuwata-Gonokami 2016). Nonetheless, dramatic differences between ferromagnetic and antiferromagnetic dynamics have been illustrated by Kimel et al. (Kimel et al. 2004) in $TmFeO_3$ orthoferrite and Fiebig et al. (Fiebig, Duong, and Satoh 2004) in NiO. The authors established that the antiferromagnetic order parameter could be reoriented upon optical excitation. This observation has been recently extended to $HoFeO_3$ and interpreted in terms of inertial dynamics (Kimel et al. 2009). As a matter of fact, these materials possess biaxial anisotropy that provides two metastable magnetic states, as well as Dzyaloshinskii-Moriya interaction that cants the magnetic order and results in a small magnetization. As a consequence, a light-induced time-dependent magnetic field (driven by inverse magneto-optical Faraday effect) triggers the dynamics of the antiferromagnetic order parameter [second term $\sim \partial_t\mathbf{H}$ at the right-hand side of Eq. (13)]. Upon inertial motion, the order parameter keeps evolving even after the light pulse is turned off, which drives the order parameter reorientation (see FIG. 8). This inertial switching has been confirmed numerically using atomistic modeling by Wienholdt et al. (Wienholdt, Hinzke, and Nowak 2012).

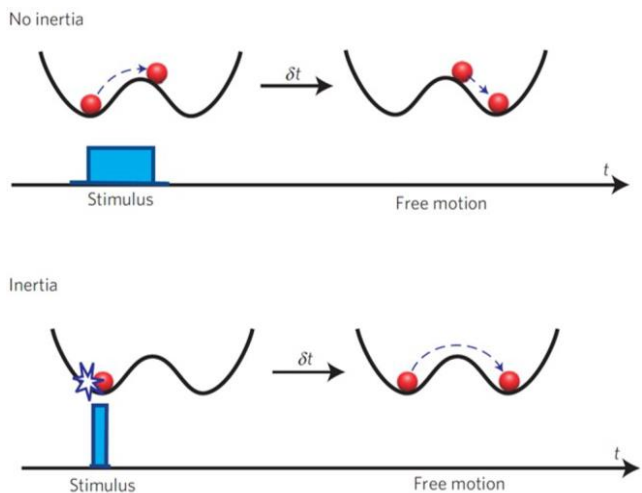

FIG. 8 (Color online). Difference between non-inertial and inertial spin reorientation, typical of ferro- and antiferro-magnetic dynamics, respectively. From (Kimel et al. 2009).

The inertial dynamics is also believed to be a key ingredient in the optical manipulation of rare earth transition metal ferrimagnets, for example in GdFeCo (Stanciu et al. 2007), FeTb (Hassdenteufel et al. 2013) and TbCo, DyCo, or HoFeCo (Mangin et al. 2014) alloys. It is however absent in ferromagnets (after an ultrashort field pulse, the magnetization falls back to the lowest energy state). Light-induced switching in transition-metal ferromagnets therefore necessitates some degree of thermal heating (Lambert et al. 2014).

More generally, the investigation of antiferromagnetic spin wave modes has received significant attention lately with the development of femtomagnetism and TeraHertz (THz) research [e. g. see review article by Kirilyuk et al (Kirilyuk, Kimel, and Rasing 2010)]. As a matter of fact, upon optical drive the very high frequency modes of antiferromagnetic insulators can be excited, such as in NiO (Kampfrath et al. 2011; Nishitani, Nagashima, and Hangyo 2013), MnO (Nishitani, Nagashima, and Hangyo 2013) or $Cr_2O_3$ (Satoh et al. 2007) as well as in rare-earth orthoferrite weak ferromagnets (i.e. canted antiferromagnets), such as $DyFeO_3$ (Kimel et al. 2005), $TmFeO_3$ (Kimel et al. 2006), $ErFeO_3$ (de Jong et al. 2011),

YFeO3 (R. Zhou et al. 2012), and HoFeO$_3$ (Mukai et al. 2014). The exploration of the interplay between such radiations and spin transport is still at its infancy and offers thrilling perspectives for ultrafast order parameter manipulation and spin current generation. The present techniques have been demonstrated only for bulk antiferromagnetic materials (or at least a few tens micron thick film) because the absorption signals or the Faraday rotation are proportional to the volume of the sample. It is still a major challenge to detect such excitations in thin films, on which usual spintronic devices are built and where the spin-current related phenomena become dominant. This will be discussed in the next section where the physics of spin transfer electronics in antiferromagnets is dealt with.

Conversely Seifert et al. (Seifert et al. 2016) took advantage of the electron spin to produce new THz emitters. In a metallic ferromagnet / non-magnet or antiferromagnet bilayer (e.g. PtMn), a femtosecond laser pulse excites electrons in the metal stack, creating a spin-polarized current. A TeraHertz electromagnetic transient is then emitted due to the conversion of the spin current into an ultrafast transverse ac charge current in the non-magnetic metal, through the spin Hall effect. The authors are currently investigating how the antiferromagnetic order influences the THz emission in antiferromagnet/non-magnet based emitters (Seifert and Kampfrath 2017).

### *Manipulation by exchange bias*

Exchange bias refers to magnetic interactions between ferromagnetic and antiferromagnetic materials (Meiklejohn and Bean 1956; Meiklejohn 1962). As a result of these interactions uniaxial anisotropy builds up and creates a hysteresis loop shift. The antiferromagnet is usually regarded as a means to manipulate the ferromagnet through magnetization pinning. In a reciprocal manner the antiferromagnetic order can also be manipulated via exchange bias. In subsequent sections in this paper, we describe demonstrations of antiferromagnetic spintronic effects where exchange bias is used, e.g. to probe current-induced antiferromagnetic order manipulation (II.A.2), to excite antiferromagnetic dynamics (II.B.2), to manipulate the antiferromagnetic order (III.B), and to combine several antiferromagnetic functionalities (III.D.4). In this section, we briefly discuss the basis of exchange bias. This presentation is intended to guide readers who may not be familiar with this topic, to make understanding of some of the following sections easier. Ferromagnetic/antiferromagnetic exchange bias is of course much more complex than the simplistic description given here. Any magnetic frustration due to roughness, grain boundaries in polycrystalline films, or stacking faults will challenge the idealized picture. Interested readers may wish to consult focused reviews where the exchange bias phenomenon is extensively described, e.g. (Berkowitz and Takano 1999; Nogués and Schuller 1999).

Here, we first discuss the intuitive picture and how exchange bias can be manipulated by cooling (domain imprint, FIG. 9). Exchange bias is nduced by raising the sample temperature above the blocking temperature ($T_B$) of the ferromagnet/antiferromagnet bilayer and cooling in a field ($H_{FC}$) which is sufficiently large to saturate the magnetization of the ferromagnet. During field cooling, coupling causes the moments in the antiferromagnet to align with those of the ferromagnet ($J_{F-AF}$). Below the blocking temperature, the moments in the antiferromagnet remain pinned, regardless of the direction of the moments in the ferromagnet. It is said that the initial ferromagnetic configuration is imprinted in the antiferromagnet during field-cooling. Due to coupling ($J_{F-AF}>0$), when sweeping the magnetic field at temperatures below the blocking temperature, the configuration when moments in the ferromagnet are parallel to moments in the antiferromagnet is energetically favored compared to the opposite configuration when moments are antiparallel. As a result, the hysteresis loop of the ferromagnet is shifted by a quantity called the exchange bias field ($H_{EB}$) (II.A.2 and III.D.4). When the exchange bias field exceeds the coercive field, only one magnetization direction is stable at zero field. This property is widely used to set a reference direction for the spin of conduction electrons in spintronic applications (Dieny, Speriosu, and Parkin 1991).

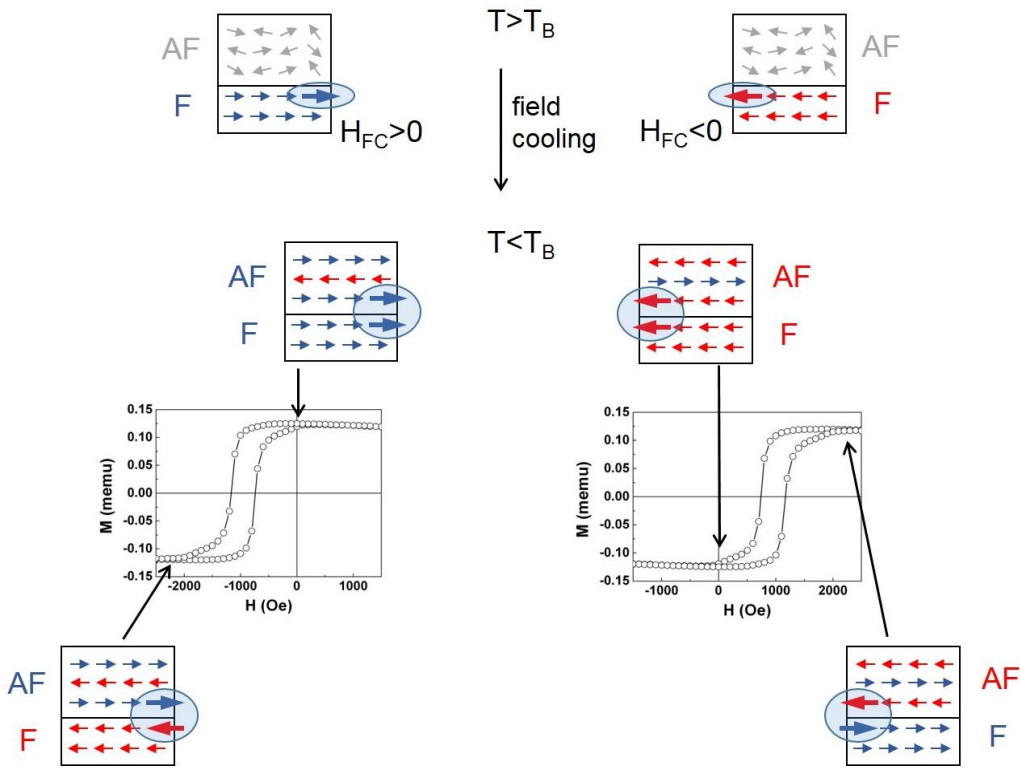

FIG. 9 (Color online) Intuitive picture of exchange bias and antiferromagnetic order manipulation by field cooling.

FIG. 9 illustrates how the direction of the antiferromagnetic order can be manipulated by exchange bias. In fact, the combination of coupling with a ferromagnet and field cooling can be used to direct the antiferromagnetic order. In other words, positive (negative) field cooling imprints positively (negatively) oriented configuration at the antiferromagnet interface. By field cooling at 90°, it is also possible to imprint configurations at 90° (III.B). Finally, imprinting multidomain states and magnetic textures in antiferromagnets can be achieved by preparing the ferromagnet in specific magnetic configurations, e. g. multidomain (Brück et al. 2005; Roshchin et al. 2005) and vortex (Salazar-Alvarez et al. 2009; J. Wu et al. 2011) states. Section I.C.4 is devoted to antiferromagnetic textures.

We now turn our attention to quasistatic manipulation by magnetic field and torques induced by exchange bias (exchange spring). In the picture presented above, the antiferromagnetic order is preserved when magnetization is reversed at temperatures below the blocking temperature. Another simple phenomenon useful for the present review was suggested by Néel (Néel 1967) and Mauri et al (Mauri et al. 1987). When the anisotropy in the antiferromagnet is small compared to the interfacial coupling, the antiferromagnet is no longer magnetically rigid. As a result of the torque exerted by the ferromagnet's magnetization on the antiferromagnet's sublattices, a domain wall parallel to the interface develops, like an exchange spring. Néel and Mauri predictions of exchange spring have been demonstrated in Co/NiO bilayers (Scholl et al. 2004). This constitutes another means to manipulate the antiferromagnetic order by combining coupling with a ferromagnet and sweeping the magnetization of the ferromagnet with a magnetic field (III.B).

Finally, ferromagnetic/antiferromagnetic exchange bias can also be used to dynamically inject and propagate spin angular momentum in antiferromagnets. This case will be specifically discussed in sections II and IV. As a preliminary example for now, in spin pumping experiments at finite temperatures the precessing magnetization in the ferromagnet pumps the oppositely-polarized magnons in the antiferromagnet differently, making magnonic spin transport possible (Rezende, Rodríguez-Suárez, and Azevedo 2016b) (also see TABLE 5).

## 4.    Magnetic textures

Ferromagnetic textures such as domain walls, vortices and skyrmions are currently attracting a lot of attention due to their rich spin physics and to their potential for three-dimensional electronic devices for storage and logic computing (Parkin, Hayashi, and Thomas 2008; Allwood et al. 2005; Fert, Cros, and Sampaio J. 2013). Magnetic textures also exist in antiferromagnetic materials and interesting differences compared to their ferromagnetic counterparts can be noted. For example, divergent vortices cannot form whenever a ferromagnetic component is present. In contrast, compensated antiferromagnetic materials can form divergent vortices in disks because they do not produce magnetic charges at the disc boundary (J. Wu et al. 2011). Another example can be found in domain wall dynamics. Since dipolar coupling is vanishingly small in antiferromagnets, antiferromagnetic domain walls do not exhibit Walker breakdown, thus a "massless" motion of the wall is observed (O. Gomonay, Jungwirth, and Sinova 2016). Furthermore, the antiferromagnetic domain walls exhibit a relativistic-like motion, which results in a Lorentz contraction of the wall when its velocity approaches the magnon group velocity (S. K. Kim, Tserkovnyak, and Tchernyshyov 2014; Shiino et al. 2016). In this paragraph, we briefly introduce some of the typical antiferromagnetic textures encountered and how these textures can be measured. Manipulation of these textures, in particular by an electron or a magnon flow, will be discussed in sections II.2, III.2 and IV.C.

### Experimental observation of antiferromagnetic textures

While antiferromagnetic textures may show some advantages over ferromagnetic analogs, because they lack net magnetization they are difficult to detect. Observation usually requires large scale facilities with element sensitive techniques like X-ray absorption spectroscopy (Weber et al. 2003; Salazar-Alvarez et al. 2009; J. Wu et al. 2011) or specific techniques with local probes like spin-polarized scanning tunneling microscopy (Bode et al. 2006; Loth et al.) and quantum sensing with single spins (nitrogen vacancies) in diamond (Gross et al. 2017).

Magnetic moments in a domain wall undergo a gradual reorientation. The domain wall width depends on the exchange, anisotropy and magnetoelastic energies in the material. FIG. 10 shows a 160 nm wide domain wall in antiferromagnetic NiO observed by X-ray linear magnetic dichroism photoelectron emission microscopy (Weber et al. 2003). In this type of material, the formation and properties of walls are dominated by magnetoelastic interactions. Local probes capable of atomic resolution like spin-polarized scanning tunneling microscopy are needed to further detect details of the spin structure in antiferromagnetic domain walls (FIG. 11). The wide variety of long-range spin structures in antiferromagnets (e.g. 3Q spin structure, TABLE 1) results in a much wider variety of possible domain wall configurations in these materials. We note also that controlling the amount of domain walls in an antiferromagnet has been demonstrated by Brück et al (Brück et al. 2005) and Roschchin et al (Roshchin et al. 2005) through domain imprinting in ferromagnet/antiferromagnet exchange-biased bilayers (section I.C.3).

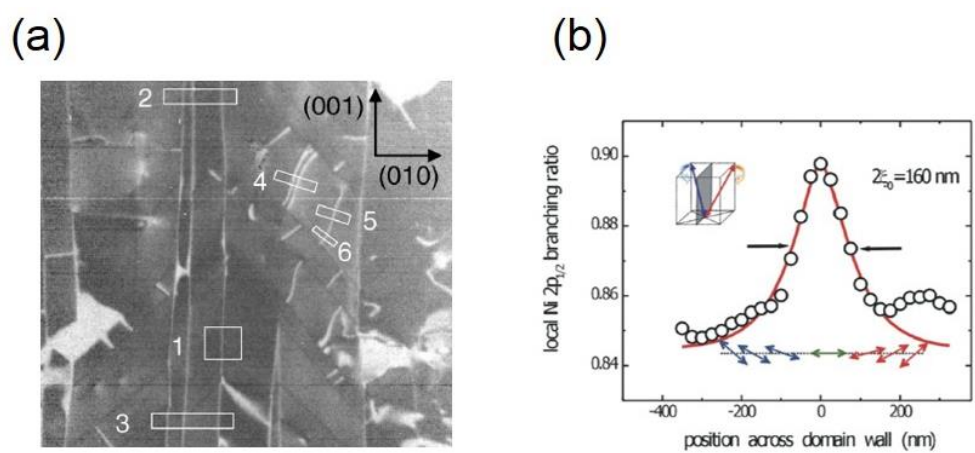

FIG. 10 (Color online). (a) X-ray linear magnetic dichroism imaging of antiferromagnetic domain walls in NiO. The width of the field of view is about 35 µm. (b) Profile of a domain wall averaged along the wall in region 1. From (Weber et al. 2003).

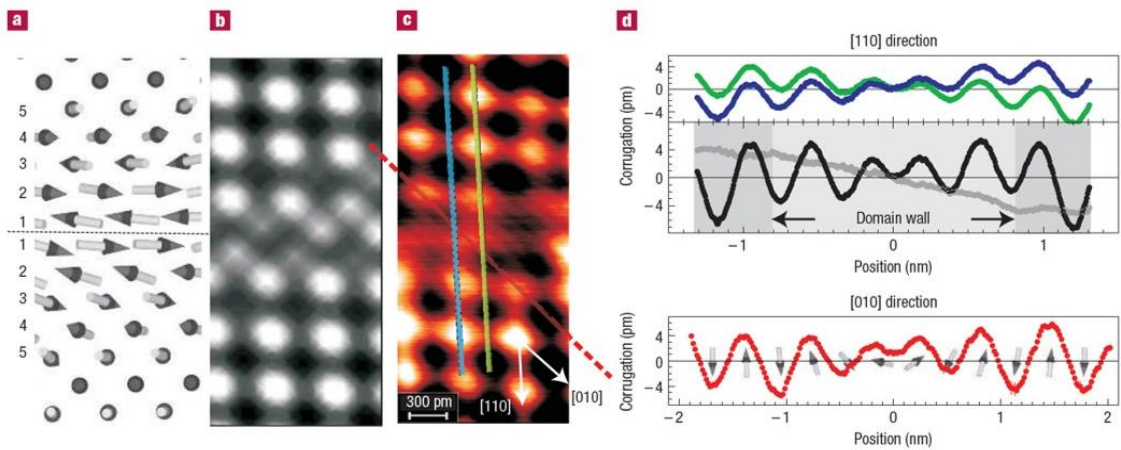

FIG. 11 (Color online). Domain wall in antiferromagnetic Fe monolayer on W(001). (a) Monte Carlo simulation of the spin structure. (b) Simulated spin-polarized scanning tunneling microscopy images based on the spin structure and (c) corresponding experimental data. (d) Height profiles from (c). From (Bode et al. 2006).

In a magnetic vortex the magnetization vector curls around the center of a confined structure (e. g. discs). The polarity and winding number of the vortex govern its gyroscopic rotation, reversal and motion. FIG. 12 shows curling and divergent vortices in the volume of antiferromagnetic NiO and CoO discs, as measured by X-ray magnetic linear dichroism (J. Wu et al. 2011). These data are complementary to circular dichroism observations of vortex states in IrMn layers (Salazar-Alvarez et al. 2009), where the signal was produced by the uncompensated moments at the IrMn interface rather than from the moments in the volume of the antiferromagnet. In both experiments, the vortex was imprinted from a ferromagnet into the antiferromagnet via exchange bias interactions (section I.C.3).

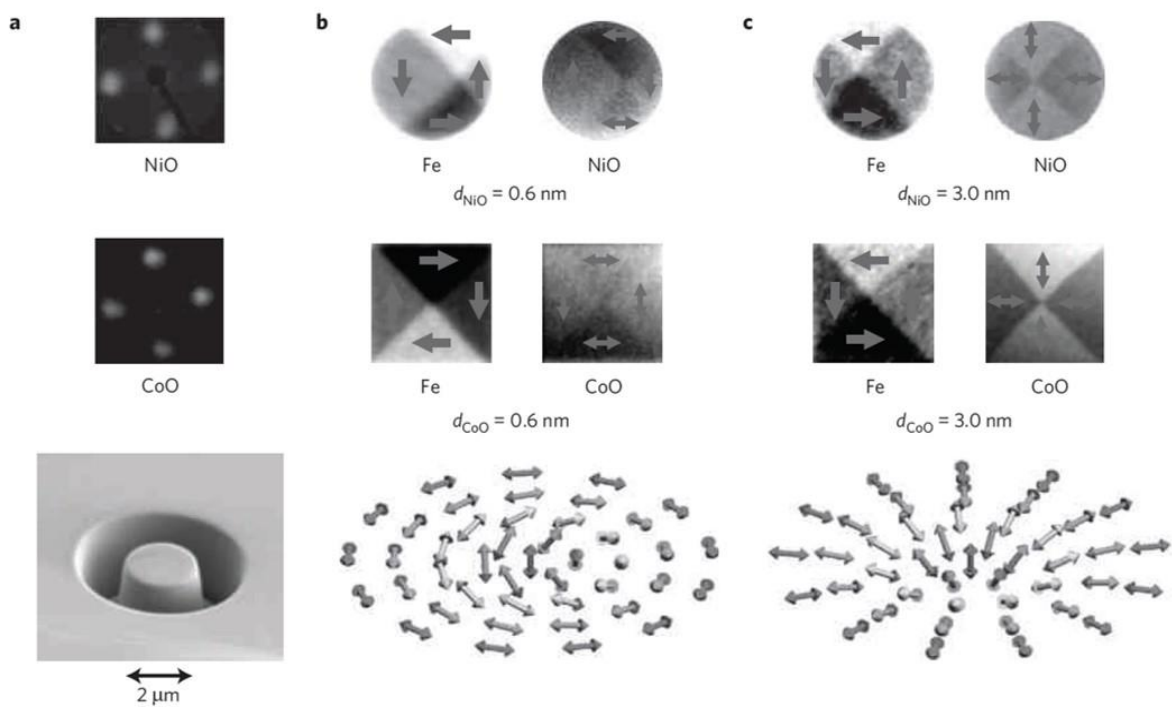

FIG. 12 (Color online). (a) Low-energy electron diffraction patterns and scanning electron microscope image of single-crystalline NiO/Fe/Ag(001) and CoO/Fe/Ag(001) discs. (b,c) Element-specific (X-ray magnetic linear dichroism) magnetic-domain images showing (b) curling and (c) divergent vortices (no ferromagnetic analog in discs) in antiferromagnetic NiO and CoO. From (J. Wu et al. 2011).

Finally, the last textures of interest are magnetic skyrmions, which are topological magnetic defects (Nagaosa and Tokura 2013). Compared to skyrmions in ferromagnets, the advantage of skyrmions in (bipartite) antiferromagnets is that the topological indices are opposite for each sublattice (FIG. 13). This opposition cancels the Magnus force and thus eliminates the unwanted transverse velocity, thereby enhancing skyrmion mobility, as computed by Barker and Tretiakov (Barker and Tretiakov 2016). Magnetic skyrmions can be found either in the bulk or at magnetic interfaces and require either bulk or interfacial Dzyaloshinskii-Moriya interaction (DMI). Since bulk DMI is more prevalent in antiferromagnetic materials, the potential for skyrmions in antiferromagnets is larger. Although antiferromagnetic skyrmions have yet to be directly observed, indirect evidence of these textures has been reported (from magnetoresistive data), for example in La$_2$Cu$_{1-x}$Li$_x$O$_4$, a La$_2$CuO$_4$ antiferromagnetic insulator doped with Li (Raičević et al. 2011).

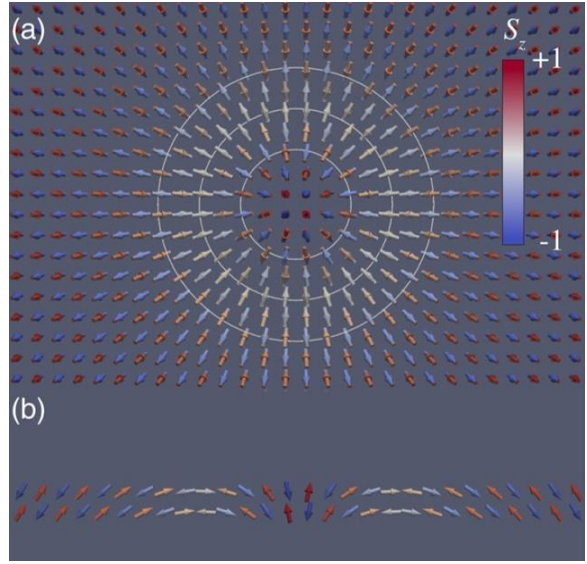

FIG. 13 (Color online). Computed Skyrmion in a G-type (checkerboard texture) antiferromagnet. (a) Top view and (b) cross section. The Skyrmion radius is 2.1 nm. From (Barker and Tretiakov 2016).

### *Relativistic dynamics*

An effective low-energy long-wavelength theory of bipartite antiferromagnets can be developed in terms of the two continuum fields introduced earlier, $\mathbf{l}(\mathbf{r},t)$ and $\mathbf{m}(\mathbf{r},t)$. In equilibrium and in the absence of external fields, $\langle \mathbf{l} \rangle \neq 0$ and $\mathbf{m} \equiv 0$, in the ordered phase. Under time reversal, $\mathbf{l} \rightarrow -\mathbf{l}$ and $\mathbf{m} \rightarrow -\mathbf{m}$, while under the space-group transformation the two sublattices are swapped, $\mathbf{l} \rightarrow -\mathbf{l}$ and $\mathbf{m} \rightarrow \mathbf{m}$ (together with a possible global rotation). The textbook Lagrangian density of the simplest isotropic cubic (Heisenberg) antiferromagnet (Auerbach 1994; Sachdev 1999) is given by

$$\mathcal{L}[\mathbf{l}, \mathbf{m}] = s\mathbf{m} \cdot \mathbf{l} \times \partial_t \mathbf{l} - \frac{A}{2}(\partial_t \mathbf{l})^2 - \frac{\mathbf{m}^2}{2\chi_\perp} - \mathbf{m} \cdot \mathbf{b} \,.$$

(14)

omitting the topological term (which is not important in two or three spatial dimensions). Here, $s = \hbar S / \mathcal{V}$ is the saturated spin density ($S$ is the local spin and $\mathcal{V}$ the volume per site), $\mathbf{b} = \gamma s \mu_0 \mathbf{H}$ the normalized magnetic field, $A$ the staggered-order stiffness, $\chi_\perp$ the (transverse) magnetic susceptibility, and $i$ sums over spatial dimensions. Both $A$ and $\chi_\perp^{-1}$ are proportional to $J_{AF} S^2$, where $J_{AF}$ is the microscopic exchange energy. This Lagrangian sets the stage for the *nonlinear sigma model* of coarse-grained antiferromagnetic dynamics (in real time $t$) and thermodynamics (in imaginary time $\tau = it$).

The coherent-state functional integration corresponding to the Lagrangian $\mathcal{L}_a(\mathbf{l})$ is Gaussian with respect to $\mathbf{m}$ and can be easily carried out to eliminate it [by completing the associated square in Eq. (14)], resulting in

$$\mathcal{L}[\mathbf{l}] = \frac{\chi_\perp}{2}(s\partial_t \mathbf{l} + \mathbf{l} \times \mathbf{b})^2 - \frac{A}{2}(\partial_t \mathbf{l})^2 \,.$$

(15)

To this, local anisotropies can be added as functions of the Néel field $\mathbf{l}, \mathcal{L}_a(\mathbf{l})$. For an easy-axis anisotropy along an axis $\boldsymbol{c}$, for example, $\mathcal{L}_a = K(\mathbf{l} \cdot \mathbf{c})^2 / 2$, where $|\mathbf{c}| = 1$ and $K > 0$. The equation governing the dynamics of the Néel order parameter is thus

$$\mathbf{l} \times \left(c^2 \partial_i^2 \mathbf{l} - \partial_t^2 \mathbf{l} - \partial_\mathbf{l} W_a\right) = \frac{1}{s^2}(\mathbf{l} \cdot \mathbf{b})\mathbf{l} \times \mathbf{b} + \frac{2}{s}(\mathbf{l} \cdot \mathbf{b})\partial_t \mathbf{l} - \mathbf{l} \times (\partial_t \mathbf{b} \times \mathbf{l}) + \frac{\alpha}{s\chi_\perp}\mathbf{l} \times \partial_t \mathbf{l} \,,$$

(16)

where $c = \sqrt{A / s^2 \chi_\perp}$ is the magnon velocity and $W_a = -L_a / s^2 \chi_\perp$. When $\mathbf{b}, \alpha \to 0$, the spatial and temporal derivatives appear in Lorentz-invariant combinations (Ivanov and Kolezhuk 1995; Bar'yakhtar and Lvanov 1983; Haldane 1983). In other words, Eq. (16) is invariant under the transformations (S. K. Kim, Tserkovnyak, and Tchernyshyov 2014)

$$t \mapsto t' = \frac{t - vx / c^2}{\sqrt{1 - v^2 / c^2}}, \quad x \mapsto x' = \frac{x - vt}{\sqrt{1 - v^2 / c^2}} \ .$$

(17)

This Lorentz invariance results in the relativistic-like dispersion of the antiferromagnetic spin waves mentioned earlier, $\omega^2 \sim \omega_0^2 + c^2 q^2$ in uniaxial antiferromagnets. Following this transformation, the profile of the domain wall, $\mathbf{l}_v(x,t)$, can be obtained from the profile of the zero-velocity domain wall, $\mathbf{l}_0(x,t)$, as follows:

$$\mathbf{l}_v(x,t) = \mathbf{l}_0\left( \frac{x - vt}{\sqrt{1 - v^2 / c^2}}, \frac{t - vx / c^2}{\sqrt{1 - v^2 / c^2}} \right) .$$

(18)

Hence, antiferromagnetic domain walls experience Lorentz contraction when their velocity approaches $c$ (Kosevich, Ivanov, and Kovalev 1990). This contraction has been detected in both spin torque and magnon-driven domain wall motion, and will be further discussed in sections III and IV.

Notice that the antiferromagnetic textures are accompanied by a small intrinsic magnetization, driven by the spatial variation of the Néel order parameter $\mathbf{m} \sim -\partial_t \mathbf{l}$, which enables manipulation of the domain wall using spatial gradients of magnetic fields (Tveten et al. 2016).

*Field-driven manipulation of domain walls*

Although the Néel order parameter cannot be readily manipulated using reasonable (i.e., <1T) static external magnetic fields, different strategies have recently been proposed to induce domain wall motion using fields varying in time or space. Gomonay et al. (O. Gomonay, Klaui, and Sinova 2016) recently proposed exploitation of magnetic pulses. They showed that a magnetic pulse couples efficiently to the antiferromagnetic domain wall via a force $F \sim \partial_t \mathbf{H} \cdot (\mathbf{l} \times \partial_t \mathbf{l})$. Therefore, two adjacent walls move synchronously in the same direction. However, the direction of motion depends on the derivative of the magnetic pulse, such that a symmetric pulse induces no overall displacement. In contrast, an asymmetric magnetic pulse can enable "ratchet" motion of the wall (O. Gomonay, Klaui, and Sinova 2016). For a typical uniaxial antiferromagnet with high Néel temperature, a train of 100 Oe nanosecond magnetic pulses produces an average velocity of 0.44 m.s$^{-1}$.

Finally, we mentioned above that Tveten et al. (Tveten et al. 2016) pointed out the potential relevance of the intrinsic magnetization that emerges from the spatial gradient of the Néel order parameter. In fact, the authors showed that such intrinsic magnetization $\mathbf{m} \sim -\partial_t \mathbf{l}$ efficiently couples to an external field $\mathbf{H}$, such that the force exerted on the domain wall reads $F \sim \partial_t \mathbf{H} \cdot \partial_t \mathbf{l}$. Simulations demonstrate that a field gradient of about 100 Oe.nm$^{-1}$ induces a velocity of about 50 m.s$^{-1}$.

## II.     SPIN TRANSFER ELECTRONICS

Spin transfer electronics encompasses the phenomena resulting from the strong exchange interactions between the spins of conduction electrons and the local moments of the lattice. These effects cover the spin transfer torque, enabling the electrical manipulation of ferromagnetic materials, for example via domain wall motion, and magnetization dynamics and reversal, as well as its Onsager reciprocal - the spin pumping effect - where a precessing magnetization pumps a spin current into an adjacent normal metal. This section reviews whether and how spin transfer effects create effective torques on antiferromagnetic devices and textures, and how antiferromagnetic dynamics promote spin pumping. Microscopic parameters quantifying the spin transfer effects, such as spin mixing conductance and spin penetration depth are also addressed. Finally, giant and tunneling magnetoresistive effects are discussed.

### A.     Spin transfer torque and spin pumping

### 1.     Principle of spin transfer torque

In ferromagnetic spin-valves, tunnel junctions or magnetic domain walls, spin transfer torque arises from the transfer of spin angular momentum from a flowing spin current to the local magnetic environment. The study of spin transfer torques in magnetic devices took off when Slonczewski (Slonczewski 1996) and Berger (Berger 1996) independently predicted current-induced magnetization switching in metallic spin-valves. Interested readers may complement their knowledge by consulting the various reviews available on this topic (Ralph and Stiles 2008; Brataas, Kent, and Ohno 2012).

In ferromagnetic systems lacking spin-orbit coupling, spin torque can be expressed as the gradient of spin current, polarized transversally to the local magnetization $\mathbf{m}$, i.e.,

$$\tau = -\mathbf{m} \times \left[ \left( \nabla \cdot J_s \right) \times \mathbf{m} \right],$$

(19)

where $J_s$ is the spin current tensor. When spin relaxation and/or spin-orbit coupling are present, this expression is no longer valid since spin angular momentum is not entirely transferred to the local magnetization. In general, it is more convenient to define the spin torque in terms of the torque between the local nonequilibrium spin density $\delta \mathbf{s}$ and the magnetization,

$$\tau = \frac{2\Delta}{\hbar} \delta \mathbf{s} \times \mathbf{m},$$

(20)

where $\Delta$ is the exchange parameter between itinerant and local electron magnetic momenta.

In antiferromagnets, the moments' layout is staggered and it is difficult to attribute the spin torque to *total* spin current absorption. Indeed, the incoming spins precess at different rates about the local magnetic moments of the different sublattices making up the antiferromagnet, resulting in local torques that are equal to the *local* transfer of angular momentum. Hence, to describe spin transfer torque in antiferromagnets it is necessary to identify the torque exerted on each individual magnetic moment making up the antiferromagnetic unit cell. For instance, in compensated bipartite antiferromagnets, injected spins precess in opposite directions on the different sublattices causing spin torque to occur locally even though there is no overall spin precession at the level of the magnetic unit cell. In general, the torque exerted on a sublattice $\alpha$ reads

$$\tau = \tau_{\parallel}^{\alpha} \mathbf{m}_{\alpha} \times \left( \mathbf{p} \times \mathbf{m}_{\alpha} \right) + \tau_{\perp}^{\alpha} \mathbf{m}_{\alpha} \times \mathbf{p} \,,$$

(21)

where $\mathbf{m}_{\alpha}$ is the local direction of the magnetic moment of sublattice $\alpha$ and $\tau_{\parallel}^{\alpha}$ ($\tau_{\perp}^{\alpha}$) is the (possibly space dependent) magnitude of the torque component that lies in (out of) the $(\mathbf{m}_{\alpha}, \mathbf{p})$ plane. The vector $\mathbf{p}$ is related to the symmetry of the system. For instance in spin-valves and tunnel junctions, it represents the direction of the order parameter of the polarizing layer (either ferro- or antiferromagnetic), while in the context of spin-orbit torques, it is related to the direction of the current, $\mathbf{j}$, with respect to the symmetry of the structure (e.g. $\mathbf{p} \sim \mathbf{z} \times \mathbf{j}$ for spin Hall or Rashba torques). In Eq. (21) the first term is usually referred to as damping-like torque, while the second term is the field-like torque.

A crucial question is *"what type of torque can reorient the Néel order parameter?"*. To address this question we consider three simple situations, depicted in FIG. 14. FIG. 14(a) displays the case where an external field $\mu_0 \mathbf{H}$ is applied perpendicular to the Néel order parameter (for the sake of simplicity, anisotropy fields are neglected). Under such a field, the magnetic moment of the two sublattices $\mathbf{m}_A$ and $\mathbf{m}_B$ cant in such a way that the torque exerted by the external field on each magnetic moment (black symbols) exactly compensates for the exchange torque exerted by one sublattice on the other (blue and red symbols). The bottom panel shows the same situation represented in terms of Néel order parameter $\mathbf{l} = \left( \mathbf{m}_A - \mathbf{m}_B \right) / 2$ and effective magnetic order parameter $\mathbf{m} = \left( \mathbf{m}_A + \mathbf{m}_B \right) / 2$: the torques $\sim \mathbf{l} \times \mathbf{m}$ and $\sim \mathbf{m} \times \mathbf{H}$ compensate each other. Hence, the Néel order parameter cannot be reoriented by the magnetic field. If we now consider a situation where the external field is *staggered*, i.e., of opposite sign on opposite sublattices (FIG. 14(b)), such a torque can be generated, e.g. in CuMnAs (see section III.D.2). In this case, the staggered field cants the magnetic moments and does not compensate the exchange field, such that the Néel order parameter $\mathbf{l}$ precesses around the induced magnetization $\mathbf{m}$. Similarly, when a damping-like spin torque is applied ($\sim \mathbf{m}_\alpha \times (\mathbf{p} \times \mathbf{m}_\alpha)$), it also cants the magnetic moments but does not compensate the exchange torque, resulting in precession of the Néel order parameter around the effective magnetization $\mathbf{m}$, see FIG. 14(c).

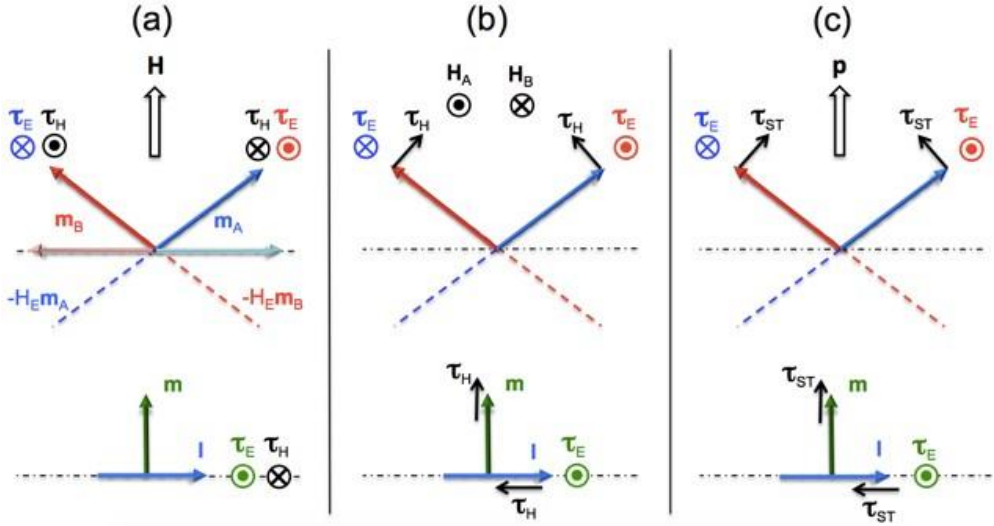

FIG. 14 (Color online). Illustration of the torques ($\tau_i$) exerted on two antiferromagnetically coupled magnetic moments $\mathbf{m}_A$ and $\mathbf{m}_B$, for three different external forces: upon application of (a) an external field ($\mathbf{H}$), (b) a staggered field ($\mathbf{H_A} = \mathbf{-H_B}$) or (c) a spin transfer torque ($\sim \mathbf{m}_\alpha \times (\mathbf{p} \times \mathbf{m}_\alpha)$). On the top panels, the blue and red arrows represent the normalized spin direction of the A and B sublattices, together with their respective exchange torques. The bottom panels show the effective (field-driven or exchange-driven) torques applied on the induced magnetization $\mathbf{m}$ and Néel order parameter $\mathbf{l}$ In the case of an external field (a), the two torques compensated each other, simply resulting in a static canted configuration, while in the two other cases, (b) and (c), the exchange torque was not compensated and induced reorientation of the Néel order parameter, $\mathbf{l}$.

In other words, the torque that enables reorientation of the Néel order parameter either takes the form of an antidamping torque $\sim \mathbf{m}_\alpha \times (\mathbf{p} \times \mathbf{m}_\alpha)$ $[\equiv \mathbf{l} \times (\mathbf{p} \times \mathbf{l})]$ or that of a *staggered* field-like torque $(-1)^\alpha \mathbf{m}_\alpha \times \mathbf{p}$ $(\equiv \mathbf{l} \times \mathbf{p})$, with $\alpha$ equal to 0 and 1 for sublattices A and B, respectively. In both cases, the torque arises from a *staggered* spin density, i.e., a spin density that changes sign on opposite sublattices. Although the present discussion provides phenomenological arguments about the symmetry the torque needed to control antiferromagnets, it is clearly oversimplified as it disregards the role of magnetic anisotropies and damping, which are crucial to understanding the actual current-driven dynamics of antiferromagnets. These aspects are discussed in more detail in the next section.

## 2.    Manipulation of the order parameter by spin transfer torque

### *Antiferromagnetic spin-valves*

Manipulating antiferromagnets efficiently and reliably is at the heart of all-antiferromagnetic spintronics applications (MacDonald and Tsoi 2011). For example a simple switching of an antiferromagnetic order parameter could represent the writing operation of a magnetic random access memory based on antiferromagnetic elements. Such a switching of an antiferromagnet may be realized via the so-called spin-transfer torque effect which has been used successfully in ferromagnetic systems to switch magnetic moments by spin currents.

The first prediction of spin transfer torques in antiferromagnets was made by Núñez et al. (Núñez et al. 2006). The authors studied spin transport in a one-dimensional metallic spin valve composed of two antiferromagnets separated by a metallic spacer. When injecting a current through the system, a staggered spin density builds up in the first antiferromagnet and is transmitted to the second antiferromagnet. When the two Néel order parameters are misaligned, the transmitted staggered spin density exerts a torque on the local magnetic moments of the second antiferromagnet. The torque computed by Núñez et al., displayed in FIG. 15, produces a large staggered in-plane component and a spatially inhomogeneous out-of-plane component. In addition to the spin torque, the authors also predict a magnetoresistive effect arising from spin-dependent quantum interferences in the metallic spacer (see II.C.1).

Noticeably, both torques extend over the length of the antiferromagnet in this model. Because of the alternating orientations of the moment in the antiferromagnetic layers, no global spin precession occurs at the level of the magnetic unit cell, resulting is a much weaker spin dephasing and therefore a much longer spin penetration length. This is in sharp contrast with ferromagnetic spin-valves, where the torque is localized at the interface due to the large spin dephasing of the incoming spin current inside the ferromagnet, e.g. (Stiles and Zangwill 2002). Because of the alternating moment orientations in antiferromagnetic layers, commensurate staggered spin density and subsequent torques occur generically. It follows that spin transfer torques in antiferromagnets can act cooperatively over a longer distance from the interface. This plus the absence of shape anisotropy in antiferromagnetic materials explain that smaller critical currents were predicted for perfect antiferromagnets compared to ferromagnets (Núñez et al. 2006). Similar results were obtained by first principle methods on FeMn/Cu/FeMn (Xu, Wang, and Xia 2008) and Cr/Au/Cr (Haney et al. 2007), which confirmed the non-locality of the spin torque in antiferromagnets. Núñez et al. (Núñez et al. 2006) suggest that the staggered in-plane component of the torque (FIG. 15 top panel) makes electrical manipulation of the Néel order possible. However, as discussed above, careful analysis of the Néel order dynamics indicates that, rather it is the inhomogeneous out-of-plane torques (FIG. 15 bottom panel) that can be used to manipulate the antiferromagnetic order parameter.

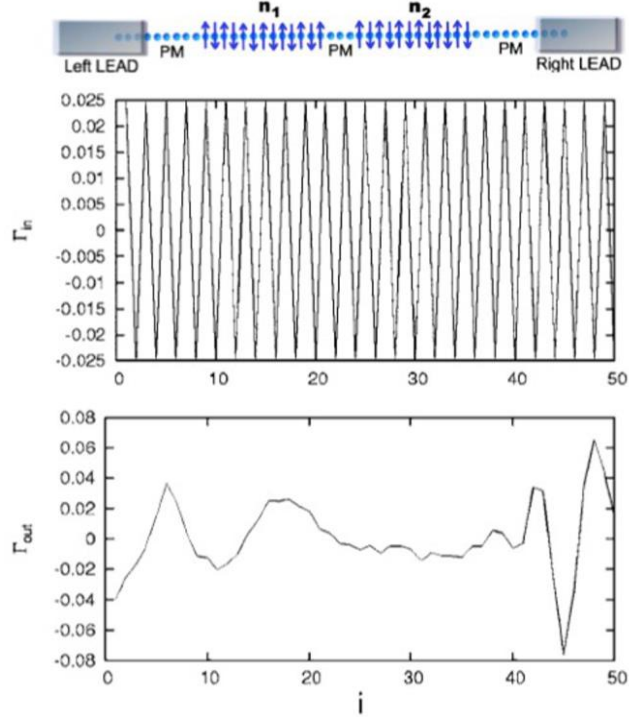

FIG. 15 (Color online). Spatial profile of the local spin torque in the free antiferromagnetic layer of a one-dimensional spin-valve. While the in-plane component is staggered, the out-of-plane component is spatially inhomogeneous, providing an effective non-vanishing torque. Adapted from (Núñez et al. 2006). The order parameters of the right and left leads are rotated by $\pi/2$ around the current direction, meaning that the out-of-plane torque points toward the current direction.

The calculations performed above considered clean metallic multilayers and idealized bipartite antiferromagnetic materials. In contrast, realistic magnetic multilayers possess dislocations, defects, grain boundaries as well as interfacial roughness, resistivity mismatch, randomly spread spin-glass like phases at antiferromagnetic interfaces, peculiar spin-structures etc. (Berkowitz and Takano 1999). These imperfections stimulate quantum decoherence and momentum scattering that dramatically impact the spin transport in antiferromagnetic spin-valves. As a matter of fact, in contrast with ferromagnetic spin-valves, which are well described within incoherent semiclassical models, quantum coherence is crucial to enable the transmission of staggered spin density from one part of the spin-valve to the other. Recent tight binding calculations (Duine et al. 2007; Saidaoui, Manchon, and Waintal 2014) have indeed demonstrated that spin dephasing and mere spin-independent disorder in the spin-valve dramatically quenches the spin torque efficiency (see FIG. 16).

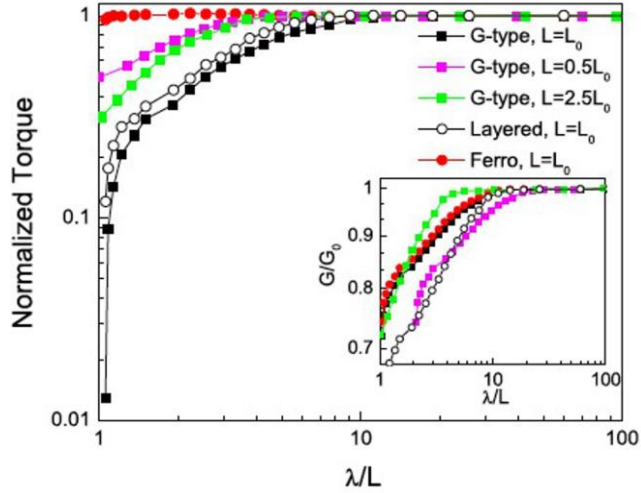

FIG. 16 (Color online). Dependence of the torque as a function of the mean free path $\lambda/L$ in the spacer in antiferromagnet/spacer/antiferromagnet spin-valves ($L$ being the length of the spacer). G-type (checkerboard) and L-type (layered) antiferromagnets where considered and the case of the ferromagnetic spin-valve is given for comparison (red symbols). From (Saidaoui, Manchon, and Waintal 2014).

Since damping-like spin torques make electrical control of the Néel order parameter possible, spin-valve configurations involving a ferromagnetic polarizer rather than an antiferromagnetic should be more promising (H. V. Gomonay and Loktev 2010). Haney and MacDonald (Haney and MacDonald 2008) revealed that the torque between the ferromagnet and the antiferromagnet vanishes when the ferromagnetic order parameter lies along one of the antiferromagnet's axes of spin-rotation symmetry. In compensated bipartite antiferromagnets, the torque has a $\sim sin(2\theta)$ angular dependence (Haney and MacDonald 2008), while in antiferromagnets with a 3Q spin structure [the 3Q spin structure is for example that of $Ir_{20}Mn_{80}$ in the $L1_2$ crystalline phase (TABLE 1)], the torque adopts the 3-fold symmetry of the antiferromagnet (Prakhya, Popescu, and Haney 2014). Interestingly, in both cases the authors showed that when coherent quantum transport is maintained, the torque exerted by the antiferromagnet on the ferromagnet stabilizes the *perpendicular* configuration of the two magnetic order parameters.

Tight-binding approaches are powerful tools to investigate spin transport in antiferromagnets but lack transparency, particularly to model disordered systems. An alternative is to describe spin transport within the framework of spin diffusion theory (Manchon 2017a). This theory assumes that the transport is incoherent (scattering is strong) and parses the spin density in two components: a uniform $\mathbf{s} = \mathbf{s}_A + \mathbf{s}_B$ and a staggered component $\delta\mathbf{s} = \mathbf{s}_A - \mathbf{s}_B$, where $\mathbf{s}_{A,B}$ is the non-equilibrium spin density on sublattice A and B, respectively. The uniform spin density is governed by an anisotropic drift-diffusion equation with respect to the order parameter,

$$\partial_t \mathbf{s} + \vec{\nabla} \cdot J_s = -\frac{1}{\tau_{sf}} \mathbf{s} - \frac{1}{\tau_\varphi} \mathbf{l} \times (\mathbf{s} \times \mathbf{l}),$$

(22)

where $J_s$ is the total spin current density (averaged over the magnetic unit cell). The first term on the right-hand side $\propto 1/\tau_{sf}$ is the isotropic spin relaxation while the second term, $\propto 1/\tau_\varphi$, only relaxes the spin component *transverse* to the Néel order parameter $\mathbf{l}$. The latter accounts for the enhanced dephasing of the transverse spin components due to precession around the magnetic moments of the two sublattices. The staggered spin density then reads $\delta\mathbf{s} = \eta\mathbf{l} \times \mathbf{s}$ where $\eta = \tau^* / \tau_\Delta$ is the ratio between the spin precession time around the magnetic moment of one sublattice $\tau_\Delta$ and the time the carrier spends on this sublattice, $\tau^*$. The torque exerted on the Néel order is therefore $\mathbf{T} = (2\Delta/\hbar)\mathbf{l} \times \delta\mathbf{s}$. Applying this theory to spin-valves and metallic bilayers involving antiferromagnets confirms the results obtained using tight-binding models (Manchon 2017a). It provides a useful tool to explicitly model spin transport and torque in such systems.

### Antiferromagnetic tunnel junctions

Since spin torque is extremely sensitive to disorder in antiferromagnetic spin-valves, one needs to find a way to prevent momentum scattering inside the spacer. This can be done by replacing the metallic spacer by a tunnel barrier (Merodio, Kalitsov, et al. 2014a).

Tight-binding models of one- (Merodio, Kalitsov, et al. 2014a) and two-dimensional antiferromagnetic tunnel junctions (Saidaoui et al. unpublished) were recently developed. The approach was extended to the case of tunnel junctions with ferrimagnetic electrodes (Merodio, Kalitsov, et al. 2014b). In the case of one-dimensional spin-valves composed of a ferromagnetic polarizer and an antiferromagnetic free layer (FIG. 17) (Merodio, Kalitsov, et al. 2014a), the in-plane torque is found to be staggered like in their metallic counterparts (FIG. 15). In contrast with metallic spin-valve though, the out-of-plane torque remains quite large. In antiferromagnet/tunnel-barrier/antiferromagnet tunnel junctions, the symmetry of the torques is found to be the same as in metallic spin-valves and their bias dependence is similar to that in ferromagnetic tunnel junctions (Saidaoui et al. unpublished) (FIG. 18). Notice that the staggered nature of the in-plane torque (in both metallic and tunneling spin-valves) is lost when considering multiple bands (i.e. multiple orbitals, realistic Fermi surface, energy integration) due to enhanced dephasing.

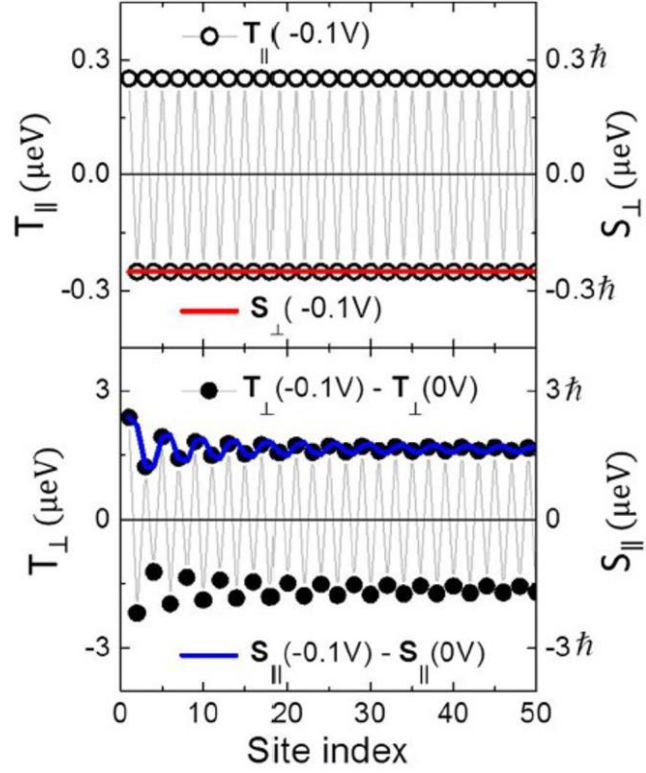

FIG. 17 (Color online). Spatial distribution of the spin torque (left axis) and associated spin density (right axis) in a ferromagnet/tunnel-barrier/antiferromagnet. From (Merodio, Kalitsov, et al. 2014a).

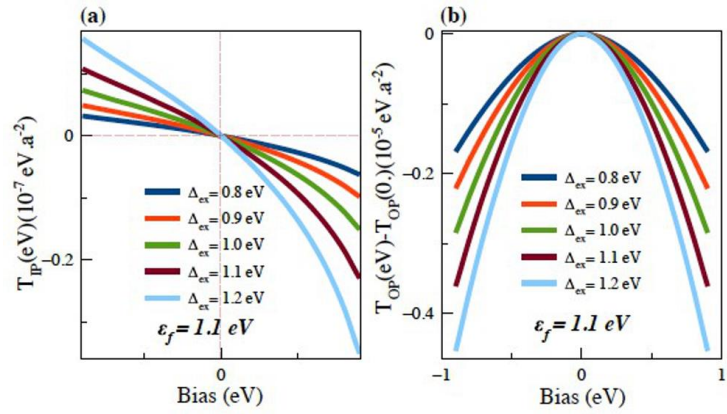

FIG. 18 (Color online). Bias dependence of the (a) in-plane and (b) out-of-plane in a two-dimensional antiferromagnet/tunnel-barrier/antiferromagnet tunnel junction for different exchange parameters. From (Saidaoui et al. unpublished).

*Seeking spin transfer torque experimentally*

The first prediction of antiferromagnetic spin-transfer torque (Núñez et al. 2006) was soon followed by experiments aiming at probing the effects of spin-polarized currents on the antiferromagnetic order parameter. A simple ferromagnet/antiferromagnet (F/AF) bilayer represents a natural test system where an electrical current can be first (spin) polarized, by being driven across the ferromagnetic layer, and then injected directly into the antiferromagnet. It is, however, a challenge to detect the resulting effect of the polarized current on the magnetic order in the antiferromagnet. Wei et al. (Wei et al. 2007) explored the exchange bias (I.C.3) at the ferromagnet/antiferromagnet interface as an indirect means to inquire about what happens with the antiferromagnetic order parameter near the interface. Indeed, as the exchange bias phenomenon is known to be associated with the interfacial antiferromagnets' moments, the observed current-induced variation of the exchange-bias field ($H_{EB}$) can be taken as the first evidence of antiferromagnetic spin transfer torque. In this experiment the action of a spin current on the antiferromagnet was demonstrated with currents flowing perpendicular to plane across a $F_s$/N/$F_{pol}$/AF polycrystalline spin valve: CoFe(10 nm)/Cu(10 nm)/CoFe(3 nm)/FeMn(8 nm). The subscripts 's' and 'pol' stand for 'sensing' and 'polarizing', respectively, and N is a 'non-magnetic' metallic spacer. Here $F_s$/N is only a probe of the exchange bias field and the corresponding layers compositions and thicknesses were on purpose designed not to perturb the $F_{pol}$/AF spin dependent interactions (e.g. N is thick enough to lower the $F_s - F_{pol}$ mutual spin transfer torque but thin enough to allow giant magnetoresistance detection). In order to achieve high enough current densities a point-contact technique was used: the current was injected into the spin valve through a point contact and propagated across the spin-valve stack directly into a back copper electrode to ensure a close to perpendicular-to-plane flow. Here the point contact is used as a local probe of the spin-valve resistance.

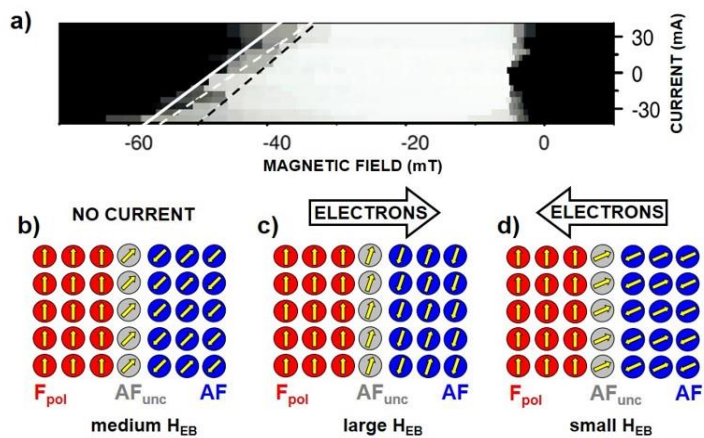

FIG. 19 (Color online). (a) Two-dimensional grey-scale plot of the spin-valve resistance as a function of the applied magnetic field and bias current. Lighter color indicates higher resistance. Linear fits (see text for details) indicate the trend of exchange-bias field $H_{EB}$ shift with current. (b-d) Intuitive picture of antiferromagnetic spin transfer torque for a ferromagnet/antiferromagnet ($F_{pol}$/AF) interface. The current gets spin polarized in $F_{pol}$. The transmitted (reflected) polarized current transfers torque to the uncompensated AF spins: $AF_{unc}$, which rotate away from (toward) the $F_{pol}$ spins. It influences the $F_{pol}$ magnetizations reversals and results in a reduction (enhancement) of $H_{EB}$. From (Wei et al. 2007).

The key experimental result of Wei et al (Wei et al. 2007) is shown in FIG. 19(a). The two-dimensional gray-scale plot shows the spin-valve resistance as a function of the applied magnetic field and bias current. The data were obtained from (down) magnetoresistance sweeps at different bias currents. White color indicates a high resistance of the spin valve in the antiparallel state − $F_s$ is antiparallel to $F_{pol}$; black color indicates a low resistance in the parallel state − $F_s$ is parallel to $F_{pol}$. The transition between black and white at a small negative field (~ -5 mT) corresponds to the reversal of $F_s$; this transition is almost independent of the applied current. The transition at a large negative field (between -60 and -30 mT) corresponds to the reversal of $F_{pol}$ and identifies $H_{EB}$; this transition shifts approximately linearly with the applied current and highlights the dependence of $H_{EB}$ on the applied current. Similar behavior was also found with IrMn alloys (Basset et al. 2008; Wei, Basset, et al. 2009).

The intuitive picture describing the consequences of spin transfer torque on an antiferromagnet is shown in FIG. 19(b-d). The cartoon shows an interface between $F_{pol}$ and AF. The surface layer of the latter contains uncompensated magnetic moments $AF_{unc}$ (gray). A fraction of these uncompensated moments are pinned, thereby inducing an energy barrier for $F_{pol}$ reversal; they do not reverse when $F_{pol}$ reverses and are responsible for the existence of exchange bias. Electrons flowing from the ferromagnet into the antiferromagnet induce spin transfer torques which alter its magnetic configuration. These torques tend to favor parallel alignment of moments at the $F_{pol}$/AF interface and will therefore tend to increase $H_{EB}$. Electrons flowing in the opposite direction will tend to have the opposite effect. The cartoon picture in FIG. 19(b-d) highlights a major issue for potential applications of ferromagnet/antiferromagnet bilayers, e.g. in antiferromagnet memory applications. The current-induced changes of antiferromagnetic order are present only as long as the current is present too. As soon as the current is removed the antiferromagnet goes back to its original state just like an exchange spring. Furthermore the state of the art experiments with ferromagnet/antiferromagnet bilayers only give a qualitative picture. In particular, the experimental evidence of the antiferromagnetic spin transfer torque mentioned above used nonuniform current flows inherent to the point-contact technique (Wei et al. 2007). Quantitative data have to be determined. In addition experiments are sometimes perturbed by unstable antiferromagnetic configurations (Urazhdin and Anthony 2007) or reconfiguration originating from Joule heating and not spin transfer torque (X.-L. Tang et al. 2007; N. Dai et al. 2008; X. L. Tang et al. 2010). New experimental geometries and stacks symmetries need to be proposed. The effects of disorder discussed earlier need to be empirically quantified too, and possibly minimized.

This first set of experiments provides interesting clues as to the physics involved in exchange-biased ferromagnet/antiferromagnet interfaces but does not constitute a demonstration of antiferromagnetic spin transfer torque *per se*. As a matter of fact, these experiments reveal the influence of spin transfer torque on exchange bias – a complex phenomenon that is very sensitive to interface details (I.C.3) – suggesting that the interfacial spin texture can be altered by spin torques. However, these systems remain quite different from the antiferromagnetic spin-valves studied theoretically by Núñez et al. (Núñez and MacDonald 2006) and their followers. The major challenge that still needs to be addressed is whether spin current can be injected while independently detecting the Néel order parameter response. Up till now, attempts to detect magnetoresistance or torques in antiferromagnetic spin-valves have remained unconvincing. As discussed in section III, most of the difficulties can be solved by considering spin-orbit torques, rather than spin transfer torques.

### *Current-driven dynamics*

While the dynamics of antiferromagnets under ultrashort field pulses has been addressed in the context of optical manipulation, as introduced in section I.C.3, their dynamics under current drive has only recently attracted interest. As mentioned in the previous section, the spin torque component that controls the antiferromagnetic order parameter arises from a staggered non-equilibrium local spin density and must have the generic form

$$\tau_A = \tau_\parallel \mathbf{m}_A \times (\mathbf{p} \times \mathbf{m}_A) + \tau_\perp \mathbf{m}_A \times \mathbf{p}$$
$$\tau_B = \tau_\parallel \mathbf{m}_B \times (\mathbf{p} \times \mathbf{m}_B) - \tau_\perp \mathbf{m}_B \times \mathbf{p}$$

$$(23)$$

which results in a torque on the order parameter

$$\partial_t^2 \mathbf{l} \times \mathbf{l} \Big|_\tau = \gamma \frac{H_E \tau_\parallel}{2} \mathbf{l} \times (\mathbf{p} \times \mathbf{l}) + \gamma H_E \tau_\perp \mathbf{l} \times \mathbf{p} \, .$$

(24)

Gomonay et al. (H. V. Gomonay and Loktev 2010; H. V. Gomonay, Kunitsyn, and Loktev 2012; H. V. Gomonay and Loktev 2014) have investigated the impact of the damping-like torque $\mathbf{l} \times (\mathbf{p} \times \mathbf{l})$ on the dynamics of an antiferromagnet considering various combinations of polarization direction vector $\mathbf{p}$ and external magnetic field $\mathbf{H}$ in the case of uniaxial and biaxial anisotropy.

This setup typically corresponds to a ferromagnetic/antiferromagnetic spin-valve or to a bilayer composed of an antiferromagnet and a heavy metal with a spin Hall effect (see section III). It is well-known that in ferromagnetic spin-valves, spin torque transfer exerts either a damping or an antidamping effect on the ferromagnetic order parameter, depending on the current direction, leading either to stabilization or destabilization of the magnetic state. In contrast, uniaxial antiferromagnets possess two degenerate excitation modes (see FIG. 6(a)) so that, above a certain critical current density $|J_{cr}|$, one of them is damped while the other is excited. Hence, uniaxial antiferromagnets are excited whatever the direction of the current. Interestingly, the antiferromagnetic order parameter tends to align perpendicularly to the polarization, i.e. $\mathbf{p} \perp \mathbf{l}$ consistently with Haney and MacDonald (Haney and MacDonald 2008) as illustrated in FIG. 20. The critical current above which excitations are triggered (H. V. Gomonay and Loktev 2010) reads

$$J_{cr} = \frac{\alpha M_s d}{\xi} \omega_{AF} \, .$$

(25)

where $\xi$ is the spin torque efficiency, $M_s$ is the saturation magnetization of one sublattice, and $\omega_{AF} = \gamma\mu_0\sqrt{H_K H_E}$ is the frequency of the excited mode, typically of the order of a few 100 GHz to 1 THz. Since this frequency is much higher than in ferromagnets (~GHz), the critical current above which such excitations are triggered is about two orders of magnitude larger. The case of biaxial or easy-plane anisotropy (e.g. NiO) presents instructive differences. Indeed, in this case the two excitation modes (acoustic and optical modes) are non-degenerate, and they behave differently in the presence of spin transfer torque (H. V. Gomonay and Loktev 2010). Depending on the configuration, the acoustic mode can be excited while the optical is damped. Cheng et al. (Cheng, Xiao, and Brataas 2016) recently demonstrated that spin Hall torque triggers self-sustained TeraHertz oscillations, as further discussed in section III.D.

The current-driven excitation of antiferromagnets has also been investigated theoretically by Cheng et al. (Cheng et al. 2015) who reported inertial switching of an antiferromagnet with uniaxial anisotropy under a current pulse. The principle of this switching is similar to that of the pulse field reversal computed by Wienholt et al. (Wienholdt, Hinzke, and Nowak 2012). During the pulse duration, the damping-like torque applied perpendicularly to the anisotropy axis cants the magnetic moment of the sublattices [e. g. FIG. 14(c)]. Notice that this canting remains very small (~0.1%) due to the extensive exchange. Thus, energy is transferred from the flowing spin current to the antiferromagnetic exchange. When the damping-like torque is turned off, the dynamics of the Néel order is triggered and may switch if a large enough current drive is applied (estimated to be about 6-7 $10^7$ A/cm$^2$ in NiO).

Spin transfer torques were also studied in less conventional structures, e.g. spin transfer torque on an antiferromagnet sandwiched between two ferromagnets (Linder 2011). In this case, the direction of the magnetization induced in the antiferromagnet depends on the relative magnetic orientation of the two ferromagnets.

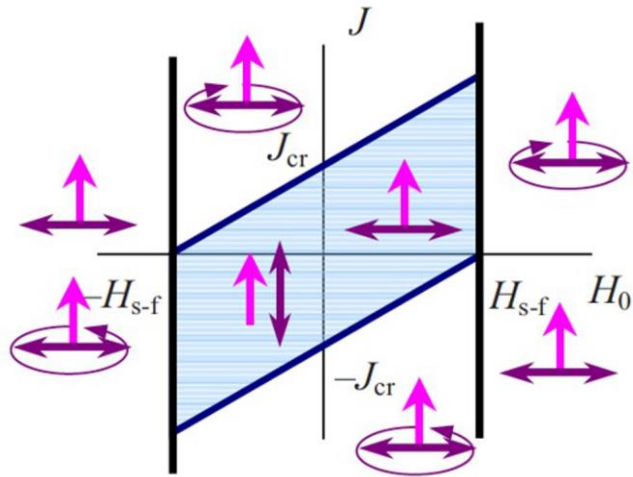

FIG. 20 (Color online). Stability phase diagram of a ferromagnet/antiferromagnet bilayer. From (H. V. Gomonay and Loktev 2010). In the central region (blue) a bistable state is found where the antiferromagnetic order parameter (purple arrows) can be either perpendicular or parallel to the spin current polarization (magenta arrow). Above the critical current $J_{cr}$ and above the spin-flop field $H_{s-f}$, the system oscillates about the polarization direction.

Technical issues are among the main reasons why a direct experimental observation of the dynamics of antiferromagnetic thin films under the influence of a spin current has not been realized in spite of many theoretical predictions. The advancement and development of THz measurement technologies suitable for antiferromagnetic thin films are definitely required (section I.C.3.). In order to get around the THz measurement circuitry, down conversion of the measuring frequency to a microwave regime by the exchange coupled ferromagnetic/antiferromagnetic bilayer is again one of the possible schemes to extract the beneficial information of the antiferromagnet dynamics (Moriyama, Takei, et al. 2015). Measurement around the spin flopping (I.C.3) of the antiferromagnetic magnetization maybe another alternative since the antiferromagnetic resonant frequency can come down to a microwave range when spin flopping occurs. For instance, MnF$_2$ alloys exhibit spin flop at around 9 Tesla (FIG. 6(a)) and an antiferromagnetic dynamics is observed at tens of GHz which is comfortably measurable with the conventional microwave circuitry (P. Ross et al. 2015).

## 3. Moving magnetic textures by spin transfer torque

Spin transfer torque can be used to efficiently manipulate antiferromagnetic domain walls and skyrmions. When the antiferromagnetic texture couples to electronic (Hals, Tserkovnyak, and Brataas 2011; Swaving and Duine 2011; Cheng and Niu 2014) or magnonic (Tveten, Qaiumzadeh, and Brataas 2014; S. K. Kim, Tserkovnyak, and Tchernyshyov 2014) spin flows $\mathbf{j}_S$, the dynamic equations derived from the Lagrangian, Eq. (14), acquire additional torques $\boldsymbol{\tau}$

$$s\partial_t \mathbf{l} = \mathbf{l} \times \mathbf{f_m} + \boldsymbol{\tau_l} \ ,$$

$$s(\partial_t \mathbf{m} + \alpha \mathbf{l} \times \partial_t \mathbf{l}) = \mathbf{l} \times \mathbf{f_l} + \mathbf{m} \times \mathbf{f_m} + \boldsymbol{\tau_m} \ .$$

$$(26)$$

Here, the effect of the torque $\boldsymbol{\tau_l} \sim (\mathbf{j}_S \cdot \nabla)\mathbf{l}$ is reduced relative to $\boldsymbol{\tau_m} \sim \mathbf{l} \times (\mathbf{j}_S \cdot \nabla)\mathbf{l}$ by the small parameters $\hbar\omega / J_{AF}, \mu_B \mu_0 H / J_{AF}$, which is rooted in the smallness of the susceptibility $\chi_\perp \propto J_{AF}^{-1}$. The physical meaning of $\boldsymbol{\tau_m}$ is self-evident: it is the (local) net transfer of the (spin) angular momentum onto the antiferromagnetic state. In systems with weak spin-orbit coupling, such transfer of spin is generally associated with hydrodynamic continuity flows. This makes it readily amenable to simple phenomenological treatments, as further discussed below. The equations of motion, Eqs. (26), have been derived and solved by several authors for the case of an electric drive. It is shown that the steady state velocity is given by the ratio $\sim \tau_{\mathbf{m}}/\alpha$, similarly to the case of ferromagnetic domain walls (Hals, Tserkovnyak, and Brataas 2011; Swaving and Duine 2011; Cheng and Niu 2014). Yamane et al. (Yamane, Ieda, and Sinova 2016) recently analytically calculated the torque efficiency using a tight-binding approach. These authors showed that a charge current predominantly couples to the Néel order parameter $\mathbf{l}$ in an exchange-dominant regime, while it couples mostly to the induced magnetization $\mathbf{m}$ in a mixing-dominant regime.

A convenient way to model the dynamics of magnetic solitons (such as domain walls and skyrmions, I.C.4) is to track the dynamics of collective coordinates parameterizing the slow modes of the system (Kosevich, Ivanov, and Kovalev 1990; Tveten et al. 2013; S. K. Kim, Tserkovnyak, and Tchernyshyov 2014). For a rigid translational texture motion, $\mathbf{l}(\mathbf{r},t)=\mathbf{l}[\mathbf{r}-\mathbf{R}(t)]$, the momentum $\mathbf{P}$ canonically conjugate to the center-of-mass position $\mathbf{R}$ is given (componentwise) by

$$P_i = \partial_{\dot{R}_i} L(\mathbf{R},\dot{\mathbf{R}}) = -\int dV\, \pi \cdot \partial_{r_i}\mathbf{l} \,,$$

(27)

where $L$ is the total Lagrangian associated with the rigid solitonic dynamics with the rigid solitonic dynamics, Eq. (14). The stochastic (solitonic quasiparticle) equations of motion (S. K. Kim, Tchernyshyov, and Tserkovnyak 2015) are then found as

$$\hat{M}\,\dot{\mathbf{R}} = \mathbf{P} \,,$$

$$\dot{\mathbf{P}} + \hat{\Gamma}\,\dot{\mathbf{R}} = \mathbf{F} + \mathbf{F}^{\text{th}}$$

(28)

Here, $\mathbf{F} \equiv -\partial_{\mathbf{R}} U$ is the deterministic force, $\hat{\Gamma}$ is the damping tensor with components $\Gamma_{ij} \equiv \alpha s \int dV (\partial_i \mathbf{l} \cdot \partial_j \mathbf{l})$, $\hat{M} = \tau \hat{\Gamma}$ is the mass tensor, where $\tau \equiv \rho / \alpha s$ is the viscous relaxation time. The stochastic force obeys the fluctuation-dissipation relation:

$$\langle F_i^{\text{th}}(t) F_j^{\text{th}}(t') \rangle = 2 k_B T \Gamma_{ij} \delta(t - t')$$

(29)

in the classical limit that is relevant for the slow dynamics. Focusing on the simplest case of an isotropic soliton, $\hat{M}$ and $\hat{\Gamma}$ become scalars, $M$ and $\Gamma$. Combining Eqs. (28), we then get a damped stochastic Newton's law:

$$\tau \ddot{\mathbf{R}} + \dot{\mathbf{R}} = \mu \mathbf{F} + \mathbf{\eta} ,$$

(30)

where $\mu \equiv \Gamma^{-1}$ is the mobility and

$$\langle \eta_i(t) \eta_j(t') \rangle = 2 D \delta(t - t') .$$

(31)

$D \equiv k_B T \mu$ (Einstein-Smoluchowski relation) gives the diffusion coefficient. Magnetic solitons thus undergo an ordinary Brownian motion of massive particles through a viscous medium.

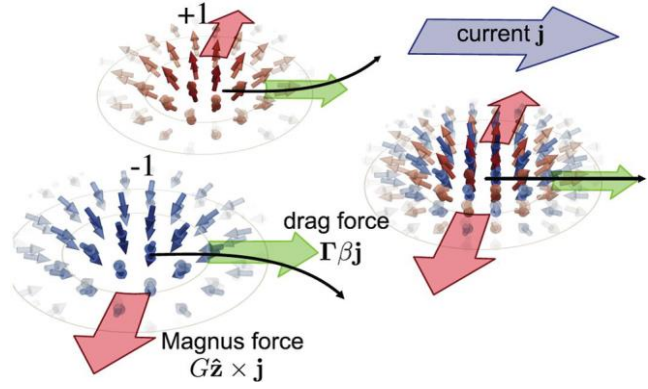

FIG. 21 (Color online). The Magnus force associated with skyrmion motion (either relative to a static background or an electronic or magnonic flow) is proportional to the topological charge in the ferromagnetic case. This transverse force is absent in antiferromagnets as the topological charge is odd under reversal of the local spin orientation, and thus cancels between the two tightly exchange-bound antiferromagnetic sublattices. From (Barker and Tretiakov 2016).

In contrast to the one-dimensional case, motion of the two-dimensional solitons (in the $xy$ plane), such as skyrmions, differs drastically in the antiferromagnets compared to the ferromagnets (Barker and Tretiakov 2016; X. Zhang, Zhou, and Ezawa 2016), as depicted schematically in FIG. 21. For ferromagnets, Eq. (28) is complemented with a gyrotropic force (Tretiakov et al. 2008; Wong and Tserkovnyak 2010):

$$(\Gamma - G\mathbf{z}\times)\dot{\mathbf{R}} = \mathbf{F} + \mathbf{F}^{\text{th}} \ ,$$

(32)

where $G \equiv s \int dx dy \, \mathbf{l} \cdot \partial_x \mathbf{l} \times \partial_y \mathbf{l} \equiv 4\pi s Q$ , in terms of the topological (skyrmion) charge $Q \in \mathbb{Z}$ . Note also that at the present level of treatment, disregarding possible internal degrees of freedom (apart from the translational motion), ferromagnetic solitons have no inertia (i.e., effectively $\rho \to 0$). The diffusion coefficient is thus reduced in the ferromagnetic case (Schütte et al. 2014)

$$D = k_B T \frac{\Gamma}{\Gamma^2 + G^2} \ .$$

(33)

In the typical case of weak damping $\alpha \ll 1$, so that, $\Gamma \ll G$, $D \propto \Gamma / G^2$ so that the diffusion coefficient is sped up by increasing the Gilbert damping $\alpha$. This is opposite to the antiferromagnetic case, $D \propto 1/\Gamma$. Brownian motion of ferromagnetic and antiferromagnetic skyrmions has been studied numerically in (Barker and Tretiakov 2016).

Including additional collective internal degrees of freedom can tremendously enrich the solitonic dynamics. A one-dimensional domain wall in an easy-axis antiferromagnet, for example, possesses not only the kinetic energy and inertia associated with its translational motion, but also rotational energy and moment of inertia associated with the precession of the order parameter at the domain-wall center about the easy axis (S. K. Kim, Tserkovnyak, and Tchernyshyov 2014). The coupled rotational and translational dynamics of such domain walls, as well as their interactions with spin waves [which can induce their motion (S. K. Kim, Tserkovnyak, and Tchernyshyov 2014; Tveten, Qaiumzadeh, and Brataas 2014)], have the Lorentz symmetry. The associated effective speed of light is given by the spin-wave velocity. Further insights on the thermal motion of antiferromagnetic textures and their interaction with spin waves are given in section IV.

## 4.    Spin pumping from antiferromagnets

We have seen in the previous section that currents can act on the antiferromagnetic order parameter. The reciprocal phenomenon exists and antiferromagnets can also be used to generate pure spin currents through spin pumping as first discussed by Takei et al. (Takei et al. 2014). The pumped currents are actually connected to current-induced torques via Onsager reciprocity relations. We remind readers that spin pumping (I.C.2) results from the non-equilibrium magnetization dynamics of a spin injector, which pumps a spin current ($\mathbf{J}_s \approx \frac{\hbar}{4\pi} g_r^{\uparrow\downarrow} \mathbf{m} \times \partial_t \mathbf{m}$) into an adjacent spin sink layer. The initial theoretical framework of spin pumping is built on adiabatic charge pumping and involves the interfacial parameter called spin mixing conductance ($g_r^{\uparrow\downarrow}$) (Tserkovnyak, Brataas, and Bauer 2002), (see also section II.B.1). More recently, a linear-response formalism was developed to complete the existing theories and describe spin pumping near thermal equilibrium (Ohnuma et al. 2014). The mixing conductance of a collinear (ferro or antiferro)magnetic system $g_r^{\uparrow\downarrow}$ is independent of the order parameter and therefore is nonvanishing in antiferromagnets. As a consequence, when antiferromagnetic order parameter precesses, it also pumps a spin current into the adjacent normal metal of the form ($\mathbf{J}_s \approx \frac{\hbar}{4\pi} g_r^{\uparrow\downarrow} \mathbf{l} \times \partial_t \mathbf{l}$) (Cheng et al. 2014), as illustrated in FIG. 22. Furthermore, upon magnetic field rf excitation, antiferromagnetic resonance produces two types of resonances related to two precession modes. These modes are accompanied by a small ferromagnetic component ($|\mathbf{m}| \sim \sqrt{H_K / H_E}$, where $H_K$ is the anisotropy and $H_E$ corresponds to the exchange interactions between moments in the antiferromagnet) that oscillates very fast and can thereby induce an additional spin current ($\mathbf{J}_s \approx \frac{\hbar}{4\pi} g_r^{\uparrow\downarrow} \mathbf{m} \times \partial_t \mathbf{m}$) (Cheng et al. 2014). The authors argue that the smallness of the magnetic moment ($H_K \ll H_E$) is compensated in part by the large precession frequency (THz). Recent spin Seebeck signal in antiferromagnet seems to confirm this concept (Seki et al. 2015; S. M. Wu et al. 2016). Sekine and Nomura (Sekine and Nomura 2016) recently highlighted nontrivial charge responses resulting from spin excitations in antiferromagnetic insulators with spin-orbit coupling. Because of the time dependences and spatial variations of the antiferromagnetic order parameter the authors calculated chiral magnetic and anomalous Hall effects, respectively. Sekine and Chiba (Sekine and Chiba 2016) took advantage of the reciprocal process to theoretically demonstrate electric-field-induced antiferromagnetic resonance.

Some of the potential pitfalls hampering experimental observation of the proposed phenomena were pointed out by Cheng et al. (Cheng et al. 2014). It turns out that the efficiency of the microwave absorption close to resonance is also proportional to $\sqrt{H_K / H_E}$, which means that spin pumping is likely to be quenched in antiferromagnets with weak anisotropy, such as MnF$_2$ (<0.1) (M. P. Ross 2013). Antiferromagnets with large $\sqrt{H_K / H_E}$ ratio such as FeF$_2$ (~0.6) are promising candidates for the experimental demonstration of spin pumping, as illustrated in FIG. 22 (bottom). Microwave absorption is also maximal when the local easy axis is perpendicular to the oscillating magnetic field excitation. In polycrystalline antiferromagnetic films, spatial dispersion in the anisotropy properties from grain to grain will take the system away from maximal absorption and therefore from maximal spin current creation. Optimization and control of the spatial variability of magnetic properties (also see discussion in I.C.2) also contributes to enhancing experimental signals. Finally, Cheng et al (Cheng et al. 2014) suggested that microwave absorption could be enhanced by increasing the resonance frequency by applying a magnetic field (I.C.3) as they noticed that combining high frequency and high absorption efficiency may be difficult to achieve. Johansen and Brataas (Johansen and Brataas 2017) showed that some of these limitations can be circumvented in uniaxial antiferromagnets such as $MnF_2$ and $FeF_2$ close to the spin-flop transition.

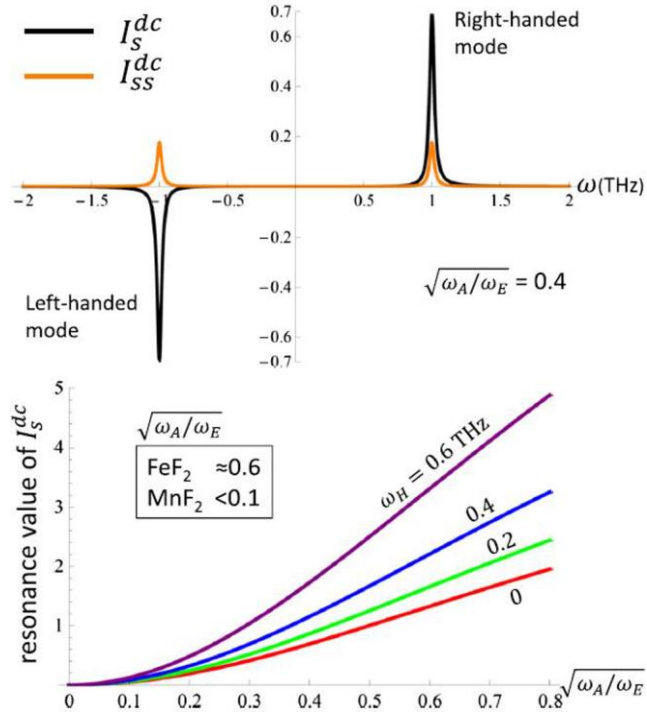

FIG. 22 (Color online). DC spin pumping calculated at the antiferromagnet's resonance. From (Cheng et al. 2014).

Another appealing way to generate pure spin current is through the spin Hall effect. The spin Hall effect in antiferromagnets is dealt with in section III.C. The spin Seebeck and Nernst effects also make it possible to generate angular momentum currents in antiferromagnets and are dealt with in section IV.A.

**B.    Spin mixing conductance and penetration depths**

Whether and how spin currents can be injected and transmitted in antiferromagnets are discussed in this part both theoretically and through the experimental data tabulated in TABLE 4-5.

**1.    Spin mixing conductance at antiferromagnetic interfaces**

The interfacial mixing conductance is a parameter quantifying the amount of spin momentum absorbed at magnetic interfaces upon reflection and transmission [see e.g. (Brataas, Tserkovnyak, and Bauer 2006)]. This concept is also particularly suitable to describe mechanisms such as spin pumping (see previous paragraph) and spin transfer torque. Cheng et al. (Cheng et al. 2014) and Takei et al. (Takei et al. 2014) have calculated the reflected mixing conductance (Brataas, Tserkovnyak, and Bauer 2006), $g^{\uparrow\downarrow} = S^{-1}\sum_{mn}\left(\delta_{mn} - r_{mn}^{\uparrow}r_{mn}^{\downarrow*}\right)$ ($r_{mn}^{\sigma}$ is the reflection coefficient of a quantum state with spin $\sigma$ from mode $m$ to mode $n$) at the interface between antiferromagnets and normal metal using tight-binding models (see FIG. 23).

At first sight, it might seem surprising that the mixing conductance is non-vanishing at such interfaces. This is particularly intriguing in the case of a compensated interface, which does not possess an overall magnetization. However, one needs to notice that the mixing conductance, originally defined for magnetic interfaces, is independent of the magnetization direction and therefore does not have to vanish in collinear antiferromagnets. As a matter of fact, although antiferromagnets do not possess time-reversal symmetry, they are invariant upon the combination of time reversal and crystal symmetry operations, such as translation in bipartite antiferromagnets. Hence, spin mixing occurs through intersublattice (or equivalently, Umklapp) scattering processes, as revealed by Haney and MacDonald (Haney and MacDonald 2008) and Takei et al. (Takei et al. 2014). Indeed, as discussed in section I.C.1 in the case of the bipartite antiferromagnet, although spin states are degenerate, their wave functions are associated with different superpositions of the two sublattice states. Therefore, a spin-flip event is associated with a flip of the sublattice state.

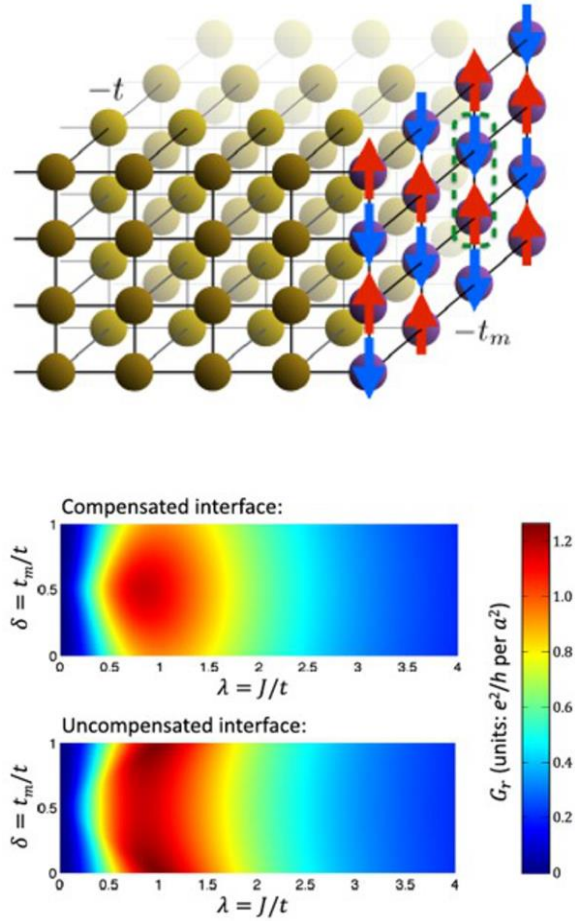

FIG. 23 (Color online). Mixing conductance calculated at the interface between an antiferromagnet and a normal metal, for two types of interfacial magnetic configurations: compensated and uncompensated. Here, $t$, $t_m$ are the hopping parameters of the normal metal and antiferromagnet, respectively and $J$ is the antiferromagnetic exchange energy. From (Cheng et al. 2014).

Saidaoui et al. (Saidaoui, Manchon, and Waintal 2014) have shown that the potential for electrically injecting a spin current from an antiferromagnet into a normal metal drastically depends on the type of antiferromagnetic texture. FIG. 24 shows three configurations, (a) ferromagnetic, (b) G-type [or checkerboard texture, as in FIG. 1(a)] and (c) L-type antiferromagnets (i.e., layered along the injection direction). Both L-type and G-type antiferromagnets result in a *staggered* spin density in reflection, while the transmitted current remains unpolarized. These results show that antiferromagnets are active materials even from the most rudimentary spin transport perspective, and call for further first-principle investigations and further investigations of non-collinear antiferromagnets.

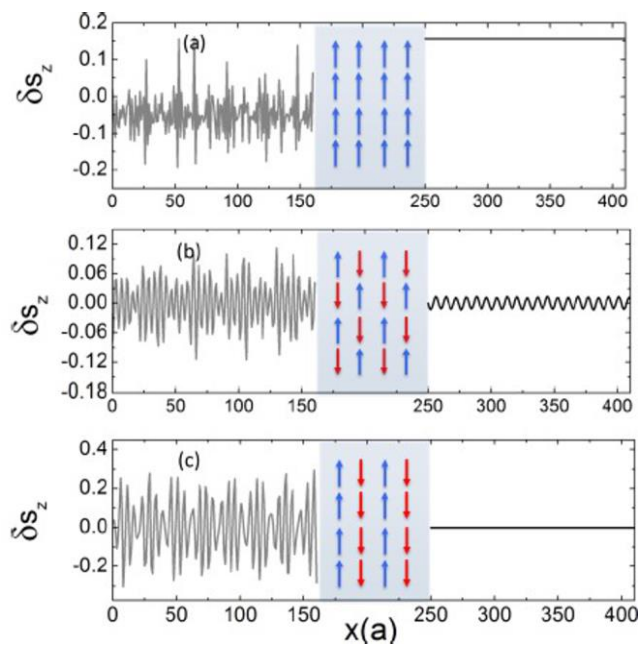

FIG. 24 (Color online). Spin polarization upon electrical injection through (a) a ferromagnet, (b) a G-type antiferromagnet and (c) a L-type antiferromagnet. From (Saidaoui, Manchon, and Waintal 2014).

Let us now turn our attention toward the experimental determination of interfacial spin mixing conductance. In a lattice, the transfer of angular momentum between incoming spins and local moments is linked to various mechanisms controlled by several parameters. When entering the antiferromagnet, electron spins experiences two types of spin memory loss mechanisms: spin-flip relaxation and spin dephasing. The former relaxes the spin angular momentum through the lattice via spin-orbit coupling and magnetic impurities. The latter occurs in magnetic materials (ferromagnets and antiferromagnets) and relaxes the component of the spin density that is transverse to the magnetic order parameter. These various mechanisms determine the manner incident spins are re-oriented upon reflection at the interface and how far they propagate inside the material without losing their memory. The former is quantified by the interfacial spin-mixing conductance while the latter is measured in term of spin penetration depth. Independently of the considerations related to relaxation mechanisms, interfacial spin mixing conductance and spin penetration depth must be as large as possible to efficiently transmit spin information.

Most frequently, magnetoresistive and dynamic experiments are used to study the parameters controlling the transfer of angular momentum. These experiments are commonly applied to ferromagnetic layers, but they are not ideal for antiferromagnetic films, which display low magnetoresistive signals and require very high frequency (THz) to induce dynamic excitation (see I.C.3.). Early attempts to determine both spin-mixing conductance and characteristic lengths in IrMn and FeMn were conducted using NiFe/Cu/AF/Cu/NiFe spin valves, cryogenic-temperatures, fits using drift-diffusion models (Valet and Fert 1993), and considerations on the magnetoresistance data (Acharyya et al. 2010; Acharyya et al. 2011). The authors of these studies indicated significant spin flipping at (IrMn,FeMn)/Cu interfaces and nanometric spin penetration depths in IrMn and FeMn, although no precise values could be determined.

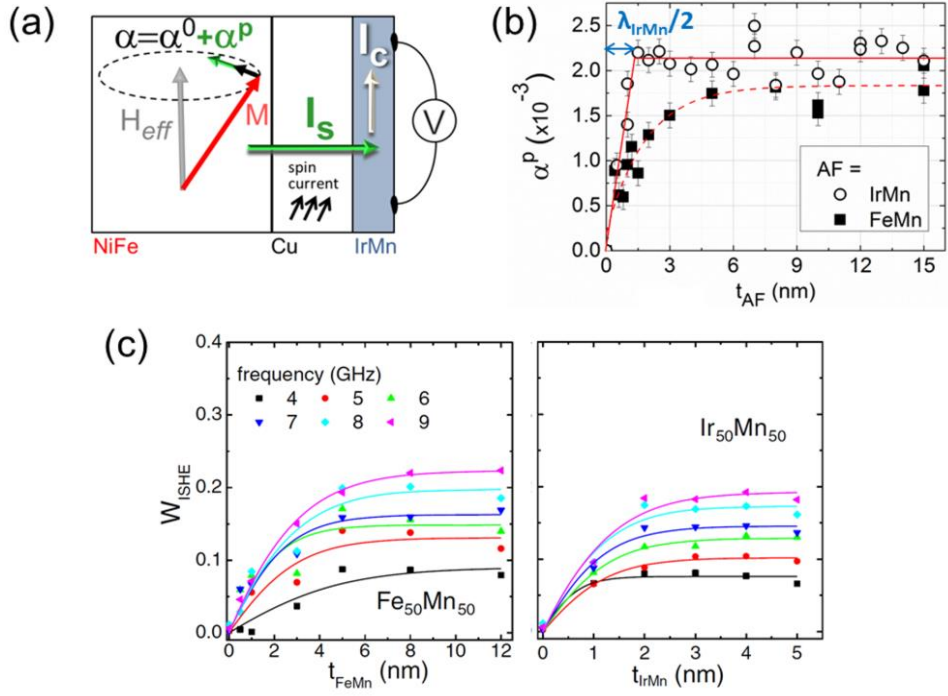

FIG. 25 (Color online). (a) Schematic representation of the experiment performed with a NiFe spin injector, a Cu spin conductor (to eliminate direct exchange interactions) and an IrMn spin sink. In reflection conditions, the NiFe damping is the sum of local intrinsic damping ($\alpha^0$) and additional non-local damping ($\alpha^p$) associated with the spin sink. In transmission conditions, the inverse spin Hall effect converts the spin current ($I_S$) to a charge current ($I_C$) in the antiferromagnetic spin sink. Adapted from (Frangou et al. 2016). (b) Dependence of $\alpha^p$ on spin sink thickness ($t_{AF}$) for $Ir_{20}Mn_{80}$ and $Fe_{50}Mn_{50}$ antiferromagnets. The spin penetration depth in the spin sink is $\lambda_{IrMn,FeMn}$. Adapted from (Merodio, Ghosh, et al. 2014). (c) Relationship between the inverse spin Hall contribution to the voltage (V) measured across the $Ir_{50}Mn_{80}$ and $Fe_{50}Mn_{50}$ spin sink layers and $t_{AF}$. Adapted from (W. Zhang et al. 2014).

An alternative method, better suited to antiferromagnetic materials, is based on the absorption of a spin current created by spin pumping from a neighboring ferromagnet. This method has attracted considerable attention owing to its versatility (Tserkovnyak et al. 2005; Ando 2014). The technique is applicable no matter the magnetic order (ferro-, ferri- and antiferro-magnetism) and the electrical state (metal, insulator, semiconductor) of the spin sink. A schematic representation of the experiment is illustrated in FIG. 25 for a NiFe spin injector and an IrMn spin sink. In this method, unlike what was described in section II.4, the antiferromagnet is no longer the spin injector but becomes the spin sink. We recall that the spin sink absorbs the current to an extent which depends on its spin-dependent properties (Tserkovnyak, Brataas, and Bauer 2002). In practice, ferromagnetic resonance generally drives the magnetization dynamics in these systems.

The interfacial spin mixing conductance can be recorded from spin pumping experiments [FIG. 25(a) and (b)] based on the difference in ferromagnetic damping ($\alpha^p$) compared to a reference sample with no spin sink ($\alpha^0$). TABLE 4 lists the spin mixing conductance measured for various interfaces containing antiferromagnets. The interpretation of TABLE 4 is certainly not straightforward. The spin mixing conductance can be very sensitive to the quality of the interface and therefore to the nature and quality of the stacking. Structure dependent spin mixing conductance were for example shown by Tokaç et al (Tokaç et al. 2015). Different measurements, furthermore, yield an effective spin mixing conductance which is convoluted with the spin-relaxation and/or spin Hall physics away from the interface. This for example limits the level of comparison between insulating-YIG/metallic-IrMn and metallic-Cu/metallic-IrMn bilayers.

| Interface X/AF | $g_{X/AF}^{\uparrow\downarrow}/S$ (nm$^{-2}$) | Reference |
|---|---|---|
| Cu/Ir$_{20}$Mn$_{80}$ | 10 | (Ghosh et al. 2012; Merodio, Ghosh, et al. 2014) |
| Cu/Ir$_{50}$Mn$_{50}$ | 12 | (W. Zhang et al. 2014) |
| Cu/Fe$_{50}$Mn$_{50}$ | 7 | (Merodio, Ghosh, et al. 2014) |
| YIG/Ir$_{20}$Mn$_{80}$ | 0.1 – 1 (0.43 x YIG/Pt) | (Mendes et al. 2014) |
| YIG/Fe$_{50}$Mn$_{50}$ | 4.9 ± 0.4 | (Du et al. 2014a) |
| YIG/Cr | 0.83 ± 0.07 | (Du et al. 2014a) |
| YIG/Mn | 4.5 ± 0.4 | (Du et al. 2014a) |
| YIG/NiO | 3.4[*] | (H. Wang et al. 2014) |
| YIG/NiO | 3.2[**] | (H. Wang et al. 2015) |
| YIG/α-NiFe$_2$O$_4$ | 1.6[**] | (H. Wang et al. 2015) |
| YIG/α-YIG | 1.0[**] | (H. Wang et al. 2015) |
| YIG/Cr$_2$O$_3$ | 0.75[**] | (H. Wang et al. 2015) |
| SrMnO$_3$/Pt | 0.34 – 0.49 | (J. H. Han et al. 2014) |

TABLE 4. Spin mixing conductance for various interfaces containing antiferromagnets. The investigation temperature was 300 K and values were measured by spin pumping. YIG stands for yttrium ion garnet (Y$_3$Fe$_5$O$_{12}$). [*]Value estimated from $g_{YIG/AF}^{\uparrow\downarrow}/S = 4\pi M_{S,YIG} t_{YIG} \alpha^p/(|\gamma|\hbar)$ with literature values taken for $M_{S,YIG}$: 140 kA/m (X. Zhang and Zou 2014), $t_{YIG}$ = 25 nm and $\alpha^p$ = 19.1 x 10$^{-4}$ (H. Wang et al. 2014). [**]Estimated with, $t_{YIG}$ = 20 nm (H. Wang et al. 2015) and $\alpha^p$ = 18, 9, 6, and 4 x 10$^{-4}$ for NiO, α-NiFe$_2$O$_4$ α-YIG and Cr$_2$O$_3$, respectively (H. Wang et al. 2015).

## 2. Spin penetration depths and relaxation mechanisms

The spin penetration depth in the spin sink can be recorded by measuring the thickness-dependence of the changes induced in the ferromagnetic damping, $\alpha^p$ [FIG. 25(b)]. Alternatively, the inverse spin Hall contribution to the transverse voltage (V) can be used to deduce spin penetration depths. This contribution results from the spin-to-charge current conversion (I$_C$) occurring directly in the antiferromagnetic spin sink (Saitoh et al. 2006) or in the topmost capping layer in the case of insulating antiferromagnets [see FIG. 25(c)]. We will deal specifically with the physics of the spin Hall effect in antiferromagnetic layers in paragraph III.2. For now, TABLE 5 lists the spin penetration depth for various antiferromagnetic materials separated into three different cases.

First, in ferromagnetic/non-magnetic-metal/antiferromagnetic metallic trilayers the transport is purely electronic through the non-magnetic-metal. In the electronic transport regime, it was shown theoretically (see II.1) that while destructive interferences due to spin precession are reduced in staggered antiferromagnets, disorder dramatically enhances spin decoherence, thereby strongly reducing the magnitude of the current-induced staggered spin density. Moreover, most metallic antiferromagnetic compounds possess heavy metal constituents, resulting in large spin-orbit-driven spin relaxation. This explains why the experimental values for the electronic penetration length are finite. In FIG. 26, it can be seen that the electronic spin penetration depth is inversely proportional to the resistivity for most XMn materials, except for one PtMn data. This proportionality suggests that spin relaxation in these XMn alloys is mostly due to diffusion mechanisms (Bass and Pratt 2007). The resulting product of resistivity and spin diffusion length is a constant, with a value of around 2.5 f$\Omega$.m$^2$, as expected for metallic films. It should be noted that these data relate to polycrystalline films. In such a case, the different direction of the moments probably averages out any anisotropic spin relaxation contribution due to the magnetic order. This type of signal averaging argument for polycrystals was discussed by Zhang et al. in the frame of the anisotropy of the spin Hall effect (W. Zhang et al. 2014; W. Zhang et al. 2015): it will be discussed in more details in section III.C.2, which is devoted to spin Hall effect in antiferromagnets. Finally, for similar reasons, in FIG. 25(b) it can be observed that the amplitude of the spin sink efficiency ($\alpha^p$) of polycrystalline IrMn layers is constant around the magnetic phase transition [at 300 K, thin IrMn films are paramagnetic below a film thickness of 2.7 nm and antiferromagnetic above (Frangou et al. 2016)]. Once again, the spin pumping contribution of the static magnetic ordering of the spin sink at low temperature probably averages out.

| AF material | Spin penetration depth (nm) | $\rho$ ($\mu\Omega$.cm) | Technique | Stack | Reference |
|---|---|---|---|---|---|
| *Metallic AF in a F/N/AF stack, electronic transport through N* | | | | | |
| $Ir_{20}Mn_{80}$ | 0.7 | 270 | SP ($\Delta H$) | NiFe/Cu/IrMn | (Ghosh et al. 2012; Merodio, Ghosh, et al. 2014) |
| $Ir_{50}Mn_{50}$ | $0.7 \pm 0.2$ | 293.3 | SP | NiFe/Cu/FeMn | (W. Zhang et al. 2014) |
| $Ir_{20}Mn_{80}$ | $\leq 1$ (4.2K) | 126 | CPP-GMR | NiFe/Cu/IrMn/ Cu/NiFe | (Acharyya et al. 2010; Acharyya et al. 2011; W. Park et al. 2000) |
| $Pd_{50}Mn_{50}$ | $1.3 \pm 0.1$ | 223 | SP | NiFe/Cu/PdMn | (W. Zhang et al. 2014) |
| $Fe_{50}Mn_{50}$ | $\leq 1$ (4.2K) | $87.5 \pm 5$ | CPP-GMR | NiFe/Cu/FeMn/ Cu/NiFe | (Acharyya et al. 2010; Acharyya et al. 2011; W. Park et al. 2000; Dassonneville et al. 2010) |
| $Fe_{50}Mn_{50}$ | $1.8 \pm 0.5$ | 167.7 | SP | NiFe/Cu/FeMn | (W. Zhang et al. 2014) |
| $Fe_{50}Mn_{50}$ | 1.9 | 135 | SP ($\Delta H$) | NiFe/Cu/FeMn | (Merodio, Ghosh, et al. 2014) |
| $Pt_{50}Mn_{50}$ | $0.5 \pm 0.1$ | 164 | SP | NiFe/Cu/PtMn | (W. Zhang et al. 2014) |
| $Pt_{50}Mn_{50}$ | 2.3 | $119 + 260/t_{AF(nm)}$ | ST-FMR (HR) | FeCoB/Hf/PtMn | (Ou et al. 2016) |
| *Metallic AF in a F/AF stack, electronic and magnonic transport regimes* | | | | | |
| $Ir_{25}Mn_{75}$ | 0.5 | 250 | ST-FMR | NiFe/IrMn | (Soh et al. 2015) |
| $Fe_{50}Mn_{50}$ | 2 | 166 | ST-FMR (HR) | NiFe/FeMn/Pt | (Y. Yang et al. 2016) |
| $Fe_{50}Mn_{50}$ | < 2 electronic | / | SP | NiFe/FeMn/W | (Saglam et al. 2016) |
| $Fe_{50}Mn_{50}$ | 9 magnonic | / | SP | NiFe/FeMn/W | (Saglam et al. 2016) |
| Cr | 2.1 | 25 - 325 | SSE | YIG/Cr | (Qu, Huang, and Chien 2015) |
| Cr | 4.5 (4.2K) | $180 \pm 20$ | CPP-GMR | Fe/Cr/Fe | (Bass and Pratt 2007) |
| Cr | 13.3 | 500 - 1200 | SP | YIG/Cr | (Du et al. 2014a) |
| Mn | 10.7 | 980 | SP | YIG/Mn | (Du et al. 2014a) |
| *Insulating AF in a F/AF stack, magnonic transport* | | | | | |
| NiO | 1.3 | >> | SSE | YIG/NiO/Ta | (Lin et al. 2016) |
| NiO | 2.5 | >> | SSE | YIG/NiO/Pt | (Lin et al. 2016) |
| NiO | 2 - 5.5 (180 - 420K) | >> | SSE | YIG/NiO/Pt | (Prakash et al. 2016) |
| NiO | 2 | >> | SP | YIG/NiO/Pt | (Hahn et al. 2014) |
| NiO | 3.9 | >> | SP | YIG/NiO/Pt | (Hung et al. 2017) |
| NiO | 9.8 | >> | SP | YIG/NiO/Pt | (H. Wang et al. 2015) |
| NiO | 10 | >> | SP | YIG/NiO/Pt | (H. Wang et al. 2014) |
| NiO | 50 | >> | ST-FMR | NiFe/NiO/Pt | (Moriyama, Takei, et al. 2015) |
| $\alpha$-$NiFe_2O_4$ | 6.3 | >> | SP | NiFe/$\alpha$-$NiFe_2O_4$/Pt | (H. Wang et al. 2015) |
| $\alpha$-YIG | 3.9 | >> | SP | NiFe/$\alpha$-YIG/Pt | (H. Wang et al. 2015) |

| Cr$_2$O$_3$ | 1.6 | >> | SP | Cr$_2$O$_3$ | (H. Wang et al. 2015) |

TABLE 5. Spin penetration depth, and resistivity ($\rho$) for various antiferromagnetic materials. Finite size effects on $\rho$ are reported in the table, whenever available. Unless specified otherwise, NiFe is close to Ni$_{81}$Fe$_{21}$, the composition of Permalloy, and YIG stands for epitaxial Y$_3$Fe$_5$O$_{12}$. When not specified the investigation temperature was 300 K. CPP-GMR = current perpendicular to plane excitation – giant magneroresistance detection, SP and SP ($\Delta$H) = ferromagnetic resonance spin pumping excitation – inverse spin Hall effect detection when not specified, and ferromagnetic resonance linewidth detection when ($\Delta$H) is specified, ST-FMR and ST-FMR (HR) = spin torque ferromagnetic resonance excitation induced by spin Hall effect as a result of an ac current flow – anisotropic magnetoresistance detection when not specified, and 2$^{nd}$ harmonic response detection of the anomalous Hall effect and/or anisotropic magnetoresistance when (HR) is specified, and SSE = longitudinal spin Seebeck excitation induced by a thermal gradient – inverse spin Hall effect detection.

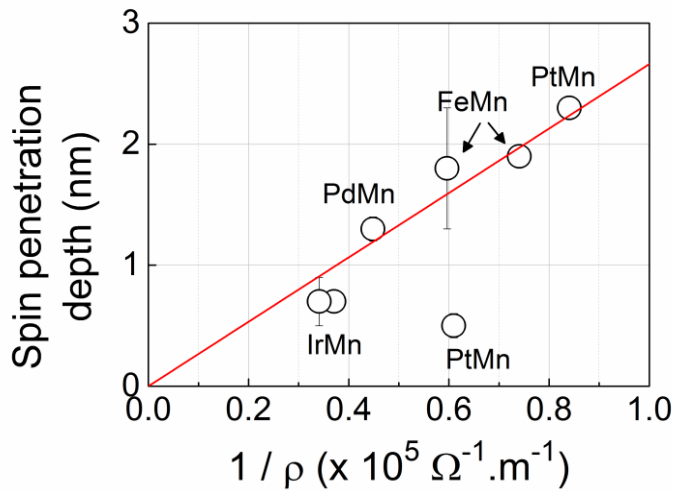

FIG. 26 (color online). Spin penetration depth is inversely proportional to bulk resistivity for some metallic antiferromagnets. Data are from TABLE 5 (metallic AF in a F/N/AF stack, electronic transport through N) and references therein. The straight line is a fit excluding the lowest data point for PtMn and constrained to pass through (0,0).

Second, in exchange biased (I.C.3) ferromagnetic/antiferromagnetic metallic bilayers transfer and propagation of spin angular momentum also involves magnonic transport, i.e., spin waves produced by the oscillating ferromagnet feed directly into the antiferromagnet. In this case, both electronic and magnonic transport regimes may coexist, which make it difficult to study. The data presented in TABLE 5 appear to suggest that, overall, spin currents propagate more easily when the metallic antiferromagnet is exchange biased to a ferromagnet. This is because spin currents carried by magnons decay more slowly than those carried by electrons. Note that the initial amplitude of the spin-angular momentum transfer contribution mediated by magnons through ferromagnetic/antiferromagnetic metallic interfaces is directly related to the interfacial exchange coupling amplitude, as demonstrated by Tshitoyan et al (Tshitoyan et al. 2015). Using spin pumping and measuring the inverse spin Hall effect in NiFe/FeMn/W trilayers, Saglam et al. (Saglam et al. 2016) managed to disentangle electronic- and magnonic-transport-related penetration length in FeMn (also see TABLE 5). They took advantage of the relatively large magnitude and opposite sign of spin Hall effects in W compared to FeMn to detect when magnonic transport takes over, i. e. when spin currents reach the W layer for FeMn thickness well above the electronic spin diffusion length. Although transmission was purely electronic through the non-magnetic-metal in the previous case of ferromagnetic/non-magnetic-metal/antiferromagnetic stacks, we believe that conversion of charge currents into magnons and vice versa cannot be excluded. These types of conversions are extensively discussed for the case of ferromagnets in the review on magnon spintronics by Chumak et al. (Chumak et al. 2015).

Third, in exchange-biased ferromagnetic/antiferromagnetic bilayers where the antiferromagnet is insulating, the transport is purely magnonic. Again, the data presented in TABLE 5 suggest/confirm that, overall, spin currents carried by magnons propagate more readily than their electronic counterparts, as previewed in the previous paragraph. In fact, a theoretical framework where the spin current in the antiferromagnet is carried by an evanescent spin-wave mode was established for the coherent coupled low-temperature dynamics (Takei et al. 2015; Khymyn et al. 2016). At finite temperatures, thermal magnons open an additional channel for spin transport (Rezende, Rodríguez-Suárez, and Azevedo 2016b). In the antiferromagnet, the two magnon modes have different frequencies and hence different thermal populations, making magnonic spin transport possible. The precessing magnetization in the adjacent ferromagnet could pump oppositely-polarized magnons differently into the antiferromagnet. These propagation mechanisms will be discussed further in section IV, which is devoted to spin-caloritronics in antiferromagnets. It should also be noted that magnons created by spin pumping or by the spin Seebeck effect have different frequencies and modes, which may contribute to data discrepancies.

Like in ferromagnetic spintronics, heterostructure engineering with conductance matching is required to increase spin transport efficiency (Du et al. 2014b; Ou et al. 2016), which necessitates thorough investigations of interfacial qualities (roughness, stacking faults, species intermixing, etc.) (Berkowitz and Takano 1999). In addition, spin absorption mechanisms in antiferromagnetic materials are not currently entirely understood. Further investigation of how spins are transmitted through these materials is therefore required, including systematic quantification of the influence of heavy scatterer content or degree of crystallinity. The extent to which the static magnetic order affects transmission is also currently unclear. A better understanding of electronic and magnonic transport in metallic antiferromagnets in highly desirable, in particular by finding efficient ways to distinguish between them, which would then potentially make it possible to control one over the other. Finite size effects will have to be studied since the antiferromagnetic order is strongly influenced by size effects, in particular those occurring at nanopillar edges due to grain size reduction (Baltz et al. 2005) and reduced spin coordinations (Baltz, Gaudin, et al. 2010).

## C.    Giant and tunnel magnetoresistance

The most popular and successful spintronics phenomena for data reading in practical applications are the giant magnetoresistance (GMR) and tunnel magnetoresistance (TMR). Their antiferromagnetic counterparts – antiferromagnetic GMR and TMR – were proposed to exist in so-called antiferromagnetic spin valves and magnetic tunnel junctions, respectively. The experimental demonstration of these antiferromagnetic phenomena, however, remains elusive.

## 1.    Giant magnetoresistance

In ferromagnets, (current-perpendicular-to-plane) giant magnetoresistance arises from the transmission of the diffusion of spin accumulation from one ferromagnetic electrode to the other. In antiferromagnets, although no such spin accumulation survives at the level of the magnetic unit cell, Núñez et al. (Núñez et al. 2006) demonstrated that the coherent buildup of a spin dependent wave function in an antiferromagnetic spin-valve results in a magnetoresistive signal (see also II.A.2 for the theoretical basis).

Several attempts have been made to demonstrate the antiferromagnetic giant magnetoresistance effect in antiferromagnetic spin valves where two antiferromagnets (AF) are separated by a nonmagnetic (N) spacer. According to the original predictions (Núñez et al. 2006), the resistance of an antiferromagnetic spin valve AF/N/AF should depend on the relative orientation of magnetic order parameters in the two antiferromagnets. To control the relative orientation of the two antiferromagnets one usually employs the exchange bias phenomenon by placing one or both antiferromagnets in contact with ferromagnets (F), e.g. in AF/N/AF/F or F/AF/N/AF/F structures. For a sufficiently thin antiferromagnet the reversal of the ferromagnet by a magnetic field should be accompanied by a reversal of the adjacent antiferromagnet thanks to the exchange coupling across the ferromagnet/antiferromagnet interface. Some experiments (Wei, Sharma, et al. 2009; L. Wang et al. 2009) observed a small (0.1-0.5%) change in resistance in such structures (AF=FeMn or IrMn, F=CoFe, N=Cu), which correlates with the reversal of ferromagnets (antiferromagnets). However, a detailed study of these resistance variations in a large number of various structures (AF/N/AF, F/AF/N/AF, F/AF/N/AF/F, AF/F/N/AF, AF/F, AF/N/F) and single (F, AF) layers showed no conclusive evidence of an antiferromagnetic giant magnetoresistance effect and could be associated with ferromagnetic contributions from either the ferromagnets or uncompensated magnetic moments in the antiferromagnets. The fact that giant magnetoresistance in antiferromagnetic spin valves has never been clearly experimentally observed may be linked the need for quantum coherence effects and minimal disorder.

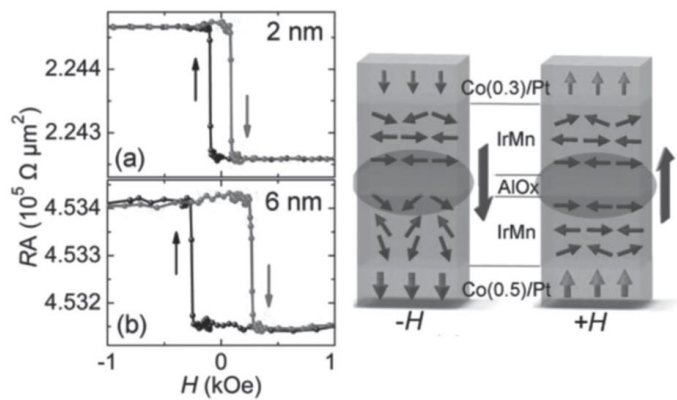

FIG. 27. (Left) Resistance times area product (RA) versus out-of-plane magnetic field (H) for IrMn-based tunnel junctions with different IrMn thickness. (Right) Scheme showing the moments arrangements in the junction for positive and negative applied magnetic field. Adapted from (Y. Y. Wang, Song, Wang, Miao, et al. 2014).

### 2. Tunnel magnetoresistance

In ferromagnets, tunnel magnetoresistance arises from spin-dependent tunneling between ferromagnetic electrodes. Its magnitude is governed by the spin polarization of the density of states at the interface between the magnetic electrodes and the tunnel barrier [e.g. see review article by Tsymbal et al. (Tsymbal, Mryasov, and LeClair 2003)]. In antiferromagnets, even though the interfacial density of states might not be spin-polarized (e.g. in collinear compensated antiferromagnets), ballistic tunneling between antiferromagnets can also lead to a magnetoresistive signal (see also II.A.2 for the theoretical basis).

The first experimental search for such an antiferromagnetic tunnel magnetoresistance was performed by Wang et al (Y. Y. Wang, Song, Wang, Miao, et al. 2014) in [Pt/Co]/IrMn/AlO$_x$/IrMn/[Pt/Co] multilayers (FIG. 27). The two IrMn antiferromagnetic layers are controlled via the exchange coupling by adjacent Pt/Co ferromagnetic multilayers with perpendicular magnetic anisotropy. For sufficiently thin antiferromagnetic layers (≤6 nm) the resistance of such tunnel junctions was found to be slightly (<0.1%) different when saturated by positive or negative magnetic fields. The difference was attributed to different magnetic configurations of the two antiferromagnetic layers associated with partial rotations of the exchange spring propelled by the applied magnetic field. Investigations of Cr(001)-based tunnel junctions are also ongoing (e.g. studies on tunnel mediated coupling, electronic structure and magnetization of the surfaces and interfaces). More specifically, Leroy et al (Leroy et al. 2013) demonstrated two 2-dimensional localized states for the Cr surface that persist at the Cr/MgO interface: a $\Delta_1$ and a $\Delta_5$ state. The $\Delta_1$ state is associated with Cr surface magnetism. The authors found that the interface moment (0.02 µB/Cr atom) is 10-fold smaller than expected from theoretical calculations and previous results (Leroy et al. 2015). This is promising for the study of truly antiferromagnetic Cr-based junctions since the ferromagnetic-like contribution should be minimal. The $\Delta_5$ state, mostly influences transport and coupling in Cr/MgO epitaxial systems. In Cr/MgO/Cr tunnel junctions, the authors demonstrated tunnel magnetic coupling between the antiferromagnetic Cr layers through the MgO insulator. They showed that this coupling can be amplified thanks to the presence of resonant states exhibiting the same $\Delta_5$ symmetry at the interface (Leroy et al. 2014).

## III.    SPIN-ORBITRONICS

The most encouraging developments in antiferromagnetic spintronics are currently building on spin-orbit interactions. Spin-orbit effects can be readily understood by considering the motion of an electron in a potential gradient $\nabla$V and how fields are transformed between inertial frames. The net electric field, created by the potential gradient becomes a magnetic induction field in the electron's rest frame, $\mathbf{B}_{SO}$. This momentum($k$)-dependent magnetic field couples to the magnetic moment of the electron, $\boldsymbol{\mu}$, through a Zeeman term, $-\boldsymbol{\mu}.\mathbf{B}_{SO}$. In magnetic materials, this interaction interconnects the direction of electron flow and the magnetic order parameter, resulting in, for example, anisotropic magnetoresistance (Mcguire and Potter 1975) or anomalous Hall effect (Nagaosa et al. 2010). Moreover, in crystals or structures lacking inversion symmetry the $k$-dependent magnetic field becomes *odd* in momentum, resulting for example in Dresselhaus or Rashba fields (Manchon et al. 2015). This provides a unique means to manipulate the order parameter in antiferromagnets. Here, we review how spin-orbit effects occur in antiferromagnetic materials and how they can represent a promising alternative to conventional spin transfer torque for the manipulation of the antiferromagnetic order. In addition, we will see how spin-orbit effects also make detection of the antiferromagnetic order possible.

### A.    Anisotropic magnetoresistance

The anisotropic magnetoresistance effect was first found in $3d$ transition metals and alloys (Thomson 1857). It is typically associated with the orientation of the material's magnetization with respect to the direction of electrical current flow. In these materials, the direction of electron orbitals is set by the magnetization direction through spin-orbit coupling and therefore, the scattering of itinerant electrons (and hence the conductivity) depends on the magnetization direction. The angular dependence of the effect is well described by cosine and sine trigonometric functions. The phenomenon has been widely used as a detection element in early magnetic recording technology. However, its amplitude is typically limited to few tens of percent.

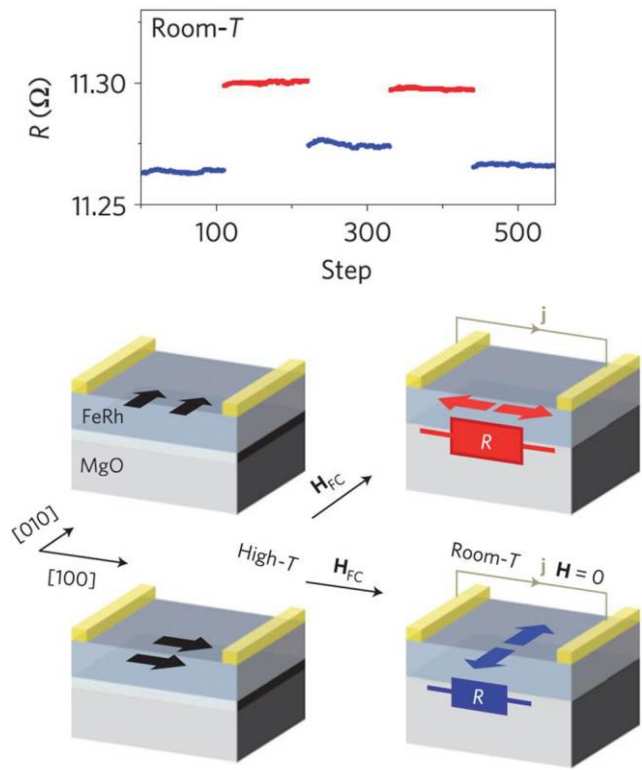

FIG. 28 (Color online). Anisotropic magnetoresistance effect measured at room temperature in the antiferromagnetic FeRh alloy. The antiferromagnetic order is alternatively set toward two orthogonal directions, [100] or [010], by raising the temperature above the metamagnetic phase transition, saturating the ferromagnetic order with a magnetic field (either along the [010] or [100] crystal axis), and field-cooling through the phase transition. From (Marti et al. 2014).

Interestingly, anisotropic magnetoresistance is *even* in magnetization, i.e. it is invariant upon magnetization reversal. Hence, such an effect also exists in antiferromagnets. Such type of (non-crystalline) anisotropic magnetoresistance was therefore first demonstrated in collinear antiferromagnets such as FeRh (Marti et al. 2014; Moriyama, Matsuzaki, et al. 2015), CuMnAs (Wadley et al. 2016), MnTe (Kriegner et al. 2016) and Mn$_2$Au (H.-C. Wu et al. 2016). The effect was later detected isothermally in IrMn, a non-collinear antiferromagnet (Galceran et al. 2016). A typical example of non-crystalline anisotropic magnetoresistance at room temperature in an antiferromagnet (FeRh) is plotted in FIG. 28.

The non-crystalline component of the anisotropic magnetoresistance effect arises from the deviation of the current direction with respect to the magnetization. There exists another anisotropic magnetoresistance component that depends on the crystal symmetries: the crystalline anisotropic magnetoresistance. Recently this component of anisotropic magnetoresistance was found to be rather significant in oxides comprising a 5$d$ transition metal (Fina et al. 2014; C. Wang, Seinige, Cao, Zhou, et al. 2014; C. Wang et al. 2015). The experiment by Fina et al (Fina et al. 2014) reported on experimental observation of the crystalline anisotropic magnetoresistance in a 6-nm thick film of antiferromagnetic semiconductor Sr$_2$IrO$_4$. The antiferromagnetic film in this experiment was exchange coupled to a ferromagnet (La$_{0.67}$Sr$_{0.33}$MnO$_3$ - LSMO) that was used to reorient the antiferromagnetic order parameter by applying an external magnetic field. Here the field was assumed to rotate the ferromagnet which in turn reorients the antiferromagnet via the exchange spring effect (see I.C.3). Resistivity measurements of this exchange-biased ferromagnetic/antiferromagnetic bilayer placed in an in-plane rotating magnetic field clearly showed correlations between the bilayer resistance and the angle between magnetization and crystal axes. Since for the in-plane field rotations the angle between the electrical current (flows along the out-of-plane c-axis) and in-plane magnetic moments remains constant, the observed ~1% effect can be associated with the crystalline component of anisotropic magnetoresistance. The experiments by Wang et al (C. Wang, Seinige, Cao, Zhou, et al. 2014) demonstrated a much larger magnetoresistive effect in a single crystal of $Sr_2IrO_4$ antiferromagnet without any ferromagnets in proximity. However, in this particular case, the results were observed at weak external magnetic fields. Part of the magnetoresistive effect was thus probably due to uncompensated moments rather than being associated with the antiferromagnetic order.

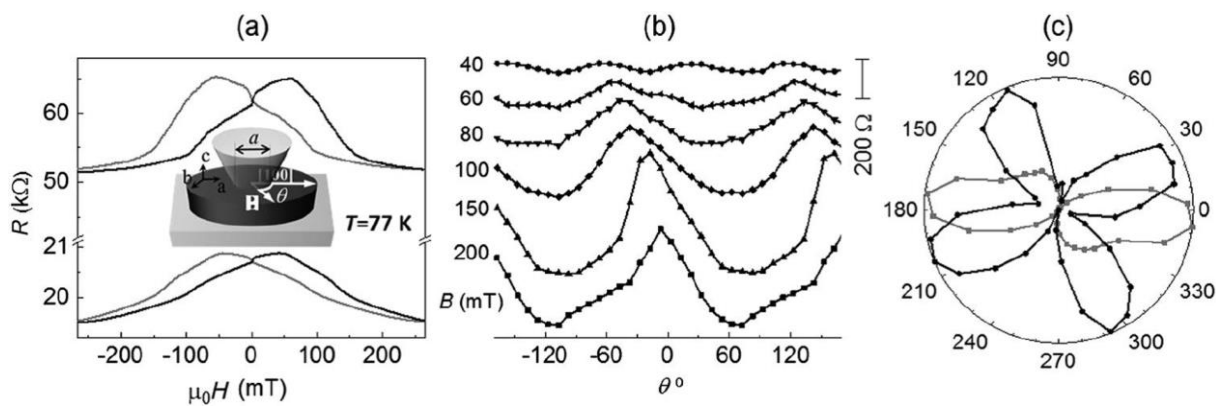

FIG. 29 (Color online). (a) Resistance versus in-plane magnetic field measured at 77K for two point contacts (3 and 1 µm) between a Cu tip and a $Sr_2IrO_4$ crystal. The insert shows a schematic of the experiment. (b) Angular dependence of a point contact resistance for different magnetic fields. (c) Polar plots of the normalized anisotropic magnetoresistance at 40 mT (black) and 270 mT (gray). From (C. Wang, Seinige, Cao, Zhou, et al. 2014).

Point-contact measurements (FIG. 29) revealed negative magnetoresistances (up to 28%) for modest magnetic fields (250 mT) applied within the $IrO_2$ a-b plane and electric currents flowing perpendicular to the plane. Here the point-contact technique was used as a local probe of magnetotransport properties on the nanoscale, as opposed to standard bulk measurements, and demonstrated the scalability of the effect for future applications. As the currents were flowing perpendicular to the in-plane magnetic moments, the observed magnetoresistance can be attributed to the crystalline component of anisotropic magnetoresistance. The angular dependence of magnetoresistance showed a crossover from four-fold to two-fold symmetry in response to an increasing magnetic field with angular variations in resistance from 1% to 14%. This field-induced transition can be associated with the effects of applied field on the canting of antiferromagnetic-coupled moments in $Sr_2IrO_4$. It should be noted that the Néel temperature in bulk $Sr_2IrO_4$ (240 K) is well below room temperature and it is even smaller in thin films (100 K) (see section I.B. and TABLE 3). The latter makes practical applications of $Sr_2IrO_4$ questionable and calls for materials science efforts to find appropriate antiferromagnet to demonstrate room-temperature crystalline anisotropic magnetoresistance.

## B.    Tunnel anisotropic magnetoresistance

Like ohmic anisotropic magnetoresistance, tunnel anisotropic magnetoresistance possesses non-crystalline and crystalline components. Tunnel anisotropic magnetoresistance is driven by the relative orientation of magnetic moments and the axes of current direction and crystalline anisotropy. The difference compared to ohmic anisotropic magnetoresistance is that the non-crystalline tunnel anisotropic magnetoresistance component is difficult to separate from the crystalline component since in multilayers the crystal is different in the in-plane and out-of-plane directions. Basically, in a ferromagnet/tunnel-barrier/non-magnet trilayer (F/B/N), the amount of current tunneling perpendicularly through the junction is proportional to the non-magnet and ferromagnet densities of states at the Fermi level and the tunneling matrix elements. These tunneling matrix elements and ferromagnet density depend on the ferromagnet's orientation with respect to the crystalline anisotropy axes. Hence, by varying the ferromagnetic orientation (e.g. by applying an external magnetic field, H), the resistance (R) of the trilayer can be changed. Some of the explanations for tunnel anisotropic magnetoresistance are based on Rashba- or Rashba and Dresselhaus-induced spin-orbit coupling (Shick et al. 2006; Matos-Abiague and Fabian 2009). The effect was first experimentally shown to exist in (Ga,Mn)As-based tunnel junctions (Gould et al. 2004) and has been widely studied since then for various ferromagnetic-based tunnel junctions. For ferromagnetic materials, only small tunnel anisotropic magnetoresistance signals have been reported at low temperature. This lack of signal is related to the fact that spin-orbit coupling in ferromagnetic transition metals which have the potential for room temperature tunnel anisotropic magnetoresistance is too weak. Thus, the tunnel anisotropic magnetoresistance reaches a few percentage points at 4 K and can rise to 10% when the spin-orbit interaction is boosted by additional heavy elements, e.g. in [Co/Pt] multilayers (B. G. Park et al. 2008).

Interestingly, in 2010 Shick et al. (Shick et al. 2010) predicted large tunnel anisotropic magnetoresistance signals for alloys such as IrMn and $Mn_2Au$ containing a heavy noble metal (Ir, Au) and a transition metal (Mn) (FIG. 30). In this case, the $5d$ shell of the noble metal offers large spin-orbit coupling and the $3d$ shell of the transition metal adds complementary large spontaneous moments. As indicated by Park et al. (B. G. Park et al. 2011): 'as Mn carries the largest moment among transition metals and most of the bimetallic alloys containing Mn order antiferromagnetically, the goals of strong magnetic anisotropy phenomena and of antiferromagnetic spintronics seem to merge naturally', although the phenomenon is mainly driven by spin-orbit interactions and the role played by the magnetic order remains to be understood.

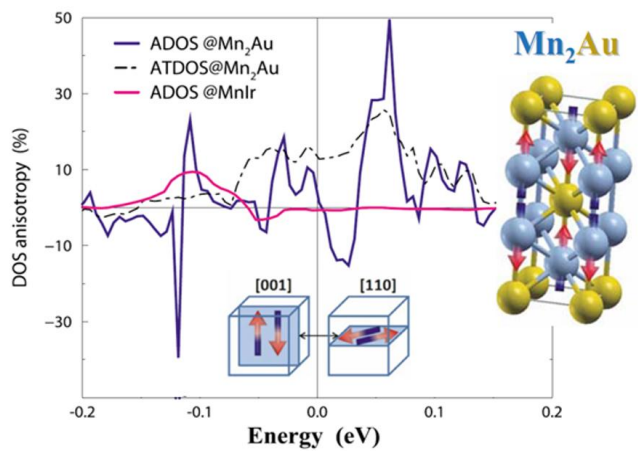

FIG. 30 (Color online). Anisotropy of the densities of state between the [001] and [110] axes for Mn₂Au and MnIr films. The inset shows a MoSi₂-type body-centered-tetragonal structure and the spin arrangement in Mn₂Au. The Fermi level is at 0 eV. From (Shick et al. 2010).

The predicted tunnel anisotropic magnetoresistance structures were experimentally confirmed for $Ir_{20}Mn_{80}$ using in-plane anisotropy (160% at 4 K) (B. G. Park et al. 2011; Martí et al. 2012). Full hysteretic R vs. H loops with bistable remanent states were obtained by forcing the antiferromagnetic configuration thanks to a neighboring exchange-biased ferromagnetic layer (see section I.C.3 for antiferromagnetic order manipulation by exchange bias). In this case, the antiferromagnetic configuration is driven by exchange bias on one side and the tunnel anisotropic magnetoresistance is detected for the other side. What matters is that the antiferromagnetic moments are dragged through the whole system when reversing the ferromagnet so as to achieve the most extensive change in the angle of the antiferromagnetic moments on the tunnel barrier side, and hence the largest possible tunnel anisotropic magnetoresistance signal. How easily and how deep the antiferromagnetic moments are dragged by the ferromagnet depends on external stimuli like temperature and is also partly system-dependent, since it depends on the amplitude of ferromagnetic/antiferromagnetic exchange energy and antiferromagnetic exchange stiffness. For NiFe/IrMn/MgO/Pt tunnel junctions, Reichlová et al. (Reichlová et al. 2016) observed tunnel anisotropic magnetoresistance at room temperature, and determined the optimal NiFe/IrMn thickness ratio to be 10/4.

Wang et al (Y. Y. Wang et al. 2012; Y. Y. Wang, Song, Wang, Zeng, et al. 2014) exploited the thermal stability of antiferromagnet-based systems with out-of-plane anisotropy. to further demonstrate room temperature (antiferromagnetic) tunnel anisotropic magnetoresistance (~0.2%) and hysteretic behavior for [Pt/Co]/IrMn,FeMn/B/N stacks (FIG. 31).

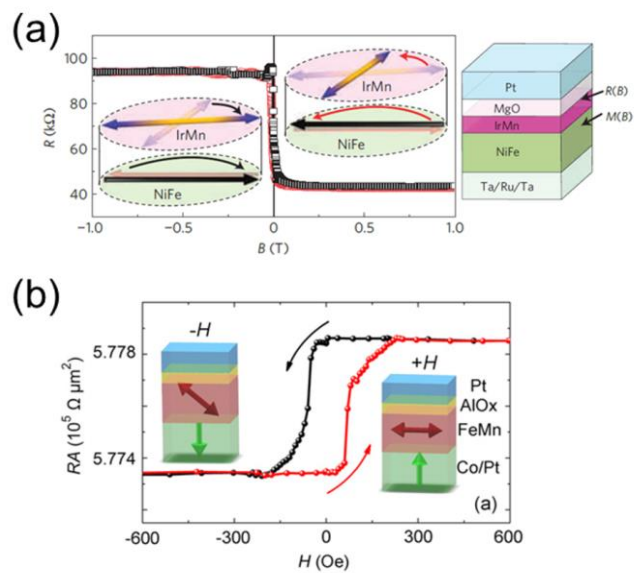

FIG. 31 (Color online). (a) Hysteretic resistance (R) vs. magnetic induction (B) showing extensive antiferromagnetic tunnel anisotropic magnetoresistance at 4 K, for IrMn. From (B. G. Park et al. 2011). (b) Hysteretic resistance area product (RA) vs. magnetic field (H) showing room temperature antiferromagnetic tunnel anisotropic magnetoresistance and binary remanence for FeMn, facilitated by the out-of-plane anisotropy of the ferromagnetic layer. From (Y. Y. Wang, Song, Wang, Zeng, et al. 2014).

Given the above results, tunnel anisotropic magnetoresistance is a promising means to analyze antiferromagnetic configurations, despite the small signals it delivers so far and even though room temperature signals can only be measured when a ferromagnetic layer with perpendicular anisotropy is present to pull the antiferromagnetic spins out-of-plane. This requirement has so far prevented the development of a '(ferro)magnet-free' device readable by means of tunnel anisotropic magnetoresistance. Another alternative would be to use antiferromagnetic alloys containing a 5$d$ shell element but with a heavier noble metal than Ir and a larger ratio of noble metal to magnetic moment-carrying transition metal. These conditions should make $Pt_{50}Mn_{50}$ or mixed antiferromagnets (Akmaldinov et al. 2014) terminated with a $Pt_{50}Mn_{50}$ interface good candidates. Antiferromagnetic semiconductors could also be used to benefit from carrier-mediated magnetism. Promising recent studies described additional examples of this type of alternative room temperature antiferromagnetic material with potential for large tunnel anisotropic magnetoresistance. These included metallic $Mn_2Au$ alloys (Barthem et al. 2013; H. C. Wu et al. 2012) and semimetallic CuMnAs (Wadley et al. 2013) (see also III.D.2.).

## C.    Anomalous and spin Hall effects

### 1.    Anomalous Hall effect in non-collinear antiferromagnets

Traditionally, the charge Hall effect exists in three main variants: (i) in the presence of an external magnetic field upon applying a Lorentz force (ordinary Hall effect), (ii) in magnetic materials with spin-orbit coupling (anomalous Hall effect) and (iii) in chiral magnetic textures (topological Hall effect) [for a review, see (Nagaosa et al. 2010)]. These three variants indicate that the charge Hall effect is produced by a combination of time-reversal symmetry breaking and some sort of "effective Lorentz force". The former is achieved by applying a non-vanishing magnetic field and the latter can be obtained through non-zero Berry curvature of the band structure, which provides anomalous velocity. Since antiferromagnetic materials are devoid of overall magnetization, it would seem logical to assume that they do not show a charge Hall effect.

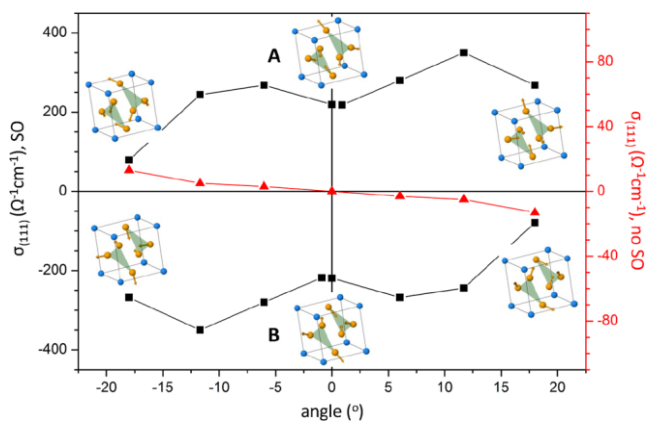

FIG. 32 (Color online). Anomalous Hall effect in IrMn$_3$ calculated from first principles in the absence (red triangles) and presence (black squares) of spin-orbit coupling. From (H. Chen, Niu, and MacDonald 2014).

Although antiferromagnets do not possess time-reversal symmetry, they are invariant when time-reversal symmetry is combined with a crystal symmetry operation (such as spatial translation by the vector connecting the two sublattices in the case of a bipartite antiferromagnet, or mirror symmetry in the case of the two-dimensional kagome lattice studied by Chen et al. (H. Chen, Niu, and MacDonald 2014)). If crystal symmetry is broken, a non-vanishing Berry curvature is produced leading to the emergence of a finite anomalous Hall effect. This property is absent in collinear antiferromagnets but has been predicted for some non-collinear antiferromagnetic compounds: distorted γ-Fe$_x$Mn$_{1-x}$, and NiS$_2$ (Shindou and Nagaosa 2001), IrMn$_3$ (H. Chen, Niu, and MacDonald 2014), Mn$_3$Ge, and Mn$_3$Sn (Kübler and Felser 2014). FIG. 32 presents the anomalous Hall conductivity calculated for IrMn$_3$. This conductivity was induced by tilting the magnetic moments out of the (111) planes in the absence (red triangles) and presence (black squares) of spin-orbit coupling (H. Chen, Niu, and MacDonald 2014). From the figure, it is clear that without spin-orbit coupling, the anomalous Hall effect is only present when the moments are tilted out-of-plane. It is therefore similar to the topological Hall effect observed in ferromagnetic textures such as vortices and skyrmions [see e.g. (Neubauer et al. 2009)]. With the spin-orbit coupling turned on, a large anomalous Hall conductivity is obtained, even when the antiferromagnet is fully compensated (zero angle).

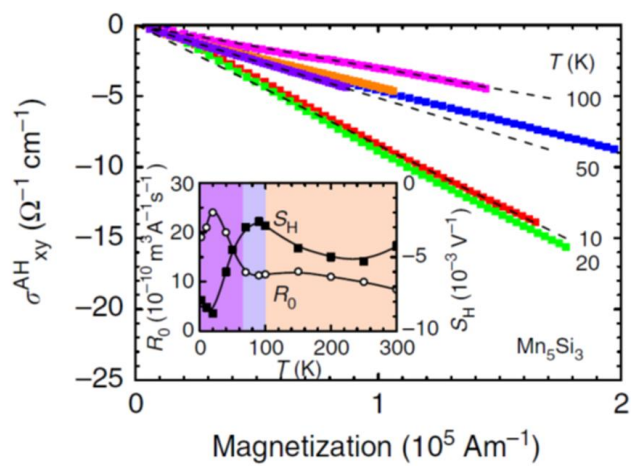

FIG. 33 (Color online). Anomalous Hall effect measured in Mn$_5$Si$_3$ layers. Adapted from (Sürgers et al. 2014).

This large anomalous Hall effect was recently experimentally confirmed in certain non-collinear antiferromagnets such as Mn$_5$Si$_3$ (Sürgers et al. 2014; Sürgers et al. 2016) (see FIG. 33), Mn$_3$Sn (Nakatsuji, Kiyohara, and Higo 2015), Mn$_3$Ge (Nayak et al. 2016), and GdPtBi (Suzuki et al. 2016) single crystals.

## 2.    Spin Hall effect

The spin Hall effect derives from coupling of the charge and spin currents due to spin-orbit interaction (see e.g. review articles (Hoffmann 2013; Sinova et al. 2015; Dyakonov 2010)). Through this interaction, a charge current induces a pure spin current orthogonal to the electric current. The inverse process - or inverse spin Hall effect - whereby the spin current induces an orthogonal charge current, is also possible. The spin Hall effect was first experimentally confirmed in semiconducting GaAs (Kato et al. 2004; Wunderlich et al. 2005); it was subsequently detected in metallic materials with large spin-orbit interaction, such as Platinum (Saitoh et al. 2006; Valenzuela and Tinkham 2006). The spin Hall effect can be induced by two classes of physical mechanism: extrinsic and intrinsic. Extrinsic mechanisms, such as skew and side jump scattering are associated with charge-to-spin conversion during scattering events. In contrast, the intrinsic mechanism is not affected by (weak) scattering and only depends on the band structure of the material (Murakami, Nagaosa, and Zhang 2003). The Berry curvature resulting from the geometric phase in the momentum space creates a local magnetic field that exerts a spin-dependent Lorentz force on the flowing spins resulting in spin Hall effect (Sinova et al. 2015). Thus, the spin Hall effect is not limited to non-magnetic materials, and indeed, a sizable spin Hall effect was recently experimentally observed in ferromagnetic (Miao et al. 2013; Du et al. 2014a) and antiferromagnetic materials (TABLE 6).

The spin Hall effect is generally quantified in term of the spin Hall angle, $\theta_{SH}$, which represents the conversion ratio between the spin current and the charge current, $\theta_{SH} = I_s/I_c$. Various experimental measurement techniques have been employed to determine the value of $\theta_{SH}$, including spin torque ferromagnetic resonance (Liu et al. 2011), spin pumping experiments exploiting the inverse spin Hall effect (Saitoh et al. 2006) [see also FIG. 25(a) and (b)], magnetization switching (Suzuki et al. 2011), domain wall dynamics (Emori et al. 2013), second harmonic measurement (Pi et al. 2010), and dc Hall measurement (Kawaguchi et al. 2013).

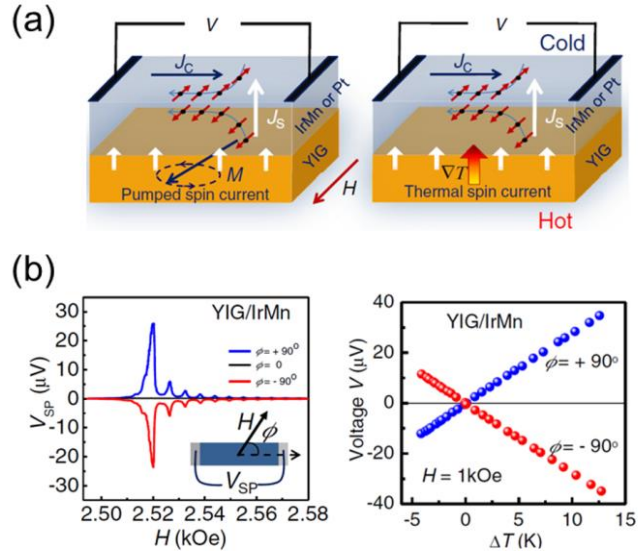

FIG. 34 (Color online). (a) Sketch showing the spin current density, $J_S$ generated by spin pumping (left) and spin Seebeck (right) effects and the resulting charge current density, $J_C$, converted by inverse spin Hall effect. (b) Voltages measured for these spin currents across the IrMn layer while varying the magnetic field ($H$) or temperature difference across the stack ($\Delta T$). From (Mendes et al. 2014).

Pioneering experimental measures of the spin Hall effect in antiferromagnets were reported by Mendes et al. (Mendes et al. 2014) in $Ir_{20}Mn_{80}$ layers, see FIG. 34. They showed that the Hall angle of $Ir_{20}Mn_{80}$ is comparable to that of Platinum: $\theta_{SH,IrMn} \sim 0.8\theta_{SH,Pt}$. Thorough investigations based on similar measurement techniques were subsequently reported for a number of other materials (TABLE 6).

| AF material | Effective spin Hall angle (%) | Technique | Stack | Reference |
|---|---|---|---|---|
| $Pt_{50}Mn_{50}$ | $6 \pm 1$ | SP | NiFe/Cu/PtMn | (W. Zhang et al. 2014) |
| $Pt_{50}Mn_{50}$ | $6.4 - 8.1$ | ST-FMR | NiFe/Cu/PtMn | (W. Zhang et al. 2015) |
| $Pt_{50}Mn_{50}$ | 8 (DL)<br>2 (FL) | MOD | NiFe/Cu/PtMn | (W. Zhang et al. 2015) |
| $Pt_{50}Mn_{50}$ (c-axis) | $4.8 - 5.2$ | ST-FMR | NiFe/Cu/PtMn | (W. Zhang et al. 2015) |
| $Pt_{50}Mn_{50}$ (a-axis) | $8.6 - 8.9$ | ST-FMR | NiFe/Cu/PtMn | (W. Zhang et al. 2015) |
| $Pt_{50}Mn_{50}$ | 10 | ST-FMS | [Co/Ni]/PtMn (oop) | (Fukami et al. 2016) |
| $Pt_{50}Mn_{50}$ | 16 − 19 (DL)*<br>4 − 0 (FL) | ST-FMR (HR) | Co/PtMn<br>and reversed | (Ou et al. 2016) |
| $Pt_{50}Mn_{50}$ | 9.6 − 17.4 (DL)*<br>4.3 − 3.6 (FL) | ST-FMR (HR) | FeCoB/PtMn<br>and reversed | (Ou et al. 2016) |
| $Pt_{50}Mn_{50}$ | 11 (DL)*<br>4 (FL) | ST-FMR (HR) | FeCoB/PtMn (oop) | (Ou et al. 2016) |
| $Pt_{50}Mn_{50}$ | 24 (DL)* | ST-FMR (HR) | FeCoB/Hf/PtMn (oop) | (Ou et al. 2016) |
| $Ir_{50}Mn_{50}$ | $2.2 \pm 0.5$ | SP | NiFe/Cu/IrMn | (W. Zhang et al. 2014) |
| $Ir_{50}Mn_{50}$ | $5.3 - 5.7$ | ST-FMR | NiFe/Cu/PtMn | (W. Zhang et al. 2015) |
| $Ir_{50}Mn_{50}$ (~poly., tentatively a-axis) | 2.3 | ST-FMR | NiFe/Cu/PtMn | (W. Zhang et al. 2015) |
| $Ir_{50}Mn_{50}$ (c-axis) | $5 \pm 0.5$ | ST-FMR | NiFe/Cu/PtMn | (W. Zhang et al. 2015) |
| $\gamma$-$Ir_{20}Mn_{80}$ | $0.8 - 6.4$ ;<br>0.8 x Pt** | SP and SSE | YIG/IrMn | (Mendes et al. 2014; Rojas-Sánchez et al. 2014) |
| $Ir_{20}Mn_{80}$ | $2.9 \pm 1.5$ (DL) | ST-FMR (HR) | CoFeB/IrMn | (Reichlová et al. 2015) |
| $Ir_{20}Mn_{80}$ | $4.3 \pm 0.1$ (DL) | MOD | NiFe/Cu/IrMn | (Tshitoyan et al. 2015) |
| $Ir_{20}Mn_{80}$ | $5.6 \pm 0.9$ | ST-FMR | NiFe/Cu/IrMn | (Tshitoyan et al. 2015) |
| $Ir_{22}Mn_{78}$ | $5.7 \pm 0.2$ (DL) | ST-FMR (HR) | CoFeB/IrMn | (D. Wu et al. 2016) |
| $Ir_{20}Mn_{80}$ | > 10.9*** | ST-FMR | NiFe/IrMn | (Tshitoyan et al. 2015) |
| $Ir_{20}Mn_{80}$ | 13.5 (DL) | MOD | NiFe/IrMn | (Tshitoyan et al. 2015) |
| $Ir_{25}Mn_{75}$ | 2 | ST-FMR | NiFe/IrMn | (Soh et al. 2015) |
| $Ir_{25}Mn_{75}$ | ~9 | ST-FMR | NiFe/IrMn | (W. Zhang et al. 2016) |
| $Ir_{25}Mn_{75}$ (111) | ~11 | ST-FMR | NiFe/IrMn | (W. Zhang et al. 2016) |
| $Ir_{25}Mn_{75}$ (100) | ~20 | ST-FMR | NiFe/IrMn | (W. Zhang et al. 2016) |
| $Pd_{50}Mn_{50}$ | $1.5 \pm 0.5$ | SP | NiFe/Cu/PdMn | (W. Zhang et al. 2014) |
| $Pd_{50}Mn_{50}$ | $2.8 - 4.9$ | ST-FMR | NiFe/Cu/PtMn | (W. Zhang et al. 2015) |
| $Pd_{50}Mn_{50}$ (c-axis) | $3.2 \pm 0.6$ | ST-FMR | NiFe/Cu/PtMn | (W. Zhang et al. 2015) |
| $Pd_{50}Mn_{50}$ (a-axis) | $3.9 \pm 0.5$ | ST-FMR | NiFe/Cu/PtMn | (W. Zhang et al. 2015) |
| Cr | $-5.1 \pm 0.5$ | SP | YIG/Cr | (Du et al. 2014a) |
| Cr (30 - 345K) | -9 (-1.38 x 20 x Cu) | SSE | YIG/Cr | (Qu, Huang, and Chien 2015) |
| Mn | $-0.19 \pm 0.01$ | SP | YIG/Mn | (Du et al. 2014a) |
| $Fe_{50}Mn_{50}$ | $0.8 \pm 0.2$ | SP | NiFe/Cu/FeMn | (W. Zhang et al. 2014) |
| $Fe_{50}Mn_{50}$ | $2.2 - 2.8$ | ST-FMR | NiFe/Cu/FeMn | (W. Zhang et al. 2015) |
| $\gamma$-Fe$_{50}$Mn$_{50}$ | -(7.4 ± 0.8) x 10$^{-3}$ | SP | YIG/FeMn | (Du et al. 2014a) |

TABLE 6. Spin Hall angles determined for various antiferromagnets. When not specified, the investigation temperature was 300 K, and the layers were polycrystalline. SP and SP ($\Delta$H) = ferromagnetic resonance spin pumping excitation – detected based on the inverse spin Hall effect, or ferromagnetic resonance linewidth (when ($\Delta$H) is specified), ST-FMR and ST-FMR (HR) = spin torque ferromagnetic resonance excitation induced by spin Hall effect subsequent to an ac current flow – detected based on anisotropic magnetoresistance, or 2$^{nd}$ harmonic response detection of the anomalous Hall effect and/or anisotropic magnetoresistance (when (HR) is specified), MOD = spin Hall effect excitations induced by a dc current flow - detection of the modulation of the ferromagnetic resonance damping, ST-FMS = spin torque ferromagnetic switching induced by spin Hall effect subsequent to a dc current flow – anomalous Hall effect detection, SSE = longitudinal spin Seebeck excitation induced by a thermal gradient – inverse spin Hall effect detection, and oop = out-of-plane magnetization. DL and FL refer to damping-like and field-like torque components, respectively. *Values of the spin torque efficiency [effective interface transparency (<1) x spin Hall angle]. **The values of the effective spin Hall angle for Pt were taken from (Rojas-Sánchez et al. 2014) and typically range between values close to 1% and 10%. ***Linear increase with the IrMn thickness.

In particular, Zhang et al (W. Zhang et al. 2014; W. Zhang et al. 2015) confirmed that 5$d$-metal-alloys, such as IrMn and PtMn, have a larger spin Hall angle than 4$d$-metal-alloys, such as PdMn, and 3$d$-metal-alloys, such as FeMn. These results highlighted the important role that spin-orbit coupling of the heavy elements plays in determining the properties of their simple alloy (H. Chen, Niu, and MacDonald 2014; Seemann et al. 2010). Further studies (Du et al. 2014a; H. L. Wang, Du, Pu, Adur, et al. 2014) demonstrated the impact of $d$-orbital filling, and highlighted the additive nature of effects due to atomic number and orbital filling as calculated by Tanaka et al. (Tanaka et al. 2008). Antiferromagnetic alloys also seem to obey this rule (TABLE 6 and corresponding FIG. 35). Thus, within the same *d*-shell, the orbital filling seems to govern the spin Hall effect. In contrast, for different materials with the same (valence) electron number, the influence of the atomic number seems to prevail. For example, PtMn and PdMn share the same electron number but show very different Hall angles since PtMn is a 5*d* alloy whereas PdMn is 4*d*. In this case, the ratio of Hall angles obeys the $Z^4$ dependence: $\theta_{SH,PtMn} / \theta_{SH,PdMn} \sim (Z_{PtMn} / Z_{PdMn})^4 \sim 4$.

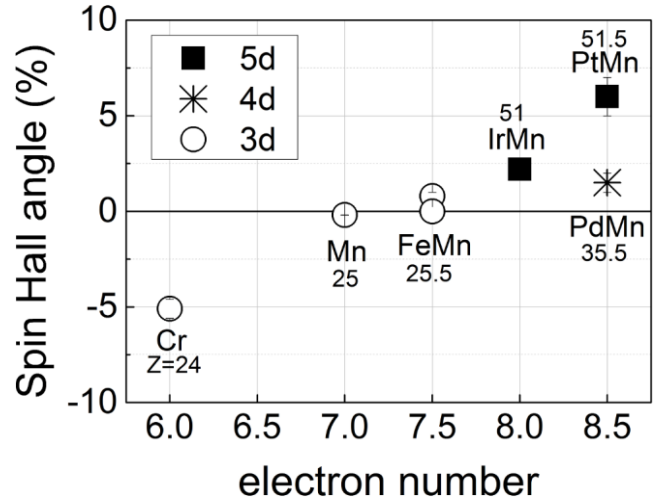

FIG. 35. Spin Hall angle *versus* electron number. The data are taken from TABLE 6 and references therein (only data measured with the SP technique are plotted so as to avoid possible misleading technique-related dispersions of data point, see text). For memory, the outer electronic shell filling patterns are as follows: Cr ($3d^5 4s^1$), Mn ($3d^5 4s^2$), Fe ($3d^6 4s^2$), Ir ($5d^9 6s^0$), Pt ($5d^9 6s^1$), and Pd ($4d^{10} 5s^0$).

These pioneering theoretical and experimental works highlighted the need for further investigations (Sklenar et al. 2016): first to quantify spin-orbit interactions prior to testing their efficiency as will be discussed later, second to experimentally determine how the magnetic order influences the spin Hall effect, and above all to maximize the spin Hall effect in antiferromagnets. In particular the role heavy elements play in determining the properties of spin-orbit-interaction in antiferromagnetic alloys needs to be scrutinized in greater detail. Future studies could involve doping the antiferromagnet with impurities up to the limits of solubility so as to investigate the resultant change in spin Hall angle and the dominant underlying mechanism. How much the static magnetic order, and related non-collinear spin texture, influence the amplitude of the spin Hall effect motivated several studies. In FIG. 25(c) it can be observed that, like spin pumping [FIG. 25(b)], the amplitude of the inverse spin Hall effect in polycrystalline IrMn layers is constant around the magnetic phase transition [i.e., around $t_{IrMn}$ = 2.7 nm (Frangou et al. 2016)]. The inverse spin Hall effect in such non-collinear polycrystalline IrMn layers (3Q spin structure in bulk, TABLE 1) is probably mainly sensitive to the nature of the elements making up the alloy, and to a lesser extent to the magnetic order where the different directions of the moments average out the spin Hall signal, as computed by Zhang et al. (W. Zhang et al. 2014). This effect will be discussed in detail in the next paragraph. The inverse spin Hall effect in polycrystalline Cr with a preferred (110) texture was also reported to be independent of the ordering of the metal (Qu, Huang, and Chien 2015). This finding was probably due to the combination of unusual spin density wave antiferromagnetic ordering and texture averaging out the spin Hall signal, although this effect remains to be accurately theoretically demonstrated.

First-principles calculations based on density functional theory actually show anisotropy of the spin Hall effect, meaning that spin Hall conductivity is sensitive to the direction of staggered magnetization with respect to the crystallographic axes (W. Zhang et al. 2014). By averaging the spin Hall conductivities computed for staggered magnetization oriented along different axes Zhang et al. (W. Zhang et al. 2014) could qualitatively reproduce their experimental findings on polycrystalline PtMn, IrMn, IrMn and FeMn films. Zhang et al. (W. Zhang et al. 2015) further showed that epitaxial CuAu-I type antiferromagnets (PtMn, PdMn, IrMn and FeMn, see also TABLE 1) could effectively be used to study how the magnitude of the spin Hall effect is influenced by the magnetic order. More specifically, CuAu-I type PtMn, PdMn, IrMn and FeMn antiferromagnets have a collinear staggered spin structure (TABLE 1), the orientation of which depends on the crystal growth. The anisotropy of the spin Hall effect arises from both the anisotropy of chemical and magnetic structure. Zhang et al. (W. Zhang et al. 2015) (TABLE 6) experimentally demonstrated the anisotropy of the spin Hall effect in antiferromagnets and corroborated there data by first-principles calculations using density functional theory.

It is noteworthy that estimation of actual spin Hall angle value in antiferromagnets faces the same problems already encountered for more common materials, such as Pt. Variations in spin Hall angle were found with various measurement methods and material combinations. In some cases, these variations were ascribed to spin-memory loss at interfaces (Rojas-Sánchez et al. 2014). With some techniques (TABLE 6), a charge-to-spin current conversion occurs in the antiferromagnet (e.g. with the ST-FMR, ST-FMS and MOD techniques). In these cases, there are alternatives to the spin Hall effect as sources of spin current (e.g. inverse spin galvanic effects). These alternative sources may explain why the effective Hall angle deduced by these techniques overestimates the actual Hall angle compared to other techniques like SP and SSE (TABLE 6). Note also that ST-FMR data can be extracted from either ratio or amplitude analysis of symmetric and antisymmetric contributions to FMR lineshape. These alternatives may also contribute to the slight data dispersion observed, although similar data has been obtained by the two methods (W. Zhang et al. 2015). While the values for ST-FMR and MOD were larger than those for SP, the results presented by Zhang et al. (W. Zhang et al. 2015) hopefully demonstrate that the trends (i. e. the overall increase of effective Hall angle from FeMn to PdMn, IrMn and PtMn for the reasons detailed in FIG. 35 and corresponding discussion) are qualitatively the same when similar samples are measured (TABLE 6).

When interpreting these data, care must be taken when the antiferromagnetic material is in direct contact with a ferromagnet. Ferromagnetic/antiferromagnetic exchange bias interactions (I.C.3) appear to increase the effective spin Hall angle compared to data obtained with ferromagnetic/non-magnetic-metal/antiferromagnetic trilayers (Tshitoyan et al. 2015) (TABLE 6). Readers should remember that for ferromagnetic/non-magnetic/antiferromagnetic multilayers no direct exchange interaction takes place between ferromagnet and antiferromagnet, meaning that the transport regime is purely electronic through the non-magnetic-metal (mostly Cu). In contrast, for ferromagnetic/antiferromagnetic multilayers, due to magnetic coupling, transfer and propagation of spin angular momentum directly involves magnonic transport, i.e. spin waves from the oscillating ferromagnet feed directly into the antiferromagnet. As the origin of the spin transport is different, spin transmission efficiency (spin mixing conductance) through interfaces may be significantly different (II.B). Tshitoyan et al. (Tshitoyan et al. 2015) showed that the enhancement of the spin angular momentum transfer through a ferromagnetic/antiferromagnetic interface is indeed directly related to the interfacial exchange-coupling amplitude. Such an effect makes calculation of the actual spin Hall angle of the antiferromagnet more difficult. However, ferromagnetic/antiferromagnetic coupling increases the spin torque efficiency, making it an advantage for current-induced switching of the ferromagnet. Section III.D.4 focuses on the use of the multifunctionalities of antiferromagnets (spin Hall torques and exchange bias) to deterministically reverse the out-of-plane magnetization of a ferromagnet in zero applied magnetic field (Fukami et al. 2016; Lau et al. 2016; Brink et al. 2016; Sklenar et al. 2016; Oh et al. 2016). Note also that Ou et al. (Ou et al. 2016) further highlighted the importance and complexity of interfacial engineering (see also section II.B.1). These authors demonstrated a two-fold increase in spin torque efficiency in FeCoB/Hf/PtMn trilayers compared to FeCoB/PtMn bilayers. In this case, the reduction of transmission inherent to the separation of the FeCoB ferromagnet and the PtMn antiferromagnet is compensated by the increased transmission following the introduction of the Hf spacer.

### 3.    Spin Hall magnetoresistance

Magnetic multilayers containing heavy metals adjacent to ferromagnets have recently been shown to display anisotropic magnetoresistance with symmetries differing from those traditionally found in bulk ferromagnets (Kobs et al. 2011). While anisotropic magnetoresistance in bulk polycrystalline films depends on the angle between the flowing current and the magnetization direction, $\sim \left( \mathbf{m} \cdot \mathbf{j} \right)^2$, in ultrathin films an additional (interfacial) anisotropic magnetoresistance emerges that depends on the angle between the magnetization and the direction transverse to the current flow, $\sim \left( \mathbf{m} \cdot \left( \mathbf{z} \times \mathbf{j} \right) \right)^2$. Several mechanisms have been identified to explain this behavior, such as interfacial Rashba spin-orbit coupling (X. Wang, Pauyac, and Manchon 2014; S. S. L. Zhang, Vignale, and Zhang 2015) and a spin Hall effect in the normal metal adjacent to the ferromagnet (Y.-T. Chen et al. 2013; Nakayama et al. 2013). This effect has been reported in a wide range of magnetic multilayers and is designated under the broad name of spin Hall magnetoresistance (SMR). Since the effect is unaffected by magnetization reversal it would appear natural to observe it in antiferromagnetic multilayers (Manchon 2017b). Indeed, it has been reported for several antiferromagnets: metallic-IrMn in YIG/IrMn stacks (X. Zhou et al. 2015), insulating $SrMnO_3$ in $SrMnO_3$/Pt stacks (J. H. Han et al. 2014) and insulating NiO in Pt/NiO/YIG trilayers (Shang et al. 2016; Hung et al. 2017; Lin and Chien 2017; Hou et al. 2017).

With metallic antiferromagnets, much attention must be paid to disentangling spin Hall magnetoresistance from anisotropic magnetoresistance induced in the antiferromagnet by potential uncompensated magnetic moments (also referred to as magnetic proximity effects).

To do so, systematic angular dependences of the magnetoresistance must be determined. For example, in YIG/IrMn bilayers, Zhou et al (X. Zhou et al. 2015) demonstrated that the competition between spin Hall magnetoresistance and magnetic proximity effects can lead to a peculiar sign change of the spin Hall-like magnetoresistance with temperature. Spin Hall magnetoresistance has also been reported in metallic FeMn/Pt bilayers (Y. Yang et al. 2016). However, in this work, the FeMn alloy was not antiferromagnetic since it presented a strong magnetization, of 250 emu.cm$^3$ for the 2 nm thick FeMn alloy and 25 emu.cm$^3$ for 7 nm. As expected, the spin Hall magnetoresistance signal scaled with the residual ferromagnetism.

With regards to insulating antiferromagnets and in particular to Pt/NiO/YIG trilayers (Shang et al. 2016; Hung et al. 2017; Lin and Chien 2017; Hou et al. 2017) NiO was demonstrated to not only convey the spin information through spin waves to the YIG (TABLE 5) but to be the source of the magnetoresistive signal. This effect was linked to enhanced spin Hall magnetoresistance at the magnetic phase transition of NiO (Lin and Chien 2017; Hou et al. 2017) and non-conventional negative signal were detected (Shang et al. 2016; Lin and Chien 2017; Hou et al. 2017), due to spin-flip coupling (90° exchange coupling) between the NiO and YIG magnetic orders.

## D.    Spin-orbit torques

### 1.    Principle of spin-orbit torques

In magnetic systems lacking inversion symmetry (globally or locally) and possessing sizable spin-orbit coupling, the transfer of angular momentum between the orbital angular momentum of carriers and the spin angular moment of the localized electrons result in so-called spin-orbit torques [see e. g. the following general articles (Brataas and Hals 2014; Gambardella and Miron 2011; Manchon 2014)]. Current-driven switching (e. g. in Pt/Co (Miron et al. 2011) and Ta/CoFeB (Liu et al. 2012) stacks), resonance (Liu et al. 2011) and domain wall motion (Miron et al. 2010) in transition metals, as well as in noncentrosymmetric magnetic semiconductors ((Ga,Mn)As, (Ga,Mn)(As,P), (Ni,Mn)Sb) (Chernyshov et al. 2009; Fang et al. 2011; Kurebayashi et al. 2014; Ciccarelli et al. 2016) have attracted massive attention in the past five years, opening fascinating venues for non-volatile memory and logic applications.

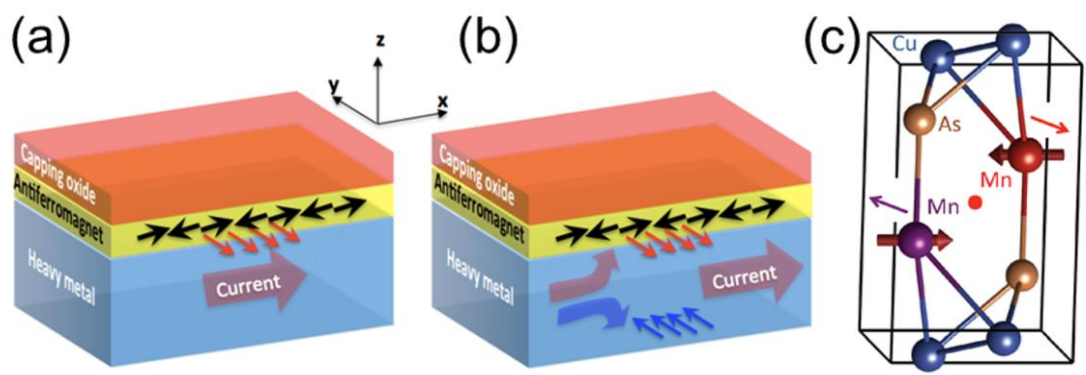

FIG. 36 (Color online). Schematics of various origins of the spin-orbit torque: (a) inverse spin galvanic effect arising at the interface with the heavy metal, (b) spin Hall effect taking place in the bulk of the heavy metal and (c) inverse spin galvanic effect emerging from bulk inversion asymmetry. From (Jungwirth et al. 2016).

Up until now, two main mechanisms have been identified as the origin of spin-orbit torque: (bulk or interfacial) inverse spin galvanic effect and spin Hall effect. The former is the electrical generation of a nonequilibrium spin density [see FIG. 36(a)]. In systems lacking (bulk or interfacial) inversion symmetry, the spin-orbit coupling becomes *odd* in momentum. Examples are linear and cubic Dresselhaus spin-orbit coupling in strained Zinc-Blende semiconductors, Rashba spin-orbit coupling in bulk Wurtzite and at interfaces between dissimilar materials as well as cubic Rashba spin-orbit coupling in oxide heterostructures (Manchon et al. 2015). Such *odd-in-k* spin-orbit coupling enables current-driven spin densities that can be used for switching the magnetization direction of ferromagnetic materials. On the other hand, a heavy metal adjacent to the ferromagnet can create spin-orbit torques driven by the spin Hall effect [see FIG. 36(b)]. Through the combination of in-plane charge current and spin-orbit interaction in the heavy metal, a pure spin current is induced orthogonal to the electric current, thereby exerting a torque on the neighboring ferromagnet. Both inverse spin galvanic and spin Hall effect produce a torque of the form

$$\tau = \tau_{\parallel} \mathbf{m} \times \left( \left( \mathbf{u} \times \mathbf{E} \right) \times \mathbf{m} \right) + \tau_{\perp} \mathbf{m} \times \left( \mathbf{u} \times \mathbf{E} \right),$$

$$(34)$$

where $\mathbf{u}$ is a unit vector determined by the symmetry of the system and $\mathbf{E}$ is the current density. In the case of a magnetic multilayer perpendicular to $\mathbf{z}$, as depicted in FIG. 36(b) $\mathbf{u} = \mathbf{z}$. Like in the case of spin transfer torque (see section II.A.1), the first term is named the *damping-like* torque, while the second is called the *field-like* torque.

　　Note that spin-orbit torques present some crucial differences compared to spin transfer torques. The latter build on transfer of spin angular momentum between flowing spins and local magnetic moment, and thereby require the existence of two ferromagnets: one of which acts as a spin-polarizer while the second acts as the free layer. In contrast, spin-orbit torques rely on momentum transfer between the spin and angular momenta mediated by spin-orbit coupling and therefore do not require an additional ferromagnetic layer. Hence, they cover both spin Hall torque and inverse spin galvanic torque, in spite of their very different origin. The inverse spin galvanic torque is really an intrinsic torque, taking advantage of the nature of the spin-orbit coupling in bulk or at the interface of the magnet (Bernevig and Vafek 2005;

Manchon and Zhang 2008; Garate and MacDonald 2009), while spin Hall torque arises from a spin current generated away from the interface, in the bulk of the heavy metal (Haney et al. 2013).

Both torques, inverse spin galvanic torque and spin Hall torque, are also present in antiferromagnetic systems lacking inversion symmetry and can provide a unique tool to manipulate the magnetic order, as predicted (Železný et al. 2014) and demonstrated recently (Wadley et al. 2016). Železný et al (Železný et al. 2017) recently analyzed the symmetries of spin-orbit torques in magnets lacking either local or global inversion symmetry, determining the tensorial forms of the torques for the different non-centrosymmetric point groups.

## 2. Manipulation of the order parameter by spin-orbit torque

The simplest model system on which such a spin-orbit torque has been proposed is the Rashba antiferromagnetic two-dimensional electron gas (Železný et al. 2014). The band structure of this system presents striking differences with the ferromagnetic Rashba system, as illustrated in FIG. 37. As a matter of fact, in a normal metal with Rashba spin-orbit coupling, the low energy dispersion of the eigenstates is given by $\varepsilon_k^s = \dfrac{\hbar^2 k^2}{2m} + s\alpha k$ [see FIG. 37(a)], while in ferromagnets, the band structure is given by $\varepsilon_k = \dfrac{\hbar^2 k^2}{2m} + s\sqrt{\Delta^2 + \alpha^2 k^2}$ [see FIG. 37(b) - we take the magnetization perpendicular to the plane for simplicity]. Here $\alpha$ is Rashba spin-orbit coupling parameter, $\Delta$ is the s-d exchange energy, and $s = \pm 1$ refer to the different spin chiralities. While Rashba spin-orbit coupling splits the electronic bands of normal metals, in ferromagnets there is a direct competition between the exchange and Rashba spin-orbit coupling and in the large exchange limit the spin splitting is mostly driven by the exchange. In contrast, in antiferromagnets, $\varepsilon_k = \eta\left(\sqrt{\gamma_k^2 + \Delta^2} + s\alpha k\right)$ [see Eq. (4), section I.C.1], i.e. the spin splitting is directly given by Rashba, no matter how strong the exchange [see Fig. FIG. 37(c)].

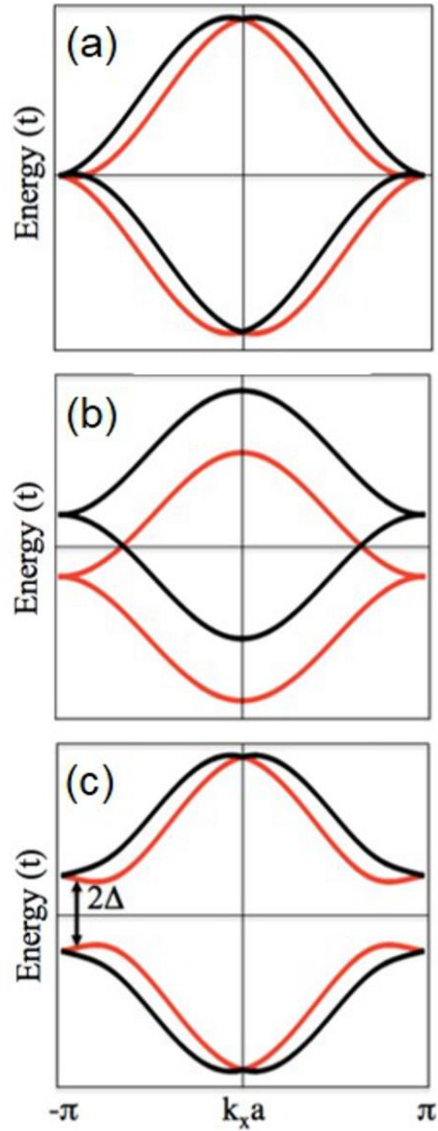

FIG. 37 (Color online). Band structure of (a) a normal metal with Rashba spin-orbit coupling, (b) a ferromagnet with Rashba spin-orbit coupling and (c) an antiferromagnet with Rashba spin-orbit coupling. These band structures were calculated for a nearest-neighbor tight-binding model with the same parameters.

Here $h = \pm 1$ refers to the valence and conduction bands. As a consequence, in the case of ferromagnetic Rashba gas with disorder broadening $\Gamma$, the two components of the torque scale $t_\wedge \propto a/\Gamma$, $t_\parallel \propto a/\Delta$ [strong exchange limit – see e.g. (Li et al. 2015)], while in the case of an antiferromagnetic Rashba gas, it reads $t_\wedge \propto a/G$, $t_\parallel \propto aD/e_F^2$ (Saidaoui and Manchon 2016).

Overall, it was demonstrated that the spin density on sublattice $i$, besides a strong angular dependence, reads

$$\mathbf{S}_i = \tau_\perp \mathbf{z} \times \mathbf{E} + \tau_\parallel \mathbf{m}_i \times (\mathbf{z} \times \mathbf{E}).$$

(35)

More specifically, the inverse spin galvanic effect arising from intraband transitions (first term) produces the same spin density on the two sublattices, $s_\perp$, while the magnetoelectric effect arising from interband transitions produces a *staggered* spin density, $s_\parallel$, which thus induces the torque efficient for the electrical control of the order parameter (Železný et al. 2014).

In the same theoretical study, Železný et al. reported that current-induced 'Néel-order' spin-orbit fields whose sign alternate between the spin sublattice can arise even in the bulk of centrosymmetric antiferromagnets such as $Mn_2Au$ [or CuMnAs, see FIG. 36(c)]. The general case of spin polarization due to atomic site asymmetries within the crystal rather than space group asymmetries was also developed by Zhang et al. (X. Zhang et al. 2014). Actually, each sublattice has broken inversion symmetry but together form 'inversion partners'. Then, each sublattice experience an opposite inverse spin galvanic effect (Železný et al. 2014). Therefore, in this case, the efficient torque enabling the manipulation of the order parameter is a *staggered field-like* torque (FIG. 14(b)). Roy et al. (Roy, Otxoa, and Wunderlich 2016) used the example of $Mn_2Au$-based devices to further compute the influence of spin-orbit field strength, current pulse properties, and damping on switching. They demonstrated robust picosecond writing with minimal risk of overshoot. These pioneering studies are very encouraging in view of triggering antiferromagnetic order reversal by current injection.

*Manipulation by inverse spin galvanic torque*

Recently, the current-driven order parameter reversal in CuMnAs was indeed observed [FIG. 36(a) and FIG. 38] [via global electrical measurements (anisotropic magnetoresistance) (Wadley et al. 2016; Olejnik et al. 2016) as well as via local direct imaging of the antiferromagnetic domains (x-ray magnetic linear dichroism photoelectron emission microscopy) (Grzybowski et al. 2017)]. CuMnAs crystal possesses local inversion symmetry breaking in bulk crystal, as explained above (see also TABLE 3 in the materials survey section), and displays anisotropic magnetoresistance allowing for the electrical detection of the order parameter orientation. The authors demonstrated that the order parameter could be switched reversibly by 90 degrees upon electrical injection. While numerous aspects still demand deeper investigations (e.g. the role of temperature and details of the reversal dynamics), these experiments constitute the first clear demonstration of current-driven Néel order manipulation.

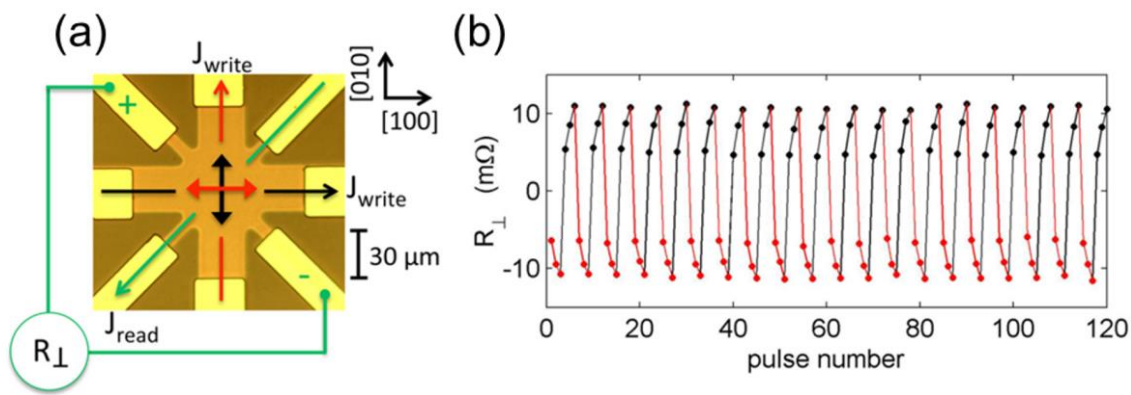

FIG. 38 (Color online). Current-induced switching of CuMnAs with the inverse spin galvanic effect and subsequent electrical detection with the anisotropic magnetoresistance effect. From (Wadley et al. 2016).

*Manipulation by spin Hall torque*

Another promising configuration is to use the spin Hall effect arising from an adjacent heavy metal [FIG. 36(b)]. Strictly speaking, this torque behaves like a spin transfer torque (section II) from a virtual ferromagnet polarized along the spin Hall polarization direction, i.e., $\sim \mathbf{z} \times \mathbf{E}$. The torque is therefore a damping-like torque with the form $\mathbf{m} \times \left( (\mathbf{z} \times \mathbf{E}) \times \mathbf{m} \right)$ FIG. 14(c). These bilayer structures present some clear advantages compared to ferromagnet/spacer/antiferromagnet spin-valves, as the spin current is directly injected from the normal metal into the antiferromagnet and survives in the diffusive regime (Manchon 2017a). Learning how to exploit spin Hall torques to reverse the antiferromagnetic order is yet another challenge to be addressed.

Experimentally, spin Hall torque has been reported by Reichlová et al. (Reichlová et al. 2015) in Ta/IrMn, using the $2^{nd}$ harmonic method. Detectability of the signal below the Néel temperature and its insensitivity to external fields supported the hypothesis that the torque relates to the antiferromagnetic order. Other attempts reported very efficient spin Hall assisted magnetic order reversal in FeMn and IrMn alloys (Y. Yang et al. 2016; J. Han et al. 2016). However, these alloys were directly grown on $SiO_2$ layers, a process which is known to hamper the production of good quality antiferromagnets. As a result, FeMn and IrMn alloys with non-zero saturation magnetization were obtained. A value of 250 emu.cm$^3$ was reported the 2 nm thick FeMn alloy and 25 emu.cm$^3$ for the 7 nm thick alloy (Y. Yang et al. 2016). The low saturation magnetization of the (non-antiferromagnetic) FeMn and IrMn alloys on $SiO_2$ readily explains the large effective fields (low critical current) obtained.

Although spin Hall torque presents opportunities in terms of system design, the dynamics of spin-Hall-driven antiferromagnets are no different from those of the spin-torque-induced systems initially studied by Gomonay et al. (H. V. Gomonay and Loktev 2010; H. V.

Gomonay and Loktev 2014). Current-driven switching and excitations are governed by the very same physics. In a recent study, Cheng et al. (Cheng, Xiao, and Brataas 2016) proposed exploitation of the spin Hall torque to trigger TeraHerz oscillations. In their study, they considered the case of a biaxial antiferromagnet and showed that when the spin Hall torque is increased, the frequency of the optical mode is progressively reduced, while the frequency of the acoustic mode is enhanced. At the critical value $J_{cr}$ both modes are degenerate, which triggers THz excitations (see FIG. 39). Uniform steady state oscillations can be sustained thanks to the dynamic feedback from the pumped spin current: the spin current backflow renormalizes the spin transfer torque in a non-linear manner, which is sufficient to stabilize the THz oscillations (Cheng, Zhu, and Xiao 2016). Notice that in this case the critical current is given by Eq. (25), with $\omega_{AF} = \gamma\mu_0\sqrt{\dfrac{H_\perp^2}{4} + \alpha^2\left(2H_\parallel + H_\perp\right)H_E}$ , i.e. it is mostly governed by the hard axis anisotropy. Using parameters for NiO, the authors predict a critical current density of about $10^8$ A.cm$^{-2}$. The oscillation threshold could be further reduced by considering a uniaxial antiferromagnet $\left(H_\perp \to 0\right)$ such as MnF$_2$. Finally, in a recent work Daniels et al. (Daniels et al. 2015) investigated the ability of spin Hall torque to drive spin wave excitations in model insulating antiferromagnets. Their results indicated that surface spin waves can be excited at much lower current density than bulk spin waves due to surface anisotropy. A similar effect had previously been observed in ferromagnetic insulators (Xiao and Bauer 2012).

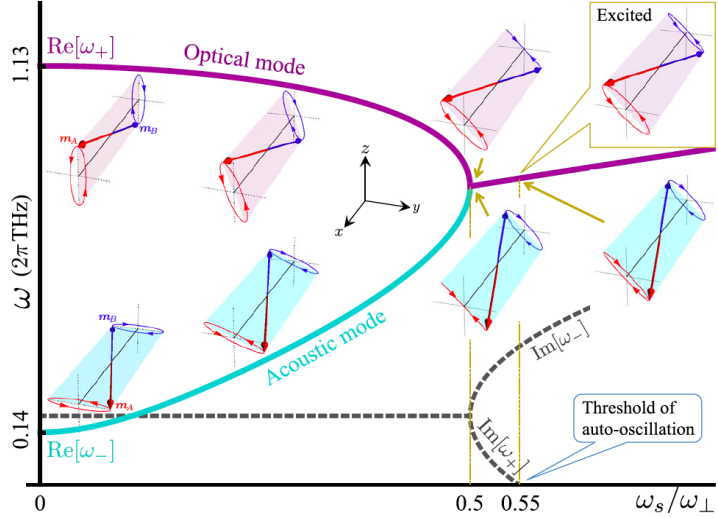

FIG. 39 (Color online). Evolution of the eigenfrequencies of NiO submitted to spin Hall torque. In the region $\omega_s < \omega_\perp / 2$, the frequency of the optical mode is reduced while the frequency of the acoustic mode increases. Note that the sublattice moments $\mathbf{m}_{A,B}$ oscillate with opposite chiralities. At $\omega_s = \omega_\perp / 2$, both modes are degenerate, whereas above $\omega_s > 0.55\omega_\perp$, auto-oscillations are triggered. In this regime, $\mathbf{m}_{A,B}$ and the Néel vector $\mathbf{l}$ oscillate with the same chiralities. Here $\omega_\perp = \gamma\mu_0 H_\perp$, where $H_\perp$ is the out-of-plane anisotropy as defined in the text. From (Cheng, Xiao, and Brataas 2016).

## 3. Moving magnetic textures by spin-orbit torque

Spin-orbit torques, such as *staggered field-like torque* and *damping-like* torque, present an interesting paradigm for the manipulation of antiferromagnetic textures. Two theoretical studies recently investigated the motion of domain walls in the presence of *staggered field-like torque* (O. Gomonay, Jungwirth, and Sinova 2016) and in the case of spin Hall (*damping-like*) torque (Shiino et al. 2016). The former modeled the current-driven motion of domain walls in bulk non-centrosymmetric antiferromagnet such as CuMnAs, while the latter aimed at modelling Néel walls in antiferromagnet/normal metal bilayers. In both cases, the authors emphasize that the azimuthal angle of the wall was not affected by the current, in sharp contrast with what occurs in ferromagnets. As a result, antiferromagnetic domain walls do not experience Walker breakdown and can reach very high velocities (as introduced in section I.C.4). The largest velocity attainable by the wall is set by the spin wave velocity, of the order of 1 to 10 km/s for current densities of about $10^7$ A/cm$^2$. This is about two orders of magnitude more efficient than can be achieved with ferromagnetic domain walls. Close to this upper limit, antiferromagnetic domain walls experience Lorentz contraction (FIG. 40) (similar to ferromagnets in the vicinity of the Walker breakdown, e. g. (Sturma, Toussaint, and Gusakova 2016)) and emit spin waves in the THz regime. These predictions are very encouraging for the development of domain wall control. Current experimental development is focused on producing current-driven domain wall motion.

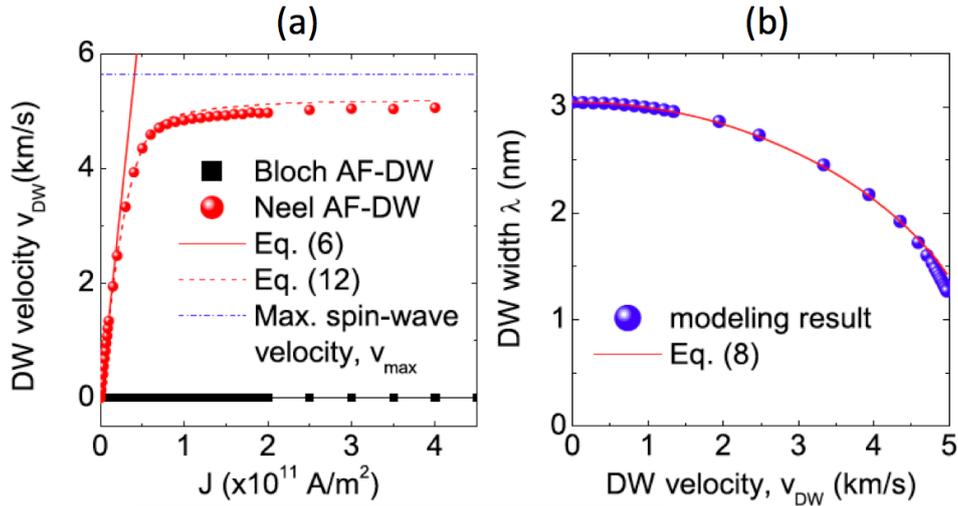

FIG. 40 (Color online). (a) Domain wall velocity as a function of current density in the presence of spin Hall torque. The blue dashed line represents the spin wave velocity which sets the upper bound for domain wall velocity. Note that only Néel domain walls can be moved by spin Hall torque. (b) Domain wall width as a function of domain wall velocity, displaying Lorentz contraction. From Shiino et al. (Shiino et al. 2016).

**4.    Current-induced switching by antiferromagnets**

More recently, researchers used spin Hall torques produced by antiferromagnets (see section III.C.2.) to excite and reverse ferromagnets without applying any external magnetic fields. The magnetization direction of ferromagnets with in-plane magnetic anisotropy can be reversed by spin Hall torques in zero applied magnetic field (Liu et al. 2012). However, ferromagnets with out-of-plane anisotropy require additional symmetry breaking - for example by applying an in-plane magnetic field (along the current flow direction) (Cubukcu et al. 2014) or by introducing geometric asymmetries (Yu et al. 2014; Safeer et al. 2015) - for spin Hall torques to switch their magnetization in a deterministic way.

Four groups (Fukami et al. 2016; Lau et al. 2016; Brink et al. 2016; Oh et al. 2016) simultaneously used the magnetic interactions between a ferromagnet with out-of-plane anisotropy (Co/Ni, CoFe and CoFe/Ni, Co/Pt, CoFeB) and an antiferromagnet (PtMn, IrMn) to create in-plane exchange bias and break the symmetry. Their results showed that the spin Hall torques either directly from the antiferromagnet (Fukami et al. 2016; Oh et al. 2016) (FIG. 41) or from an additional heavy metal (Lau et al. 2016; Brink et al. 2016) could then deterministically reverse the out-of-plane magnetization of the ferromagnet in zero applied magnetic field. In all these experiments, it was also confirmed that the damping-like torque contribution prevails over the field-like torque contribution (see section III.2 and TABLE 6 where the spin Hall effect in antiferromagnets is discussed, regardless of its further impact on neighboring ferromagnets). For example, FIG. 42 reproduces the findings of Zhang et al (W. Zhang et al. 2015) for NiFe/Cu/antiferromagnet trilayers. The signatures of damping-like and field-like torques were determined by measuring the changes in NiFe resonance linewidth and resonant frequency, respectively, when a dc current is injected in the plane of the antiferromagnet. For NiFe/Cu/PtMn trilayers, the authors found that the damping-like torque contribution is 4 times more efficient than the field-like torque (see also TABLE 6).

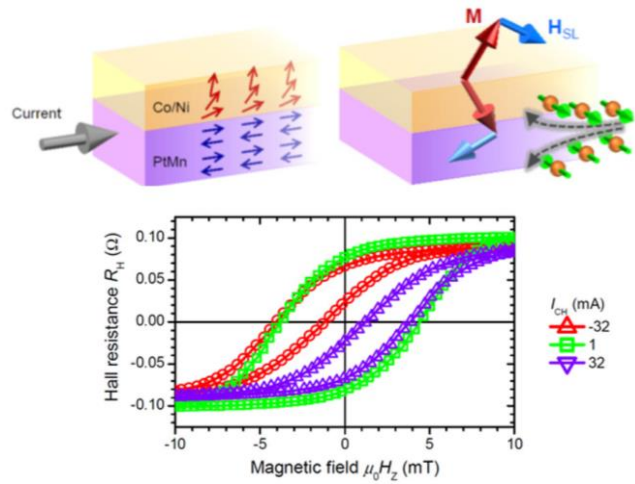

FIG. 41 (Color online). Ferromagnetic/antiferromagnetic exchange bias and current-induced spin Hall torques from the antiferromagnet combine to reverse the magnetization direction of a ferromagnet with out-of-plane anisotropy. Adapted from (Fukami et al. 2016).

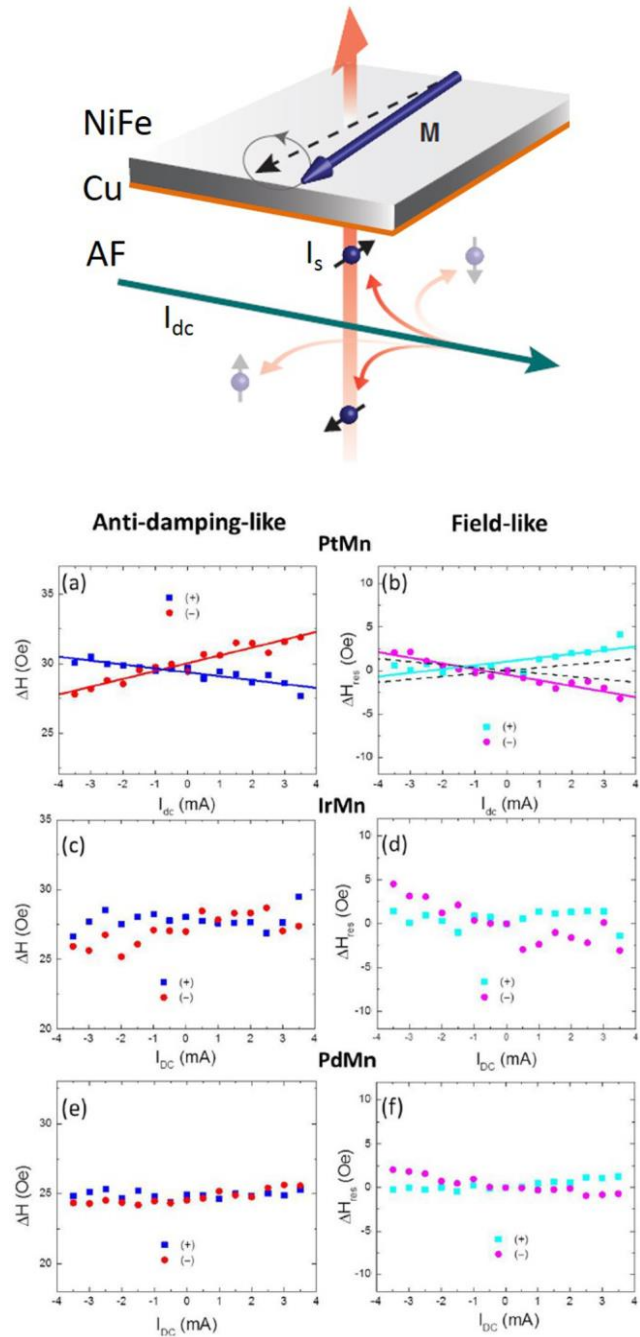

FIG. 42 (Color online). (Left) Illustration of the experimental setup. An in-plane dc current ($I_{dc}$) flowing in the antiferromagnet (AF) generates transverse spin currents that modulate the magnetization dynamics (damping and/or resonance frequency) of a ferromagnetic NiFe layer. Adapted from (Ando et al. 2008). (Right) Signatures of damping-like and field-like torques based on measurement of the changes in resonance linewidth and resonant frequency with $I_{dc}$ for positively (+) and negatively (-) magnetized NiFe. From (W. Zhang et al. 2015).

## IV.    SPIN-CALORITRONICS

Spin-caloritronics aims at using thermal gradients to generate and manipulate spin currents [for review articles, see for example (Bauer, Saitoh, and van Wees 2012) and (Boona, Myers, and Heremans 2014)]. This possibility is particularly interesting in insulating magnets, where spin currents are transported by spin waves rather than itinerant electrons. While most of the research in this area has focused on ferromagnets (and in particular on the yttrium iron garnet ferromagnetic insulator), progress has recently been made in the field of antiferromagnets. This section reviews the latest achievements in this field in terms of thermal generation of spin current and spin transport (II.B) driven by antiferromagnetic spin waves.

### A.    Thermally-induced spin currents

Spin currents can be readily induced in magnetic insulators by the application of a thermal gradient. The prototypical example of this was established by the demonstration of the spin Seebeck effect at the interface between a heavy metal and ferrimagnetic yttrium iron garnet (K. Uchida et al. 2010; K. I. Uchida et al. 2010). The thermal gradient applied normal to the interface is believed to induce a magnonic spin current in the magnetic insulator, which is converted into an electronic spin current via the exchange coupling between ($s$-like) itinerant and the ($d$-like) localized spins at the interface. The latter is then detected as a transverse voltage via the inverse spin Hall effect (see section III.C.2). Phenomenologically, the spin Seebeck physics is thus manifested as a Nernst effect of the heavy metal/magnetic insulator bilayer (FIG. 43).

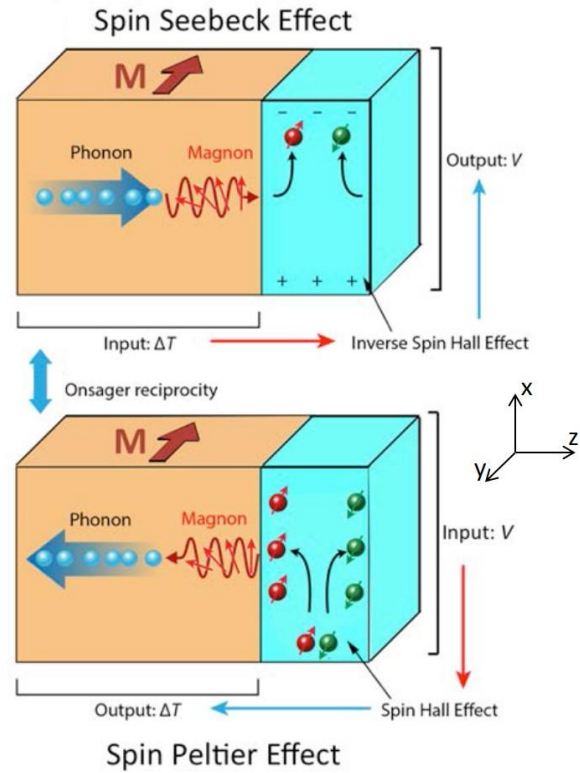

FIG. 43 (Color online). The reciprocal spin Peltier and spin Seebeck effects, manifested in a heavy metal/magnetic insulator bilayer. A mutual viscous drag can significantly impact the coupled thermal transport of magnons and phonons in the insulator. From (Hellman et al. 2016).

The essential physics at the interface is the exchange induced spin-magnon transmutation: itinerant electrons scattering off the interface can inelastically flip their spin $\hbar/2$ while producing a spin-$\hbar$ magnon. The magnons, in turn, can decay at the interface, while reemitting a spin-$\hbar$ electron-hole pair (FIG. 44). Since magnons carry energy as well as spin, these inelastic processes clearly reflect the intimately coupled nature of the spin and heat flows in magnetic insulators. Macroscopically, these magnon creation and decay processes are manifested through the Onsager-reciprocal spin-transfer torque and spin pumping (Tserkovnyak et al. 2005; Brataas et al. 2012).

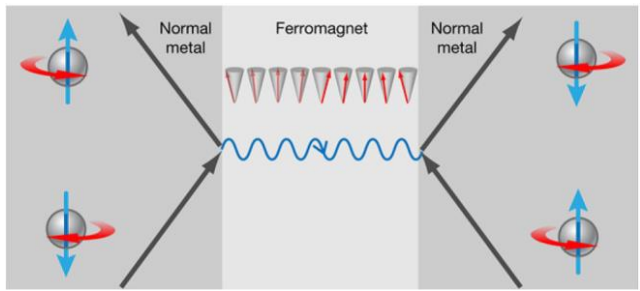

FIG. 44 (Color online). Generation of a magnon (directional wavy line and snapshot of the magnetization) associated with an electron spin flip indicated by spheres, at the left interface between normal metal and ferromagnet, and the reciprocal magnon decay at the right interface. Such electron spin-flip processes accompanied by creation and annihilation of magnons are expected for any collinear magnetic system, including antiferromagnets. From (Bauer and Tserkovnyak 2011).

The theory of such interfacially induced thermal spin currents produced by a ferromagnetic insulator is well established, at least qualitatively. With the early efforts being based on a semiclassical stochastic Landau-Lifshitz-Gilbert treatment (Xiao et al. 2010; Hoffman, Sato, and Tserkovnyak 2013), a systematic quantum-kinetic description has been formulated by Bender and Tserkovnyak (Bender and Tserkovnyak 2015). In general, the spin and the associated heat flow at a magnet/metal interface are driven by the effective temperature as well as chemical potential drops over the interface (with the counterpart of the magnonic chemical potential on the metallic side being provided by the spin accumulation, i.e., the vectorial difference of the spin-up and down electronic chemical potentials) (Bender and Tserkovnyak 2015). For example, when the spin accumulation $\boldsymbol{\mu}_s$ is collinear with the magnetic order parameter, the interfacial spin current (which is polarized along the same direction) is given by (Bender, Duine, and Tserkovnyak 2012; Bender et al. 2014)

$$j_x = \frac{\hbar g^{\uparrow\downarrow}}{\pi s} \int_{\hbar\omega}^{\infty} d\epsilon D(\epsilon)(\epsilon - \mu_s)\left\{ n_{\text{BE}}\left[(\epsilon - \mu)/k_B T_m\right] - n_{\text{BE}}\left[(\epsilon - \mu_s)/k_B T_e\right]\right\},$$

$$(36)$$

where $\mu$ is the magnonic chemical potential, $T_m$ and $T_e$ are the magnon and electron temperatures, respectively, $g^{\uparrow\downarrow}$ is the (real part of the) spin-mixing conductance (Tserkovnyak et al. 2005; Brataas, Bauer, and Kelly 2006) parameterizing the exchange-coupling strength at the interface, $D(\epsilon)$ is the magnon density of states, $\hbar\omega$ is the magnon gap, and $n_{\text{BE}}(x) \equiv (e^x - 1)^{-1}$ is the Bose-Einstein distribution function. The energy flux is given by a similar expression but with an additional factor of $\epsilon/\hbar$ in the integrand. Together, when linearized with respect to $T_m - T_e$ and $\mu - \mu_s$, these coupled spin and heat flows mimic the Onsager-reciprocal thermoelectric transport of electrons (Mahan 2000).

A likely dominant route to establish such a bias at the interface upon subjecting the bilayer to a normal thermal gradient is the magnonic spin Seebeck effect in the magnetic bulk, which leads to a magnon pile up at the interface (Flebus et al. 2016). For a small thermal gradient, the latter results in the linear-response interfacial spin and heat flow, while for a large enough thermal bias a Bose-Einstein condensation of magnons can ensue near the interface (Tserkovnyak et al. 2016), which is schematically depicted in FIG. 45. This condensation leads to a spontaneously coherent precession of the magnetic order parameter.

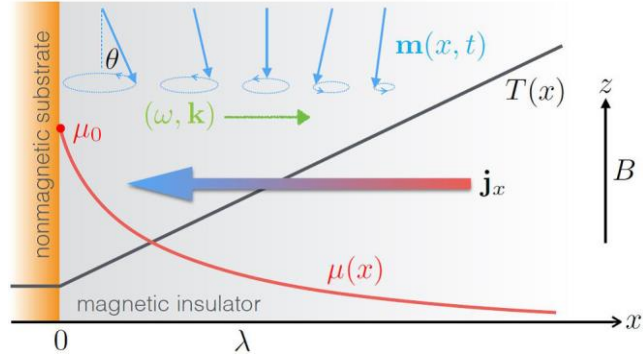

FIG. 45 (Color online). A monodomain ferromagnet with uniform equilibrium spin density pointing in the -z direction (in the presence of a magnetic field B pointing up along z). A positive thermal gradient, $\partial_x T > 0$, induces magnonic flux $\mathbf{j}_x$ toward the interface, where an excess of thermal magnons accumulates over their spin-diffusion length $\lambda$. When the corresponding nonequilibrium interfacial chemical potential $\mu_0$ reaches a critical value (exceeding the magnon gap), the magnetic order undergoes a $H_{opf}$ bifurcation toward a steady precessional state, whose Gilbert damping and radiative spin-wave losses are replenished by the thermal-magnon pumping $\propto \mu_0$. The coherent transverse magnetic dynamics decays away from the interface as $m_x - im_y \propto e^{i(kx-\omega t)}$, where Im$k > 0$. From (Tserkovnyak et al. 2016).

In simple antiferromagnets, the spin Seebeck effect in the bulk vanishes due to the sublattice symmetry (which is also true for the interfacial spin flow, if the interface is fully compensated and effectively respects this symmetry). For example, for the case of the easy-axis bipartite antiferromagnet, according to the axial symmetry, the magnon bands should be double degenerate, corresponding to spin projections $\pm\hbar$ along the easy axis. These bands will carry finite but opposite spin currents in response to a temperature gradient. However, when applying a magnetic field, e.g., along the easy axis, the subband symmetry is lifted (I.C.3) and a spin Seebeck effect emerges, with the spin current polarized along the magnetic field. This has been recently demonstrated experimentally (S. M. Wu et al. 2016; Seki et al. 2015) (FIG. 46) and studied theoretically (Rezende, Rodríguez-Suárez, and Azevedo 2016a).

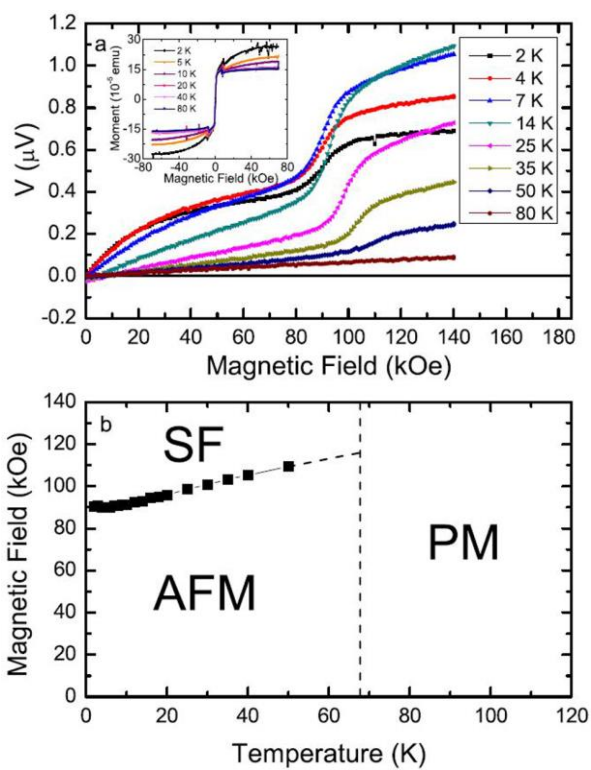

FIG. 46 (Color online). The spin Seebeck voltage (top panel) measured as a function of the magnetic field along the easy axis for a MgF$_2$//MnF$_2$/Pt/MgO/Ti multilayer. The sharp upturn in the signal corresponds to the spin-flop transition. The bottom panel shows the phase diagram of MnF$_2$ according to the existence and location of this upturn. AFM stands for the (collinear) antiferromagnet, SF the spin-flopped (canted) antiferromagnet, and PM the paramagnet (above the Néel temperature). From (S. M. Wu et al. 2016).

A net spin current in a magnetically compensated antiferromagnet can be injected by coherent spin pumping from an adjacent ferromagnet or by spin Hall effect from an adjacent normal metal (see sections II.B and III.C.2.). Both can be demonstrated in the same ferromagnet/antiferromagnet/metal trilayer, where the ferromagnet and the metal can serve interchangeably as either injector or detector of spin currents. A theoretical framework for antiferromagnet-mediated spin currents in such trilayers was established for the coherent coupled low-temperature dynamics (Takei et al. 2015; Khymyn et al. 2016), where the spin current in the antiferromagnet is carried by an evanescent spin-wave mode. At finite temperatures, thermal magnons open an additional channel for spin transport (Rezende, Rodríguez-Suárez, and Azevedo 2016b). The precessing magnetization in the adjacent ferromagnet, for example, would pump differently the oppositely-polarized magnons in the antiferromagnet. Spin conduction through magnetic insulators was first experimentally demonstrated in 2014, simultaneously by Wang et al. (H. Wang et al. 2014) and by Hahn et al. (Hahn et al. 2014) using NiO in YIG/NiO/Pt trilayers (see also TABLE 5). Wang et al. (H. Wang et al. 2014) observed efficient dynamic spin injection from YIG into NiO as a result of strong coupling. Tshitoyan et al (Tshitoyan et al. 2015) went on to demonstrate the direct relation between spin-angular momentum transfer efficiency and ferromagnetic/antiferromagnetic coupling. Wang et al. (H. Wang et al. 2014) also confirmed robust spin propagation in NiO up to 100-nm thick (i. e. a characteristic decay length of around 10 nm, see TABLE 5). This effect was mediated by its antiferromagnetic spin correlations. In contrast, Hahn et al. (Hahn et al. 2014) reported poor interface transparency, of the order of 10%, possibly due to non-optimal interface quality. The propagation length they measured was also not consistent with a picture of mobile antiferromagnetic spin correlations. Unlike Wang et al. (H. Wang et al. 2014) and Hahn et al. (Hahn et al. 2014), Lin et al. (Lin et al. 2016) used the longitudinal spin Seebeck effect to inject magnons into NiO. Since longitudinal spin Seebeck results in dc injection, they were able to confirm the dominant role of thermal magnons in the propagation process. Finally, in 2015 Wang et al. (H. Wang et al. 2015) demonstrated systematic long-distance spin transport in several antiferromagnetic insulators in YIG/antiferromagnet/Pt trilayers (TABLE 5). They found a strong correlation between spin propagation and antiferromagnetic ordering temperatures highlighting the critical role of magnetic correlations. They also found a correlation between spin propagation and extrinsic contributions to YIG damping, thus further highlighting the important role of the ferromagnetic/antiferromagnetic interface for spin transport in antiferromagnetic insulators. Spin conduction has also recently been demonstrated to be maximized, as a function of temperature, in the vicinity of the Néel temperature of NiO and CoO (Qiu et al. 2016) (spin pumping injection), (Lin et al. 2016) (spin Seebeck injection), and NiFeOx (Frangou et al. 2017) (spin pumping injection). Enhanced spin pumping efficiency due to magnetic fluctuations of antiferromagnetic spin absorbers was demonstrated at the same time in NiFe/Cu/IrMn trilayers where the spin current is purely electronic through Cu (Frangou et al. 2016) (see also section I.C.2). This enhanced efficiency is related to the fact that interfacial spin mixing conductance depends on the transverse spin susceptibility of the spin absorber (Ohnuma et al. 2014; K. Chen et al. 2016), which is known to show a maximum around the antiferromagnetic-to-paramagnetic phase transition.

We note that thermal magnons in an antiferromagnet could similarly serve as spin conduits in other heterostructures and transport scenarios. One example related to the earlier discussion in this section could be an antiferromagnet-mediated spin Seebeck effect between a ferromagnet and a normal metal. Interestingly, the spin Nernst effect can generally be expected to be present in antiferromagnets as well, due to spin-orbit interactions, in analogy to spin Hall effect in metals [spin Hall review articles, e.g. (Hoffmann 2013; Sinova et al. 2015)]. This effect is permitted even for the highest-symmetry classes of the most featureless materials. In fact, the spin Nernst effect is expected to be observed equally well in paramagnets and spin liquids. This effect does not rely on the presence of time-reversal symmetry breaking, unlike the thermal Hall effect. Several recent papers (Zyuzin and Kovalev 2016; S. K. Kim et al. 2016; Cheng, Okamoto, and Xiao 2016) demonstrated the spin Nernst effect in models of honeycomb lattice magnets with Dzyaloshinski-Moriya interactions.

With regards to magnetic textures, the interaction of thermally-induced magnon spin currents with magnetic textures, particularly topological solitons, along with the accompanying thermal-gradient-induced entropic forces, have been explored in ferromagnetic insulators (S. K. Kim and Tserkovnyak 2015a; Hinzke and Nowak 2011; Kovalev 2014; Schlickeiser et al. 2014; J. Wang, Lian, and Zhang 2014). The key problem of interest is thermophoresis, i.e., the net soliton drift induced by a thermal gradient. Depending on various details related to the interplay between entropic and thermomagnonic forces, magnetic solitons will move toward either hot or cold regions (S. K. Kim and Tserkovnyak 2015a; Yan, Cao, and Sinova 2015). A comparable effort has yet to be devoted to antiferromagnets, with the exception of the study of Brownian motion of small solitons, which are subject to a thermal gradient, as discussed in section IV.C. At finite temperatures, we may expect the collective spin and heat flows to be carried in a spin superfluid state by the coupled two-fluid dynamics, in analogy with the ferromagnetic case (Flebus et al. 2016).

## B. Superfluid spin transport

At low temperatures, thermal spin waves freeze out and cease to contribute to spin transport. The spin currents can still be transmitted by collective order-parameter dynamics. To understand this, let us return to our basic starting with equations in section II, which we rewrite as

$$\mathbf{m} = \chi_\perp \mathbf{l} \times (s \partial_t \mathbf{l} + \mathbf{l} \times \mathbf{b}) \ ,$$

$$s(\partial_t \mathbf{m} + \alpha \mathbf{l} \times \partial_t \mathbf{l}) = \partial_i (\mathbf{l} \times A \partial_i \mathbf{l}) - \mathbf{l} \times \partial_\mathbf{l} \mathcal{F}_a + \mathbf{b} \times \mathbf{m} + \boldsymbol{\tau_m} \ .$$

$$(37)$$

The second equation is readily interpreted as the continuity equation for the spin density

$\rho_s \equiv s\mathbf{m}$ :

$$\partial_t \boldsymbol{\rho}_s + \partial_i \mathbf{j}_{s,i} = \boldsymbol{\tau}_\alpha + \boldsymbol{\tau}_a + \boldsymbol{\tau}_\mathbf{b} + \boldsymbol{\tau}_\mathbf{m} , \qquad (38)$$

where $\mathbf{j}_{s,i} \equiv -\mathbf{l} \times A \partial_i \mathbf{l}$ is the collective spin current, $\boldsymbol{\tau}_\alpha \equiv -\alpha s \mathbf{l} \times \partial_t \mathbf{l}$ is the damping-like torque describing the leakage of the spin angular momentum into crystalline environment, $\boldsymbol{\tau}_a \equiv -\mathbf{l} \times \partial_\mathbf{l} \mathcal{F}_a$ is the reactive anisotropy torque, $\boldsymbol{\tau}_\mathbf{b} \equiv \mathbf{b} \times \mathbf{m}$ is the classic Larmor torque of the external field $\mathbf{b}$ on the small magnetization $\mathbf{m}$, and $\boldsymbol{\tau}_\mathbf{m}$ is the spin-transfer torque associated with any other (itinerant) degrees of freedom, such as electrons (either in the bulk or at an interface).

In the generic case of an anisotropic antiferromagnet, such as the common easy-axis scenario, injecting a collective spin current at low (subgap) frequencies would decay over a healing length $\sim \sqrt{A/K}$ governed by the anisotropy $K$ (Takei et al. 2015). This contrasts the thermal spin current that decays in magnetic insulators over a temperature-dependent spin-diffusion length of magnons (Cornelissen et al. 2015). At higher frequencies, the spin signals can be transmitted resonantly via the coherent spin-wave modes, which can propagate over much larger distances governed by the Gilbert damping (Takei and Tserkovnyak 2015).

In the more interesting easy-plane case which could be realized, for example, when the $z$-axis anisotropy $K < 0$ or in the absence of anisotropies (i.e., the pure Heisenberg limit) but with a magnetic field $\mathbf{b} \parallel \mathbf{z}$, we have an easy $xy$-plane magnet, mapping at low energies onto the $XY$ model (Sonin 1978; Sonin 2010; König, Bønsager, and MacDonald 2001). In this limit, the spin density is approximately collinear with the $z$ axis, $\boldsymbol{\rho}_s \approx \rho_s \mathbf{z}$, while $\mathbf{l}$ swings predominantly in the $xy$ plane (which we can parameterize by the azimuthal angle $\varphi$). The corresponding equations of motion for these low-energy degrees of freedom follow by projecting Eqs. (37) onto the $z$ axis:

$$\partial_t \varphi = \rho_s / \chi_\perp s^2,$$

$$\partial_t \rho_s + \partial_i j_{s,i} = \tau_\alpha + \tau_m$$

$$(39)$$

Here, $j_{s,i} = -A\partial_i \varphi$, $\tau_\alpha = -\alpha s \partial_t \varphi$ and $\tau_m$ is the $z$-axis projection of the spin-transfer torque. We are still allowing $\tau_m$ to be of a general form, either of bulk or interfacial nature, with the latter relevant for establishing the appropriate boundary conditions for bulk dynamics. These coupled equations reflect a damped spin superfluid hydrodynamics (Takei and Tserkovnyak 2014; Takei et al. 2014), associated with the canonically conjugate (coordinate/momentum) pair of variables ($\varphi, \rho_s$), with the Hamiltonian density

$$\mathcal{H} = \frac{\rho_s^2}{2\chi_\perp s^2} + \frac{A}{2}(\partial_i \varphi)^2,$$

$$(40)$$

and the Rayleigh function

$$\mathcal{R} = \frac{\alpha s}{2}(\partial_t \varphi)^2 .$$

$$(41)$$

The destruction of the associated (topologically stable) superfluid spin currents by thermal and quantum phase slips has been studied by Kim et al (S. K. Kim, Takei, and Tserkovnyak 2016) and (S. K. Kim and Tserkovnyak 2016), respectively. Quantum phase slips, which dominate at low temperatures, are particularly interesting in the case of easy-plane antiferromagnets, and exhibit a topological character depending on whether the constituent spins are integer or half-odd-integer valued (affecting profoundly the superfluid-insulator quantum phase transition). While such superfluid spin flows were believed to be severely compromised by even a small parasitic anisotropy within the easy plane (Sonin 1978; König, Bønsager, and MacDonald 2001; W. Kohn and Sherrington 1970), it gets effectively restored at finite temperatures by topological spin transport carried by Brownian diffusion of chiral domain walls (S. K. Kim, Takei, and Tserkovnyak 2015). The relevant temperature scale for this is set by the total energy of an individual domain wall, $A\sqrt{A\kappa}$, where $\kappa \ll |K|$ is the parasitic anisotropy and $A$ is the geometric cross section of the wire.

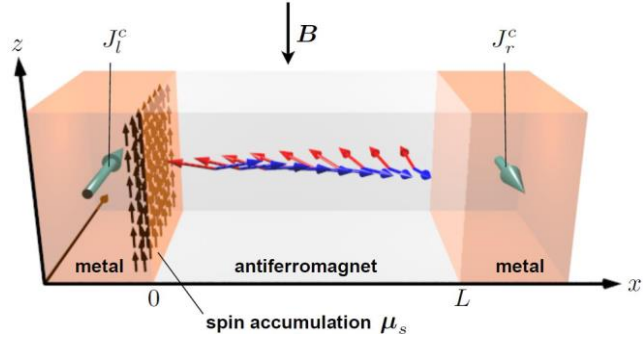

FIG. 47 (Color online). Negative electron drag between two metals mediated by an antiferromagnetic spin superfluid. Left metal injects a spin current, by producing the interfacial torque $\tau_m = g^{\uparrow\downarrow}\mu_s/(4\pi)$, which then propagates along the $x$ axis according to Eq. (36). $\mu_s$ is in the first place induced by the spin Hall effect associated with the electrical current $J_l^c$ flowing in the $y$ direction. The spin-current detection by the right metal proceeds in the Onsager-reciprocal fashion: the Néel dynamics pump spin current [associated with the torque $\tau_m = -g^{\uparrow\downarrow}\hbar\partial_t\varphi/(4\pi)$ ], which finally produces a detectable voltage by the inverse spin Hall effect. From (Takei et al. 2014).

Injection and detection of such superfluid spin currents in easy-plane antiferromagnets has been proposed by Takei et al (Takei et al. 2014) using the direct and inverse spin Hall effects at interfaces with heavy metals. This is enabled by a spin-mixing conductance $g^{\uparrow\downarrow}$ associated with the Néel order, which is finite even in the case of a compensated magnetization at the interface (Takei et al. 2014; Jia et al. 2011; Cheng and Niu 2014), thus allowing for an electronically induced interfacial spin torque $\tau_m$. In the spirit of our previous discussion [see section II, regarding the smallness of $\mathbf{m}$ ], it is sufficient here to retain only the spin torques and pumping associated with the rigid dynamics of the Néel order $\mathbf{l}$ alone. FIG. 47 shows a proposed detection of a spin superflow via a long-ranged negative electron drag that it mediates. Refined detection schemes have been discussed in (Takei and Tserkovnyak 2015). Similar setups could be used also for experimental detection of thermal and quantum phase slips, as proposed in (S. K. Kim, Takei, and Tserkovnyak 2016; S. K. Kim and Tserkovnyak 2016).

To conclude this section, we note that, in contrast to what is observed in ferromagnets, spin superfluidity in antiferromagnets is unaffected by detrimental dipolar interactions (Skarsvåg, Holmqvist, and Brataas 2015). The wealth of insulating antiferromagnetic materials, particularly of the easy-plane character, bodes well for the search for a useful spin-superfluid medium. Furthermore, using the magnetic-field-induced antiferromagnetic spin-flop transition offers a practical tool to mitigate parasitic easy-axis anisotropies (Qaiumzadeh et al. 2016).

### C.    Thermally-induced domain wall motion

The interplay between spin waves, antiferromagnetic domain walls and temperature gradient has been investigated in various ways. Tveten et al. (Tveten, Qaiumzadeh, and Brataas 2014) and Kim et al. (S. K. Kim, Tserkovnyak, and Tchernyshyov 2014) addressed the motion of a uniaxial antiferromagnetic wall submitted to a magnon flow. They identified three main mechanisms responsible for driving domain wall motion: transfer of angular momentum ($=2\hbar$), reflection against the wall and redshift of the magnon frequency corresponding to a net transfer of momentum (S. K. Kim, Tserkovnyak, and Tchernyshyov 2014). When circularly polarized, the transfer of angular momentum causes the wall to precess *away* from the magnon source, resulting in reflection and redshift at high and low magnon intensities, respectively. With linear polarized magnons, no spin transfer occurs (both circularly polarized magnon states are equally populated) and instead a viscous force associated with magnon damping drags the wall toward the magnon source (Tveten, Qaiumzadeh, and Brataas 2014).

Using the collective coordinate approach presented in section II, Kim et al. (S K Kim, Tchernyshyov, and Tserkovnyak 2015) demonstrated that domain-wall and other stochastic solitonic dynamics under the influence of thermal random forces exhibit Brownian thermophoresis. This effect tends to induce a mean drift toward colder regions, as illustrated in FIG. 48.

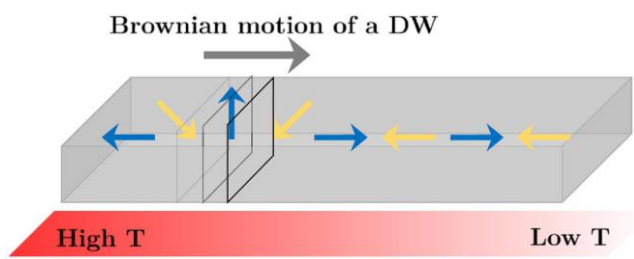

FIG. 48 (Color online). Stochastic thermal forces associated with an inhomogeneous temperature profile push an antiferromagnetic domain wall to a colder region. From (S. K. Kim, Tchernyshyov, and Tserkovnyak 2015).

Indeed, averaging Eq. (30) over an ensemble of solitons undergoing stochastic Brownian motion leads to the Fokker-Planck equation in the form of a continuity relation:

$$\partial_t \varrho + \nabla \cdot \mathbf{j} = 0, \tag{42}$$

where $\varrho$ is the soliton density and $\mathbf{j}$ their net flux. The latter is given by (S. K. Kim, Tchernyshyov, and Tserkovnyak 2015)

$$\mathbf{j} = \mu \varrho (\mathbf{F} - k_B \nabla T) - D \nabla \varrho. \tag{43}$$

The average velocity of (a uniform ensemble of) skyrmions subjected to a constant force and thermal gradient is thus

$$\mathbf{v} = \mu (\mathbf{F} - k_B \nabla T), \tag{44}$$

which shows the tendency to drift toward the colder regions where the solitons are less agitated and their Brownian motion freezes. This thermophoretic torque competes with thermomagnonic torques ($\sim \mathbf{F}$), whose entropic component pushes the localized magnetic textures toward hotter regions (S. K. Kim and Tserkovnyak 2015a). In the special case of a one-dimensional domain-wall motion in an easy-axis antiferromagnet (S. K. Kim, Tchernyshyov, and Tserkovnyak 2015), the mobility is given by $\mu = \lambda_{\text{dw}} / 2$, where $\lambda_{\text{dw}} \equiv \sqrt{A/K}$ is the domain-wall width and $s$ is the (saturated) spin density per unit length. The dissipation of energy associated with a steady rigid domain-wall motion is $P = \dot{R} \cdot F = \dot{R}^2 / \mu$. Since for the ferromagnetic dynamics, according to the Landau-Lifshitz-Gilbert phenomenology, the Rayleigh function associated with the (directional) magnetic order parameter $\mathbf{m}$ (constrained by $|\mathbf{m}| \equiv 1$) coincides with that of the antiferromagnetic $\mathbf{l}$ (with $|\mathbf{l}| \equiv 1$), the above expression for the mobility remains (with $\alpha$ being the usual ferromagnetic Gilbert damping). When the physical cross section A of the magnetic wire increases and the mobility $\mu \propto A^{-1}$ decreases, eventually thermomagnonic torques (including spin transfer and entropic forces) start dominating, pushing domain walls generally toward the hotter side (S. K. Kim and Tserkovnyak 2015b), in the presence of a thermal gradient $\nabla T$.

Stochastic motion of topological solitons is qualitatively different (and faster) in antiferromagnets compared to ferromagnets in two dimensions, due to the absence of the gyrotropic Magnus force in the former, as discussed in section II.A.3 (Schütte et al. 2014; S K Kim, Tchernyshyov, and Tserkovnyak 2015; Barker and Tretiakov 2016). Finally, Selzer et al. (Selzer et al. 2016) investigated the motion of antiferromagnetic domain walls in response to thermal gradients using an atomistic model. Like Shiino et al. (Shiino et al. 2016) and Gomonay et al. (O. Gomonay, Jungwirth, and Sinova 2016), they obtained massless motion of the domain wall due to the absence of azimuthal tilting during the motion; they also confirmed the hypothesis proposed by Kim et al. (S K Kim, Tchernyshyov, and Tserkovnyak 2015) and Tveten et al. (Tveten, Qaiumzadeh, and Brataas 2014).

## V.    CONCLUSION AND PERSPECTIVES

Originally considered as mere curiosities and up till recently limited to a passive role in spintronic devices, antiferromagnetic materials could represent the future of spintronic applications. The recent experimental achievements (spin pumping, spin-orbit torque, anisotropic magnetoresistance, anomalous Hall effect, spin Seebeck effect etc.) bear the promises for future outstanding developments. The wide diversity under which antiferromagnetism appears in nature (from metals to insulators, not mentioning the vast richness of magnetic textures) offers a fascinating playground for physicists, materials scientists and engineers.

Nonetheless, significant challenges still need to be thoroughly addressed before antiferromagnets can become active elements of real spintronic devices. First and foremost, collecting systematic and reproducible data represents a major difficulty due to significant sample-to-sample variability and uncontrolled size effects. Since the magnetic texture of antiferromagnets is greatly sensitive to the layer thickness, temperature and growth conditions, the impact of the magnetic order (compensated versus uncompensated interfaces, collinear versus non-collinear texture etc.) on the spin transport properties remains to be accurately understood. To this end, the development of new characterization techniques, ranging from optical imaging to TeraHertz methods (Urs et al. 2016) will be a seminal milestone in order to overcome the difficulties faced by generations of researchers working on antiferromagnetic materials.

A particularly thrilling aspect of antiferromagnetic materials is their wide variety in nature. The growth of novel materials with high spin-polarization and low damping will also be an important step (Hu 2012; Sahoo et al. 2016). A recent striking and stimulating example of how two fields of condensed matter physics may envision a common future is the prediction that Dirac quasiparticles (with the example of Dirac quasiparticles in the CuMnAs semimetal (P. Tang et al. 2016)) can be controlled by spin-orbit torque reorientation of the Néel vector in an antiferromagnet (Šmejkal et al. 2017). In a more long-term vision, we can also imagine to capitalize on the broad knowledge acquired by decades of fundamental research conducted on strongly correlated antiferromagnets (Georges, de' Medici, and Mravlje 2013; Behrmann and Lechermann 2015) and high-$T_c$ superconductors (Scalapino 2012), where the intermingling between spin transport and antiferromagnetic order requires deeper investigations. Emerging materials with strong spin-orbit coupling such as antiferromagnetic topological insulators (Mong, Essin, and Moore 2010; Bansil, Lin, and Das 2016) or Weyl semimetals (Wan et al. 2011) could also be an inspiring research direction in a near future (Šmejkal, Jungwirth, and Sinova 2017).

## ACKNOWLEDGMENTS

We would like to acknowledge all the colleagues who contributed by publishing the results shown in this review. We thank all our colleagues from our respective laboratories and within the scientific community for stimulating discussions and for motivating our curiosity and interest in this topic, and for useful comments on the manuscript, in particular: T. Jungwirth, J. Sinova, J. Wunderlich, O. Gomonay, X. Marti, A. Hoffmann, S. Maekawa, A. Brataas, W. E. Bailey, M. Chshiev, H. Béa, A. Schuhl, G. Gaudin, M. Miron, O. Boulle, S. Auffret, O. Klein, D. Givord, A. Mougin, A. Bataille, L. Ranno, M. Viret, H. Saidaoui, C. Akosa, P. Merodio, L. Frangou, G. Forestier, O. Gladii, P. Wadley, F. Lechermann, J. Linder, R. Cheng, W. Lin, A. Sekine, and J. Heremans. We thank M. Gallagher-Gambarelli for critical reading of the manuscript. V. B. acknowledges the financial support of the French National Agency for Research [Grant Number ANR-15-CE24-0015-01]. A. M., V. B., and M. T. acknowledge the financial support of the King Abdullah University of Science and Technology (KAUST) through the Office of Sponsored Research (OSR) [Grant Number OSR-2015-CRG4-2626]. M. T. acknowledges the financial support of C-SPIN, one of six centers of STARnet, a Semiconductor Research Corporation program, sponsored by MARCO and DARPA, and by the NSF [Grant Number DMR-1207577]. T. M. and T. O. were supported by the Japan Society for the Promotion of Science KAKENHI [Grant Numbers 26870300 and 15H05702], and the Grant-in-Aid for Scientific Research on Innovative Area, "Nano Spin Conversion Science" [Grant Number 26103002]. Y. T. acknowledges the financial support of the ARO [Contract Number 911NF-14-1-0016], and of the NSF-funded MRSEC [Grant Number DMR-1420451].